\documentclass[traditabstract]{aa}

\usepackage{graphicx}
\usepackage{natbib}
\usepackage{ulem}
\usepackage{siunitx}

\usepackage{xcolor}

\newcommand{\hp}[0]{\mbox{HAT-P-32}}
\newcommand{\hpb}[0]{\mbox{HAT-P-32}\,b}
\newcommand{\hei}[0]{\ion{He}{i}\,$\lambda$10833}
\newcommand{\heiair}[0]{\ion{He}{i}\,$\lambda$10830}
\newcommand{\heion}[0]{\ion{He}{i}}
\newcommand{\ha}[0]{\mbox{H$\alpha$}}
\newcommand{\carm}[0]{CARMENES}
\newcommand{\mf}[0]{\texttt{molecfit}}
\newcommand{\cairt}[0]{\ion{Ca}{ii}~IRT}
\renewcommand{\ha}[0]{H$\alpha$}
\newcommand{\lya}[0]{Lyman\,$\alpha$}
\newcommand{\ms}[0]{m\,s$^{-1}$}
\newcommand{\kms}[0]{km\,s$^{-1}$}
\newcommand{\ergcmcms}[0]{erg\,cm$^{-2}$\,s$^{-1}$}
\newcommand{\ergs}[0]{erg\,s$^{-1}$}
\newcommand{\gs}[0]{g\,s$^{-1}$}

\newcommand{\xmm}[0]{\textit{XMM-Newton}}
\newcommand{\stau}[0]{\sigma_{\tau}}

\newcommand{\rpp}[0]{$R_{\rm p, \bot}$}

\newcommand{\rpa}[0]{$R_{\rm p, \parallel}$}

\newcommand{\mlp}[1]{\textcolor{red}{#1}}
\renewcommand{\mlp}[1]{}

\def\hes{He~(1$^{1}$S)}
\def\het{He~(2$^{3}$S)}
\def\mlr{$\dot M$}
\def\rp{$R_{\rm P}$}

\usepackage{txfonts}

\begin{document}

\title{\ha\ and \ion{He}{I} absorption in \hpb\ observed with \carm} 
\subtitle{Detection of Roche lobe overflow and mass loss}
\titlerunning{Transmission spectroscopy of \hpb}
\authorrunning{S. Czesla et al.}

\author{S. Czesla \inst{\ref{instHS},\ref{instTLS}},
    M. Lampón \inst{\ref{instIAA}},
    J.~Sanz-Forcada \inst{\ref{instCAB}},
    A. García Muñoz \inst{\ref{instAIM}},
    M. López-Puertas \inst{\ref{instIAA}},
    L.~Nortmann \inst{\ref{instIAG}},
    D.~Yan \inst{\ref{instYO},\ref{instSoASS}},
    E.~Nagel \inst{\ref{instHS}, \ref{instTLS}},
    F.~Yan \inst{\ref{instIAG}},
    J.~H.~M.~M.~Schmitt \inst{\ref{instHS}}
   \and J.~Aceituno\inst{\ref{instIAA}, \ref{instCAHA}}
  \and P.~J.~Amado\inst{\ref{instIAA}}
  \and J.~A.~Caballero\inst{\ref{instCAB}}
  \and N.~Casasayas-Barris\inst{\ref{instIAC}, \ref{instLaguna}}
  \and Th.~Henning\inst{\ref{instMPIA}}
  \and S.~Khalafinejad\inst{\ref{instLSW}}
  \and K.~Molaverdikhani\inst{\ref{instMPIA},\ref{instLSW},\ref{instLMU}}
  \and D.~Montes\inst{\ref{instUCM}}
  \and E.~Pallé\inst{\ref{instIAC}, \ref{instLaguna}}
  \and A.~Reiners\inst{\ref{instIAG}}
  \and P.C.~Schneider\inst{\ref{instHS}}
  \and I.~Ribas\inst{\ref{instICE},\ref{instIEEC}}
  \and A.~Quirrenbach\inst{\ref{instLSW}}
  \and M.~R.~Zapatero Osorio\inst{\ref{instCAB2}}
  \and M.~Zechmeister\inst{\ref{instIAG}}
  }      

\institute{Hamburger Sternwarte, Universit\"at Hamburg, Gojenbergsweg 112, D-21029 Hamburg, Germany\\
  \email{sczesla@hs.uni-hamburg.de}\label{instHS}          \and
        Thüringer Landessternwarte Tautenburg, Sternwarte 5, D-07778 Tautenburg, Germany \label{instTLS} 
        \and
        Instituto de Astrofísica de Andalucía (IAA-CSIC), Glorieta de la Astronomía s/n, 18008 Granada, Spain\label{instIAA}
        \and
        Centro de Astrobiología (CSIC-INTA), ESAC, Camino bajo del castillo s/n, 28692 Villanueva de la Cañada,
        Madrid, Spain\label{instCAB}
        \and
        AIM, CEA, CNRS, Universit\'e Paris-Saclay, Universit\'e de Paris, F-91191 Gif-sur-Yvette, France
        \label{instAIM} 
        \and
        Institut für Astrophysik, Friedrich-Hund-Platz 1, D-37077 Göttingen, Germany\label{instIAG} 
        \and
        Yunnan Observatories, Chinese Academy of Sciences, P.O. Box 110, Kunming 650011, People’s Republic of China \label{instYO} 
        \and
        School of Astronomy and Space Science, University of Chinese Academy of Sciences, Beijing, People’s Republic of China \label{instSoASS}
        \and
        Centro Astronómico Hispano Alemán, Sierra de los Filabres, 04550 Gérgal, Almería, Spain\label{instCAHA}
        \and
        Instituto de Astrof\'{\i}sica de Canarias, c/ V\'{\i}a L\'actea s/n, E-38205 La Laguna,
        Tenerife, Spain\label{instIAC}
        \and
        Departamento de Astrof\'{\i}sica, Universidad de La Laguna, E-38206 Tenerife, Spain\label{instLaguna}
        \and
        Max-Planck-Institut für Astronomie, Königstuhl 17, D-69117 Heidelberg, Germany\label{instMPIA}
        \and
        Landessternwarte, Zentrum für Astronomie der Universität Heidelberg, Königstuhl 12, D-69117 Heidelberg,
        Germany\label{instLSW}
        \and
        Universitäts-Sternwarte, Ludwig-Maximilians-Universität München, Scheinerstrasse 1, D-81679 München, Germany
        \label{instLMU}  
        \and
        Facultad de Ciencias F\'{\i}sicas, Departamento de F\'{\i}sica de la Tierra y Astrof\'{\i}sica; IPARCOS-UCM (Instituto 
        de F\'{\i}sica de Part\'{\i}culas y del Cosmos de la UCM), Universidad Complutense de Madrid,
        E-28040 Madrid, Spain\label{instUCM}
                \and
        Institut de Ciències de l’Espai (ICE, CSIC), Campus UAB, c/ de Can Magrans s/n,
        08193 Bellaterra, Barcelona, Spain \label{instICE}
        \and
        Institut d’Estudis Espacials de Catalunya (IEEC), 08034 Barcelona, Spain
        \label{instIEEC}
        \and
        Centro de Astrobiología (CSIC-INTA), Carretera de Ajalvir km 4, E-28850 Torrejón de Ardoz, Madrid, Spain
        \label{instCAB2}
               }

\date{}

\abstract
{
We analyze two high-resolution spectral transit time series of the
hot Jupiter \hpb\ obtained with the CARMENES spectrograph. Our new
\xmm\ X-ray observations of the system show that the fast-rotating F-type
host star exhibits a high X-ray luminosity
of $2.3\times 10^{29}$~\ergs\ 
(5--100~\AA), corresponding to a flux of $6.9\times 10^4$~\ergcmcms\ at the planetary orbit,
which results in
an energy-limited escape estimate of about $10^{13}$~\gs\ for the planetary mass-loss rate.
The spectral time series show significant, time-dependent absorption in the \ha\ and \hei\ triplet lines
with maximum depths of about $3.3$\,\% and $5.3$\,\%.
The mid-transit absorption signals in the \ha\ and \hei\ lines are consistent with results from
one-dimensional hydrodynamic modeling, which also yields mass-loss rates on the order of $10^{13}$~\gs.
We observe an early ingress of a redshifted component of the transmission signal, which
extends into a redshifted absorption component, persisting until about the middle of the optical transit.
While a super-rotating wind can explain redshifted ingress absorption, we find that an up-orbit stream, transporting
planetary mass in the direction of the star, also provides a plausible explanation for the pre-transit signal.
This makes \hp\ a benchmark system for exploring atmospheric dynamics via transmission
spectroscopy.
}


\maketitle

\section{Introduction}
\label{sec:intro}
Transmission spectroscopy is among the most
successful techniques for studying the atmospheres
of extrasolar planets. In particular, strong
atomic transitions such as the \ion{Na}{I} D~lines provide large cross sections, which favor observation in the form of transmission signals \citep[e.g.,][]{Seager2000}.
To date, transitions from various
atomic and molecular species have
been studied and detected in a range of
planetary atmospheres such as the \ion{Na}{I} D~lines \citep{Redfield2008, Snellen2008, Khalafinejad2017, CasasayasBarris2017}, \ion{Ca}{II} \citep[][]{Yan2019, Turner2020}, CO \citep[][]{Snellen2010}, water \citep[][]{Tinetti2007, Brogi2018, AlonsoFloriano2019a}, or \ion{Mg}{ii} and \ion{Fe}{ii} \citep[][]{Sing2019}.
Because the planetary
atmosphere is observed against the background of the stellar disk, stellar effects, for example, caused by
activity \citep{Barnes2016} or the limb-angle-dependent 
spectral properties of the disk,
are common complications that need to be properly taken into account in transmission spectroscopy
\citep[][]{Czesla2015, Yan2015, CasasayasBarris2020}.

Hydrogen is
thought to be the dominant species in the atmospheres of hot Jupiters and there are several reports on detections in
the \lya\ line \citep[e.g.,][]{Vidal2003, Lecavelier2010, Ehrenreich2015}, which were
complemented by detections of the optical \ha\ line, also accessible with ground-based instrumentation, later
on \citep[e.g.,][]{Yan2018}. There are reports of \ha\ absorption for the benchmark
system HD~189733 \citep[][]{Jensen2012, Cauley2015, Cauley2016}, but the observational
evidence is controversial \citep[][]{Barnes2016, Cauley2017, Kohl2018}. Further reports of 
planetary H$\alpha$ line absorption have been made based on observations of
the hot Jupiters WASP-12\,b, KELT-9\,b, KELT-20\,b, WASP-52\,b, and WASP-121\,b (see Table~\ref{tab:contrastcomp}
for details and references). A tentative detection of H$\alpha$ line absorption exists
for WASP-76\,b but remains inconclusive \citep{Tabernero2021}.

A relatively late but productive addition to the collection of observed planetary 
transmission signals are the \hei\footnote{We use vacuum wavelengths in units of Angstrom throughout the paper. In air, the
lines are often referred to as the \heiair\ triplet.} triplet lines.
These lines are well-known indicators of
activity in the Sun and other stars \citep[e.g.,][]{Zarro1986, Sanz2008, Fuhrmeister2019}.
Because they originate from a metastable helium level
with an excitation energy of about $20$~eV, their formation
is intimately related to stellar extreme-ultraviolet (EUV) radiation.
The EUV photons with wavelengths $\lesssim 504$~\AA\ first
ionize neutral helium, and the subsequent recombination cascade populates the metastable level, which produces
the absorption. This
so-called photoionization--recombination (PR) process is thought to
work in the outer atmospheres of stars and planets.
Consequently,
the \hei\ lines had long been identified as promising tracers of
the outer planetary atmosphere \citep[][]{Seager2000}.

The first observational detections of \hei\ absorption
from planetary atmospheres were obtained
for WASP-107\,b with space-based instrumentation \citep[][]{Spake2018}, later confirmed and refined
by high-resolution ground-based spectroscopy \citep{Allart2019}, and for WASP-69\,b and HAT-P-11\,b using the
ground-based CARMENES spectrograph\footnote{Calar Alto high-Resolution search for M dwarfs with Exoearths with NIR and optical Echelle Spectrographs \citep[\url{https://carmenes.caha.es/},][]{Quirrenbach2018}} \citep[][]{Nortmann2018, Allart2018}.
 Meanwhile, highly resolved \hei\ transmission signals have been studied in 
other planetary atmospheres,
including those of HD~209458\,b, HD~189733\,b, GJ~3470\,b, and WASP-107\,b \citep[][]{Allart2018, Salz2018, AlonsoFloriano2019, Kirk2020, Ninan2020, Palle2020}.

Stellar EUV and X-ray emission, absorbed high in the planetary atmosphere, is thought to be the
main source of energy, driving planetary winds and, thus,
mass loss in planets orbiting low-mass stars \citep[e.g.,][]{Watson1981, GarciaMunoz2007, Sanz2011, Salz2016, Salz2016b}. 
In planets orbiting hotter stars, it has been shown 
that a complementary process
based on energy extracted from the near-ultraviolet (NUV) radiation field by hydrogen atoms in the lower state of
the Balmer series
can be the dominant route of energy deposition into the planetary atmosphere \citep[][]{GarciaMunoz2019}.
This establishes an intimate relation between
the stellar radiation, and in particular
stellar activity levels, and the planetary atmosphere,
which is thought to impact the overall planet population
\citep[e.g.,][]{Fulton2017, Fulton2018}.

\section{The \hp\ system}

The discovery of \hpb\ was reported by \citet{Hartman2011}, based on photometric transit light curves
from HATNet and
radial velocity (RV) measurements obtained with
HIRES \citep[High-Resolution Echelle Spectrometer, ][]{Vogt1994}.
The planet is a highly inflated hot Jupiter transiting a late-F-type host
star with a period of $2.15$~d \citep{Hartman2011}.
The pertinent parameters of the system are listed in Table~\ref{tab:pars}.
\citet{Hartman2011} considered both a circular and an eccentric orbital solution.
According to both fits, the \hp\ system
shows a large RV jitter of around $80$~\ms. \citet{Hartman2011} conclude that the eccentricity
is poorly constrained and that the data are consistent with the circular solution, which is also consistent
with the $3$~Myr circularization timescale estimated by the authors. The secondary eclipse timing, later
reported by \citet{Zhao2014} and \citet{Nikolov2018}, strongly backs the circular orbit solution.

\citet{Seeliger2014} and \citet{Wang2019} present follow-up photometry of \hpb, and
\citet{Albrecht2012} and \citet{Knutson2014} obtained follow-up RV measurements also with HIRES.
As \citet{Albrecht2012} studied the Rossiter-McLaughlin effect in \hp,
their data are concentrated around the transit.
\citet{Knutson2014} give values for an eccentric orbital solution and
report a significant
long-term acceleration of $-0.097 \pm 0.023$~\ms\,d$^{-1}$.
The host star
\hp\ has a resolved, wide M1.5V-type companion, \hp\,B
at an angular separation of $2.9''$ \citep[$\approx 850$~au projected distance,][]{Zhao2014}.
However, this stellar companion cannot explain the acceleration, which 
may rather be attributable to a hypothetical long-period
low-mass companion \hp\,c \citep[][]{Zhao2014}.

Based on a differential spectrophotometric analysis carried out in the optical regime (520 to 930~nm),
\citet{Gibson2013} reported a featureless planetary transmission spectrum, indicative of
a gray cloud absorber in the atmosphere or low elemental abundances.    
Optical photometry presented by \citet{TregloanReed2018} favors the presence of a cloud deck on
\hpb, comprising both a gray absorber and a Rayleigh scattering component.
By means of spectrophotometry, \citet{Mallonn2016} found a flat transmission spectrum with a possible
Rayleigh slope. Also, \citet{Mallonn2016b} report a flat spectrum, which is indicative of
a cloud deck and evidence for atmospheric Rayleigh scattering, and report no indication for gas-phase TiO. Marginal
evidence was given for excess absorption in the K lines, and no indication of excess absorption was found in the
Na or H$\alpha$ lines.  
Consistently, \citet{Nortmann2016} found a
flat transmission spectrum, and 
NUV photometry presented by \citet{Mallonn2017} favors the presence of a Rayleigh
scattering component in the atmosphere of \hpb.

Space-based infrared spectroscopy with \textit{HST}/WFC3 analyzed by \citet{Damino2017} favors the
presence of water vapor and probably a cloud deck. 
Spitzer photometry of the secondary eclipse presented by \citet{Zhao2014} supports a planetary dayside
atmosphere with a high-altitude absorber and temperature inversion as well as inefficient
day-to-night side heat redistribution. Also the
secondary eclipse observations with \textit{HST} WFC3 presented by \citet{Nikolov2018} indicate either an isothermal
or inverted temperature structure in the dayside atmosphere of \hpb. In particular,
dayside atmospheric structures with temperatures decreasing with height can be ruled out with high
confidence. \citet{Mallonn2019} derive an upper limit of $0.2$ for the $z'$-band geometric albedo
of \hpb, which is consistent with a temperature inversion.
Clearly, the atmosphere of \hpb\ has come under considerable scrutiny already.

\begin{table}
\centering
\caption{System parameters of \hp
\label{tab:pars}}
\begin{tabular}{l l l} \hline\hline
Parameter & Value & Ref\\ \hline
\multicolumn{3}{c}{Stellar parameters} \\
$T_{\rm eff}$ [K] & $6269 \pm 64$ & Z\\
$\log\left(g_{\star}\right)$ & $4.33\pm 0.01$ & H \\
$[$Fe/H$]$ & $-0.04 \pm 0.08$ & H\\
$v\sin{i}$ [km\,s$^{-1}$] & $20.7 \pm 0.5$ & H\\
$M_{\star}$ [$M_{\odot}$] & $1.160 \pm 0.041$ & H\\
$R_{\star}$ [$R_{\odot}$] & $1.219 \pm 0.016$ & H\\
$d$ [pc] & $292\pm 5$ & G \\
\hline
\multicolumn{3}{c}{Orbital and transit parameters} \\
$T_0$ [BJD$_{\rm TDB}$] & $2455867.402743 \pm 0.000049$ & W \\
$P_{\rm orb}$ [d] & $2.15000820 \pm 0.00000013$ & W \\
$i_{\rm orb}$ & $88.9 \pm 0.4$ & H \\
$a$ [au] & $0.0343 \pm 0.0004$ & H\\
$T_{14}$ [d] & $0.1295 \pm 0.0003$ & H \\
$T_{23}$ [d] & $0.0172 \pm 0.0002$ & H \\ \hline
\multicolumn{3}{c}{Planetary parameters} \\
\rpp\ [$R_{\star}$] & $0.1508 \pm 0.0004$ & H \\
\rpp\ [$R_{\rm Jup}$] & $1.789 \pm 0.025$ & H\\
$M_p$ [$M_{\rm Jup}$] & $0.585\pm 0.031$ & T\\
$\rho_p$ [g cm$^{-3}$] & $0.135\pm 0.016$ & T \\
$K_{\star}$ [m\,s$^{-1}$] & $83.4 \pm 4$ & T \\
$q=M_p\, M_{\star}^{-1}$ & $(4.81 \pm 0.31)\times 10^{-4}$ & T \\
\hline
\end{tabular}
\tablebib{G:~\citet{Gaia2018};
H:~\citet{Hartman2011} (values correspond to circular solution); W:~\citet{Wang2019};
T:~This work; Z~\citet{Zhao2014}.
}
\end{table}

\subsection{Stellar parameters and planetary orbit}
\label{sec:orbit}

\begin{figure}[ht]
    \includegraphics[width=0.49\textwidth]{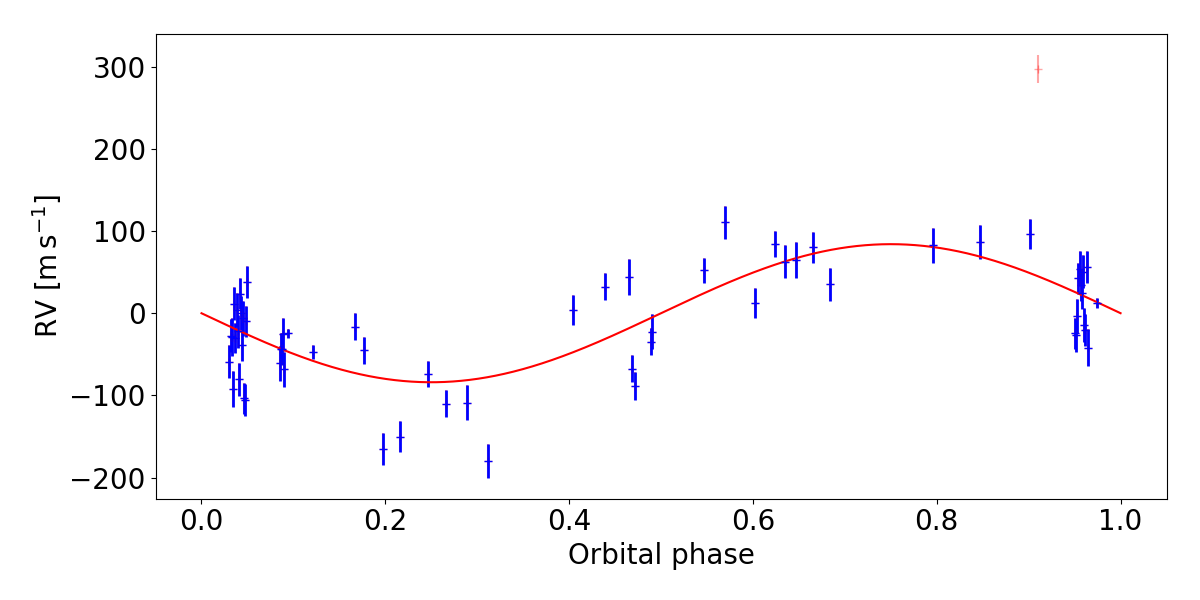}
    \caption{Phase-folded RV data (blue) obtained with HIRES along with best-fit circular orbit model (solid red).
    One outlier is marked in red.
    \label{fig:rv}}
\end{figure}

The shape of the planetary orbit is crucial to interpret RV shifts seen in transmission
spectroscopy because the orbit defines the planetary rest frame. 
We reanalyzed the RV data here under the premise
of a circular orbit solution. In Fig.~\ref{fig:rv}, we
show the phase-folded RV data presented by \citet{Hartman2011}, \citet{Albrecht2012}, and \citet{Knutson2014}
along with the minimum-$\chi^2$ circular orbit solution. Measurements taken during the optical transit are disregarded
because of the Rossiter-McLaughlin effect, and
one data point is neglected as an outlier. In our modeling,
we adopted the transit timing from \citet{Wang2019}, who considered the most comprehensive photometric data set.
The actual orbit solution comprises a free parameter for the RV zero point, which is treated as a nuisance parameter,
and the acceleration of the \hp\ system. For the latter we obtained a value of
$-0.095\,(-0.101, -0.089)$~m\,s$^{-1}$\,d$^{-1}$, consistent with the value
derived by \citet{Knutson2014} (ranges in parentheses indicate $68$\,\% credibility intervals).
For the semi-amplitude of the stellar orbital velocity, $K_{\star}$, we obtained a value
of $83.4\,(79.9, 87.9)$~\ms. This value is somewhat smaller than the values of
$122.8 \pm 23.2$~\ms\ and $110\pm 16$~\ms\ determined by
\citet{Hartman2011} and \citet{Zhao2014}, but consistent with the value
of $77 \pm 26$~\ms\ derived by \citet{Albrecht2012} from their
data alone. The RMS of the residuals of our fit is $42$~\ms.

Our
determination of $K_{\star}$ yields a planet mass of $0.585 \pm 0.031$~M$_{\rm Jup}$, which is about $30$\,\% smaller than
the value reported by \citet{Hartman2011} or \citet{Zhao2014}. However,
the main contribution to the uncertainty of the semimajor axis of the planetary orbit and, therefore,
the planet orbital velocity comes from the stellar mass estimate.

\subsection{Roche geometry}
\label{sec:Roche}

\begin{figure}
    \includegraphics[trim={0 1.5cm 0 3cm}, clip, width=0.49\textwidth]{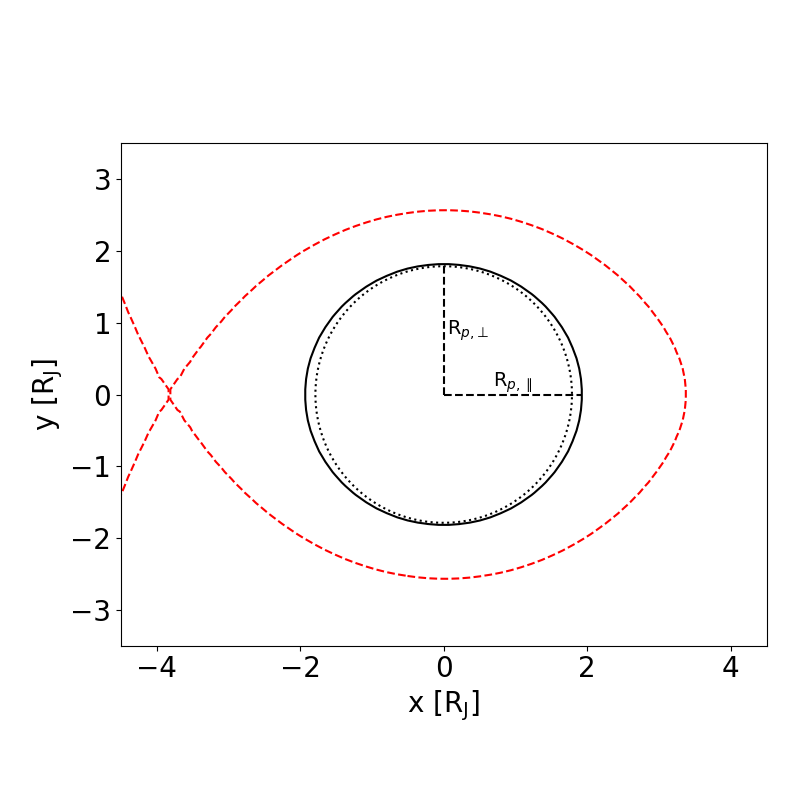}
    \caption{Roche geometry of the planetary body (solid black) and Roche lobe (red dashed) of \hpb. The dotted black line indicates a
    sphere with the perpendicular radius. The star is
    to the left.
    \label{fig:roche}}
\end{figure}

The Roche potential depends on the positions and the mass ratio, $q$, of
the bodies \citep[e.g.,][]{Hilditch2001}\footnote{See \texttt{PyAstronomy} for a Python implementation \citep{pya}.}.
For \hpb, we obtained a mass ratio
of $(4.81 \pm 0.31)\times 10^{-4}$ (Table~\ref{tab:pars}), and
the resulting geometry of the planetary Roche potential is shown
in Fig.~\ref{fig:roche}.

The planetary radius perpendicular to the star-planet axis is known from primary transit photometry and determines
the equipotential surface defining the planetary surface. The planet is slightly elliptical with its radius along
the symmetry line, \rpa, being about $8$\,\% larger than the perpendicular radius. The effective radius, $R_{\rm p, eff}$,
of a sphere with the same volume is $1.03$~\rpp. We use this value to derive the mean planet density of
$0.135\pm 0.016$~g\,cm$^{-3}$ quoted in Table~\ref{tab:pars}.

The height of the first Lagrange point above the surface is $1.140\pm 0.015$~\rpp. The second Lagrange point is
slightly further up at $1.219\pm 0.016$~\rpp. In the perpendicular direction, the limit of the Roche lobe
is closer to the planetary surface at a height of $0.435\pm 0.010$~\rpp\ in the orbital plane.
Consequently, the annulus of
a Roche-lobe filling planetary atmosphere would cover
\begin{equation}
    \frac{(R_{\rm p, \bot} \times (1+0.435))^2}{R_{\star}^2} - \frac{R_{\rm p, \bot}^2}{R_{\star}^2} \approx 2.4\,\%
\end{equation}
of the stellar disk during lower conjunction.
The effective radius of the Roche lobe
is $1.526\pm 0.034$~\rpp\ \citep[][]{Eggleton1983}, so that
the planetary Roche lobe filling factor (by volume) becomes $30$\,\%.

\section{\carm\ observations and data reduction}
\label{sec:obsreduc}
We obtained two transit time series of \hpb\ on the nights of 1 September and 9 December 2018, in the following
referred to as night~1 and night~2, with the \carm\ spectrograph. The \carm\ instrument is installed at the
$3.5$\,m telescope of the Calar Alto Observatory in Spain and features a visual (VIS) and a near-infrared (NIR) arm,
covering the $560-960$~nm and $960-1710$~nm range with a
spectral resolution of $94\,600$ and $80\,400$, respectively \citep{Quirrenbach2018}. 
In accordance with its prime purpose as a planet finder, both arms are highly stabilized.
\carm\ is fed with two fibers, of which the first was pointed at the target, \hp, and the second at the sky
during our observations.

On nights~1 and 2, 23 and 26 usable science spectra were obtained, respectively, with
an exposure time of about $900$~s. On night~2, two observations had to be aborted after 
$640$~s and $220$~s because of technical difficulties.  
These are not considered in our
analysis. In Tables~\ref{tab:logn1} and \ref{tab:logn2}, we list a log number (LN), the central time of
exposure, the corresponding temporal offset from the center of the optical transit,
the exposure time, and the overlap fraction of
the planetary with the stellar disk averaged over the exposure for all spectra analyzed here.

\subsection{Data reduction}

\begin{figure}[ht]
    \includegraphics[width=0.49\textwidth]{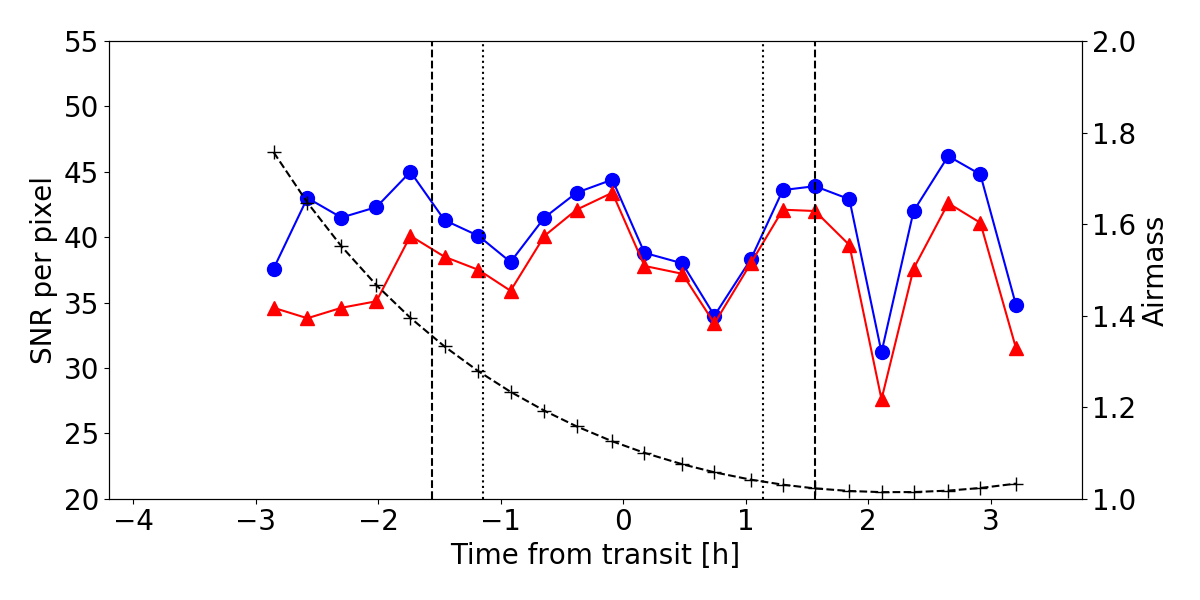}\\
    \includegraphics[width=0.49\textwidth]{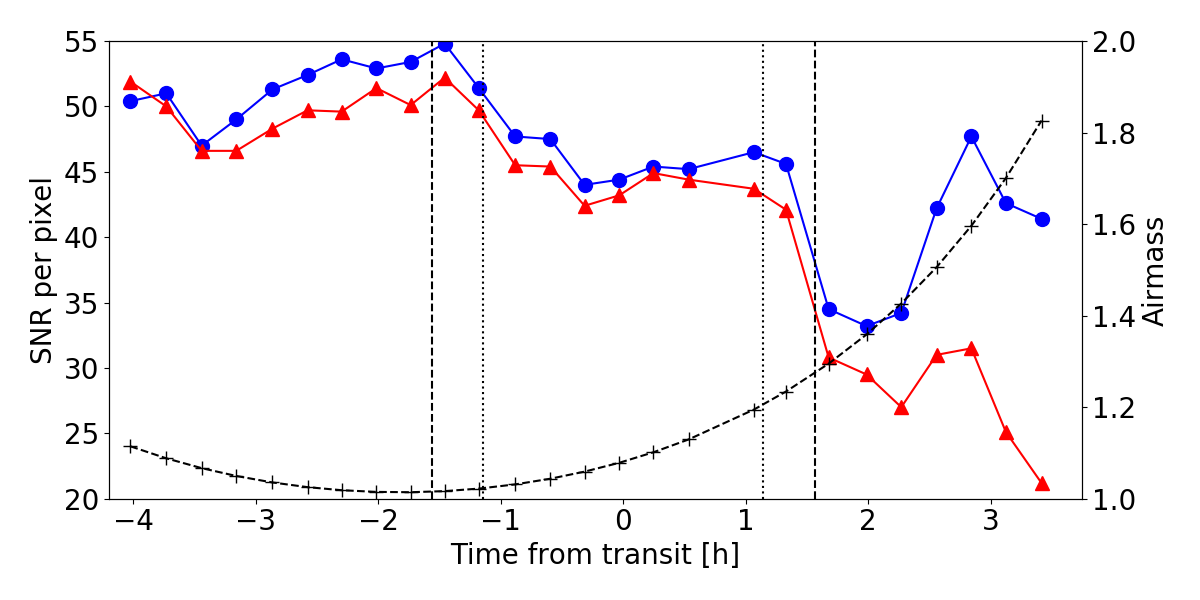}
    \caption{Signal-to-noise ratio per pixel in order 93 of the VIS arm (blue circles) and
    order 56 of the NIR arm (red triangles) along with airmass (black, crosses) as a function
    of time for night~1 (\textit{top}) and night~2 (\textit{bottom}). Dashed, vertical lines indicate first and fourth
    contact and dotted, vertical lines the second and third contact.
    \label{fig:snr}}
\end{figure}

The CARMENES data were reduced with the {\tt caracal} pipeline
\citep[CARMENES Reduction And CALibration,][]{Zechmeister2014, Caballero2016}. 
In Fig.~\ref{fig:snr}, we show the time evolution of the S/N in the spectral
orders covering the H$\alpha$ line in the VIS arm and the \hei\ triplet lines in the NIR arm
as computed by {\tt caracal}.
While the S/N remained rather stable during night~1, it shows a decrease during night~2,
which is particularly pronounced in the NIR data.  

The stellar M1.5V-type companion, \hp\,B, is located at a distance of $2.9''$ from \hp\,A.
Although the CARMENES fibers provide a $1.5''$ diameter acceptance angle \citep[][]{Quirrenbach2018},
which separates the sources,
some contamination from overlap of the seeing disks may be expected.
On the basis of the results by \citet{Zhao2014},
we estimate that the maximum possible flux contribution
of \hp\,B in the H$\alpha$ and \hei\ spectral regions, obtained by assuming
perfect overlap of the seeing disks, is $0.4$\,\% and $2.5$\,\% given (continuum)
contrasts of about $6$~mag and $4$~mag.
We expect the true
contamination to be considerably smaller due to the angular separation of the targets.
In early-type M~dwarfs, the stellar \hei\ lines tend to be in absorption and rather
stable \citep{Fuhrmeister2019, Fuhrmeister2020}.
In the hypothetical case of strong line contamination, for instance, owing to flaring, we
expect of course no relation to the actual planetary transit timing. We, therefore,
conclude that \hp\,B is of little concern to our spectral analysis.  
 
The regions around the H$\alpha$ line and \hei\ triplet lines are affected by telluric
water absorption lines. The \hei\ infrared triplet region is additionally affected by OH emission lines.
We corrected the water absorption lines in the optical and infrared using the \mf\ package \citep{Smette2015, Kausch2015}.
To remove the OH emission lines, we subtracted a semi-empirical synthetic model of the sky emission spectrum from the science spectra.
This model was calibrated using a sample of 1836 observations of sky emission obtained by \carm.
A detailed account of our telluric correction is given in App.~\ref{sec:TC}. 

\subsection{Radial velocity of \hp}
\label{sec:stellarRV}
To determine RV shift of the \hp\ system, we fitted the spectral range from $8480$ to $8700$~\AA,
which contains the strong lines of the \cairt, using a PHOENIX model spectrum calculated for an effective
temperature of $6200$~K, surface gravity $\log(g)$ of $4.5$, and solar metallicity from the grid presented by
\citet{Husser2013}. Prior to the fit, the template was broadened to account for stellar rotation
(Table~\ref{tab:pars}) and instrumental
resolution. We then fitted all individual spectra,
varying the stellar RV and the parameters of a second degree polynomial representing
the continuum in the fit.

With this procedure, we determined an RV shift with respect to the PHOENIX model spectrum for
all observed spectra.
Fitting a circular model for the planetary orbital motion with fixed semi-amplitude ($K_{\star}$, Table~\ref{tab:pars})
and free RV offset to these values, we found a mean value of $\bar{v}_{\star} = -22.10 \pm 0.03$~\kms\ for the RV shift of the spectrum
of \hp\ with respect to the PHOENIX model spectrum at orbital phase zero. 
In this step, we neglected the contribution of the Rossiter-McLaughlin
effect, which is small in amplitude in \hp\ \citep[$\approx 100$~\ms, ][]{Albrecht2012}.
The quoted uncertainty is statistical.
The value may be compared to the measurement of $-23.21 \pm 0.26$~\kms\ obtained with the Digital Speedometer \citep[e.g.,][]{Latham1996}
and reported by \citet{Hartman2011},
pointing to a true uncertainty including systematics on the order of 1~\kms.

To cross-check our result, we fitted the \num{8130} to $8175$~\AA\ range in the VIS channel and the
\num{11040} to \num{11075}~\AA\ range in the NIR channel with a synthetic, telluric water absorption model. For both channels
and nights, we obtained offsets $<75$~\ms\ with no considerable shifts within or between the nights. We conclude
that the absolute wavelength calibration of the instrument is accurate to at least $75$~\ms\ during our observations.
The accuracy is probably better \citep{Lafarga2020}, but this value is sufficient for our analysis.

\section{Stellar activity and planetary irradiation}

Stellar activity plays an important role in
exoplanetary research. In the planetary atmosphere, activity-induced irradiation acts as a
driving force for the physics and chemistry. In planetary transmission spectroscopy, the role of activity phenomena
is mostly one of a nuisance, in particular,
because short-term variations in activity sensitive lines can mask or even mimic a planetary
signal. 

\subsection{X-ray observations and planetary irradiation}
\label{sec:irradiation}

We observed \hp\ with \xmm\ on 30\,August\,2019 through the DDT proposal ID~85338 (P.I. J. Sanz-Forcada)
for an overall exposure time of 20.1~ks. 
The \xmm\ satellite is equipped with three X-ray telescopes \citep{Jansen2001} with three
CCD cameras at their focal planes, provided by the European Photon Imaging Camera (EPIC) consortium.
The assembly consists of two metal oxide semi-conductor (MOS) CCD arrays and one array of so-called pn-CCDs
\citep[][]{Stueder2001, Turner2001}, whose fields of view largely overlap, so that
they are usually operated simultaneously.

During our observation, the
EPIC~pn and MOS X-ray detectors were partly affected by high background, which was removed
prior to the spectral analysis. The EPIC cameras do not provide sufficient spatial resolution to separate the stellar
A and B components, but the optical monitor (OM) on board XMM-Newton indicates that UV emission in the UVW2 filter
($\lambda_c$ = 2120\,\AA) comes from the F-type main component, with no emission detected from the M~dwarf companion.
EPIC light curves indicate the presence of flares (Sanz-Forcada et al., in prep.), which most likely
come from the primary star in the system.

We performed a simultaneous spectral fit to the three EPIC spectra, with a S/N=11.3.
A single coronal plasma model component with a temperature, $\log(T~\mbox{[K]})$, of
$6.61\pm0.08$ and an emission measure, $\log($~EM~[cm$^{-3}$]), of $51.90\pm0.10$
and its metal abundances fixed to the slightly subsolar photospheric value
of [Fe/H]=$-0.04$ suffices to fit the spectra.
The X-ray luminosity of the system is $L_{\rm X}=2.3 \times 10^{29}$~erg\,s$^{-1}$ in the 5--100~\AA\ range.
Assuming that most X-ray flux originates from the primary, this translates into
$\log L_{\rm X}/L_{\rm bol}=-4.5$, which is a relatively high value, especially for an F~star \citep[e.g.,][]{Pizzolato2003}.
The X-ray light curve shows a moderate level of short-term variability most likely attributable
to X-ray flaring on the main component; a detailed analysis will be published elsewhere (Sanz-Forcada et al. in prep.).

We extended the coronal model toward cooler temperatures following \citet{Sanz2011}.
This model was then used to predict the spectral energy distribution at wavelengths up to 1600~\AA.
The model flux in the XUV spectral ranges relative to the He and H ionization edges are
$3.1^{+4.1}_{-1.4} \times10^{29}$~\ergs\ (100--504 \AA) and $1.2^{+2.1}_{-0.7}\times10^{30}$~\ergs\ (100--920 \AA).
From our modeling, we obtain a total XUV ($<\,912$~{\AA})
flux of $\approx 4.2 \times 10^5$~\ergcmcms\ at the planetary orbit.
To further extend the model and calculate the NUV irradiation level, we employ a photospheric model \citep{Castelli2003}. 
We here define the NUV as the $912-3646$~{\AA} range, bracketed by the onset of Balmer continuum absorption at its red boundary. For the NUV irradiation at the planetary orbital distance, we obtain $\approx 6 \times 10^7$~\ergcmcms. 

The XUV emission of {\hp} is about 100 times solar and results in an extremely high-energy irradiation environment for the planet, which
is expected to produce
substantial atmospheric mass loss from
\hpb. In this particular case, the loss rate is increased by mass flow through the Roche lobe
\citep{Erkaev2007}. Under an energy-limited escape model \citep[][and references therein]{Sanz2011},
we estimate a mass-loss rate of $10^{14}$~\gs. The statistical uncertainty
of this value is driven by the irradiating flux and amounts to about a factor of two, which
does not account for variability in the XUV flux. 
Adopting an evaporation efficiency of $0.1$
for a gravitational surface potential of $\log(\Phi\mbox{[erg\,g$^{-1}$]})=12.8$ from
\citet{Salz2016b} reduces the mass-loss rate to $10^{13}$~\gs, which remains an
extremely high value.

\subsection{Variability analysis of the \cairt\ triplet lines}

Both the \hei\ triplet lines and the \ha\ line are sensitive to activity-related changes in
the chromosphere \citep[e.g.,][]{Fuhrmeister2018, Fuhrmeister2019}. Likewise, the \cairt\ lines are well-known tracers of chromospheric
activity \citep[e.g.,][]{Martin2017}.
While planetary atmospheric absorption has been observed in these lines also \citep{Yan2019},
this has only been in planets with considerably higher surface temperatures so far.

\begin{figure}[ht]
            \includegraphics[width=0.49\textwidth]{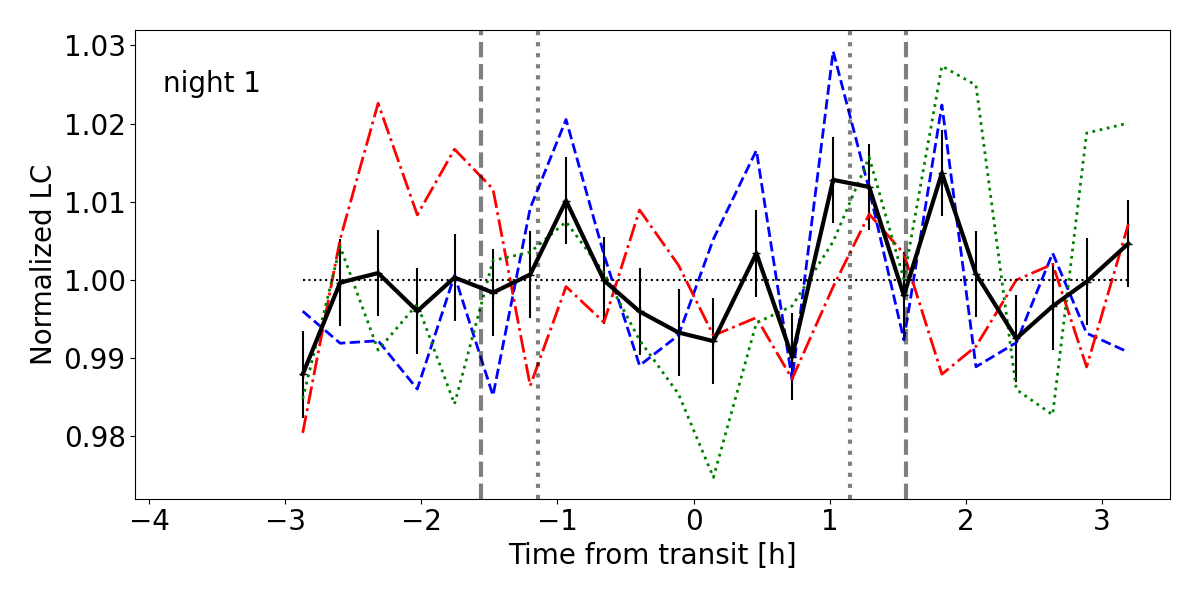}\\
    \includegraphics[width=0.49\textwidth]{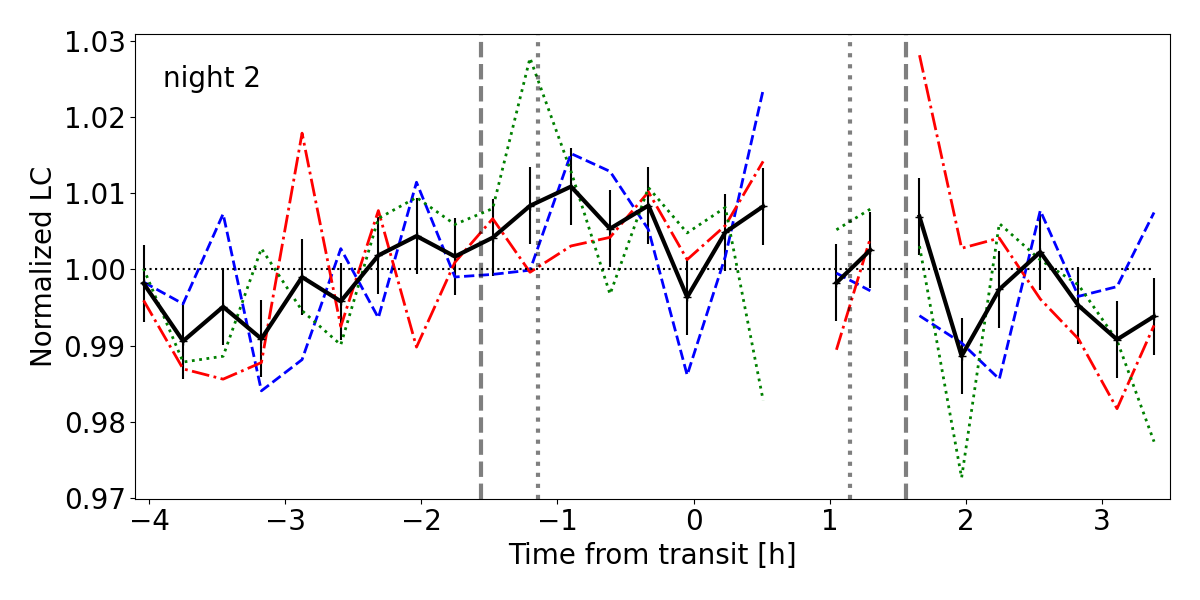}
    \caption{Normalized \cairt\ light curves of individual triplet components (blue dashed,
    red dash-dotted, and green dotted in order of ascending wavelength) along with the mean light curve
    (black solid). First to fourth contact times are indicated, and the neutral line is at one.
    \label{fig:cairtlc}}
\end{figure}

In Fig.~\ref{fig:cairtlc}, we show the light curve obtained for the three components of the
\cairt\ triplet as well as a mean \cairt\ light curve. To obtain these light curves, we
normalized the spectra in the region of the
lines and summed the normalized flux in bands with half-widths of $0.2$~\AA\ centered on the
nominal positions of the \cairt\ lines.  

No flaring activity, which would be characterized by a fast
rise phase and longer decay phase in the \ion{Ca}{ii} lines \citep[e.g.,][]{Klocova2017},
can be distinguished in these light curves. Also no signal associated with the transit itself is observed.
The \cairt\ lines are
consistent with a constant level of activity during night~1. On night~2, some systematic evolution
may be present in the \cairt\ lines, indicating a slightly elevated activity level during the transit
compared to before and after the transit. However, the amplitude of variability remains within the limits
of the scatter observed on night~1. We conclude that stellar activity, as far as it manifests itself in
the \cairt\ lines, is not a strong interference in our data sets.

\section{\ha\ and \hei\ transmission spectroscopy}
\begin{figure}[ht]
        \includegraphics[width=0.49\textwidth]{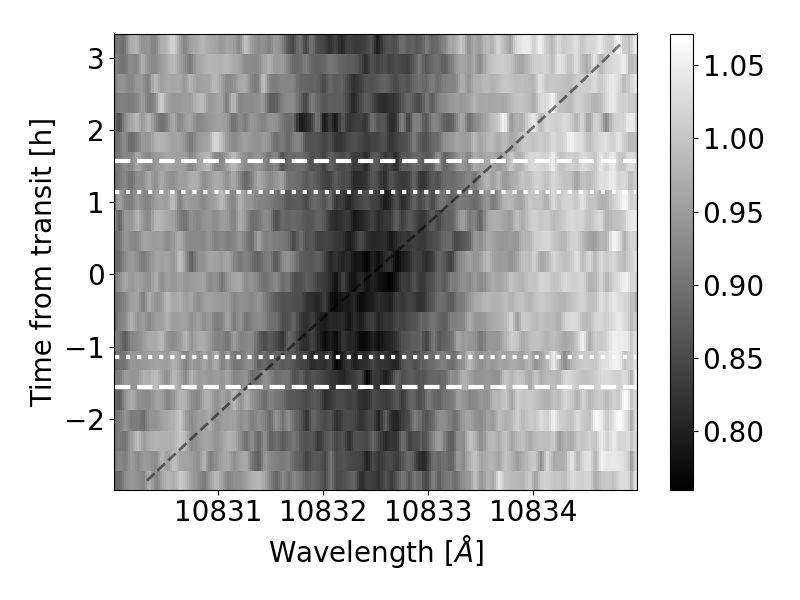}
    \caption{Heat map showing the time series of normalized spectra around the \hei\ triplet lines on night~1. 
    Optical transit contacts (horizontal dashed and dotted lines) and the planetary
    RV track (black dashed)
    are indicated.
    \label{fig:maphen1pure}}
\end{figure}

To study the planetary transmission spectrum, we started with the spectra corrected for
telluric line contamination and shifted them according the barycentric motion of the Earth.
We then normalized the spectral ranges covering the \hei\ and \ha\ lines using a linear fit to the
surrounding continuum. The spectra thus obtained are referred to as $f_{n,i}(\lambda)$, where
$n = \{1,2\}$ specifies the night and $i$ denotes the LN.
In Fig.~\ref{fig:maphen1pure}, we show a heat map\footnote{Variously these diagrams are
also called matrix plots or similar, the idea always being to
represent the magnitude of an effect across a two-dimensional field by color.}
of the spectral time series around
the \hei\ line on night~1. Already in this map, an absorption signal
aligned with the planetary track both in RV and transit timing
can be discerned.

\subsection{Reference, transmission, and residual spectra}
\label{sec:refspecmain}

\begin{table}
\centering
\caption{Log numbers of spectra used to construct reference spectra.
\label{tab:refsns}
}
\begin{tabular}{l l l}
\hline \hline
Night ($n$) & Line ($L$) & $\mathcal{R}_{n,L}$ \\ \hline
1 & \ha   & \{$1-3$, $20-23$\} \\
1 & \hei\ & \{$1-3$, $20-23$\} \\
2 & \ha\  & \{$1-6$, $24-28$\} \\
2 & \hei\ & \{$1-6$\} \\
\hline
\end{tabular}
\end{table}

To work out the signature of the planetary atmosphere more clearly, we constructed reference spectra by
averaging a number of suitable out-of-transit spectra. In particular, we chose subsets of our spectra 
$2.2$~h or more before and after the center of the optical transit for each night. 
If $\mathcal{R}_{n,L}$ denotes a suitable subset of LNs to be considered for night $n$ and spectral line $L$,
which either refers to \ha\ or \hei\ here,
we construct a reference spectrum, $R$, by averaging
\begin{equation}
    R_{n, L}(\lambda) = \frac{1}{\#\mathcal{R}_{n,L}} \sum_{i \in \mathcal{R}_{n,L}} f_{n,i}(\lambda) \; ,
\end{equation}
where $\#\mathcal{R}_{n,L}$ is the number of spectra in the set.
As the stellar rotational line width is much larger than the effect caused by the stellar reflex motion,
we neglect the effect here.

After careful examination of the spectral time series on nights~1 and 2, we constructed reference spectra
for the \ha\ and \hei\ line regions for both nights, using the ranges of spectra listed
in Table~\ref{tab:refsns} (see Sect.~\ref{sec:refspec} for details). In all but one case, we opted for
a combination of pre- and post-transit spectra. Only for the \hei\ line region on night~2, we prefer to
use only pre-transit spectra to avoid residual effects due to strong contamination by OH emission lines,
which is otherwise well accounted for (Sect.~\ref{sec:TC}).
Using the reference spectra, individual empirical transmission spectra, $t_{n,i}$, are obtained according to
\begin{equation}
    t_{n,i} (\lambda_L) = \frac{f_{n,i} (\lambda_L)}{R_{n,L}(\lambda_L)} \; ,
\end{equation}
where $\lambda_L$ indicates a suitable wavelength range around the spectral line indicated by $L$.
Finally, we call $r_{n,i} = t_{n,i} - 1$ a residual spectrum. All of $f_{n,i}$, $r_{n,i}$, and $t_{n,i}$
refer to the barycentric frame. The systemic velocity of the \hp\ system is not accounted for.

A signal originating in the planetary atmosphere is shifted in RV by
the orbital motion of the solid planetary body, potential velocity fields within the planetary atmosphere, and the
velocity of the system's barycenter, which we assume to be constant (Sect.~\ref{sec:stellarRV});
the effect of the secular acceleration
of the \hp\ system is around $-10$~\ms\ between the two nights (see Sect.~\ref{sec:orbit}), which we consider negligible.
To study the behavior of the signals in the co-moving planetary frame (i.e., the planetary rest frame moving along with
the planetary system), we obtained
shifted residual spectra, $r^p_{n,i}$, by applying a Doppler shift offsetting the planetary orbital motion \citep[see, e.g.,][]{Wyttenbach2015}.
The applied RV shift corresponds to that during the middle of the respective observation; the treatment
of finite integration times is discussed in Sect.~\ref{sec:smearing}.

\subsection{Residual maps}

\begin{figure}[ht]
        \includegraphics[width=0.49\textwidth]{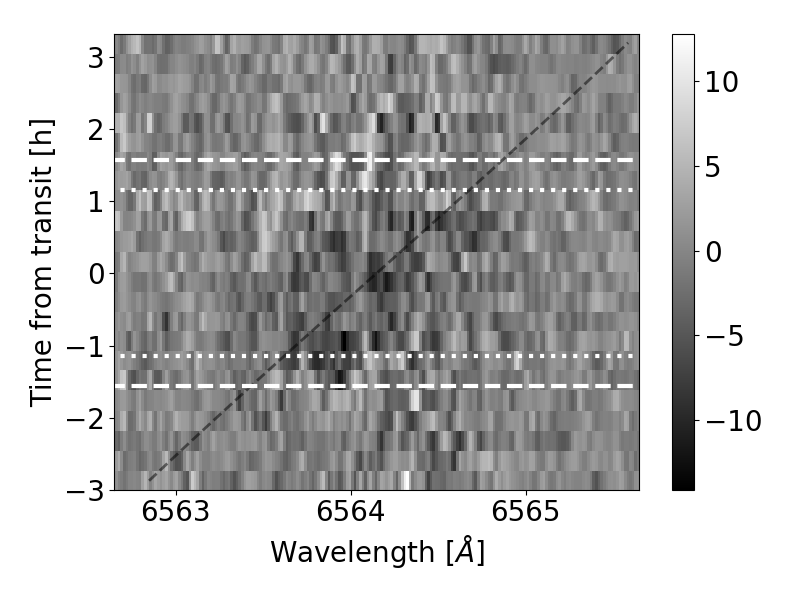}\\
    \includegraphics[width=0.49\textwidth]{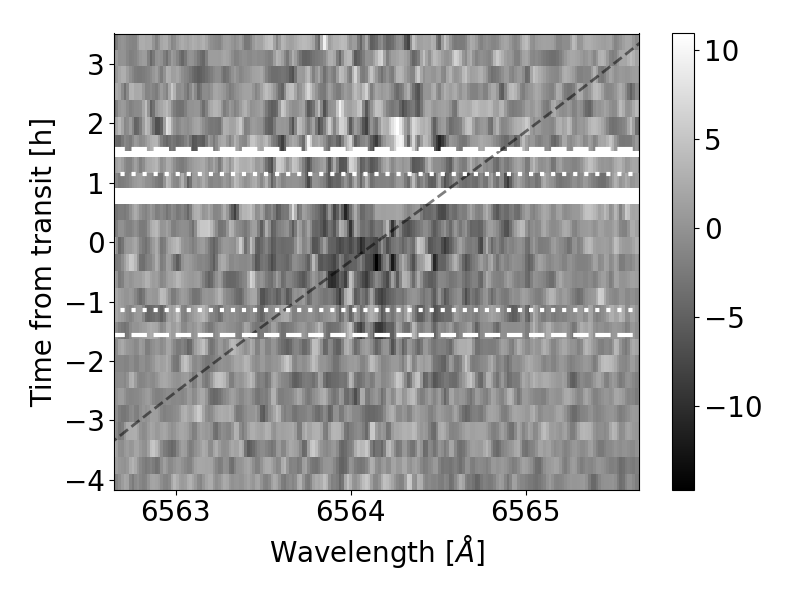}\\
    \includegraphics[width=0.49\textwidth]{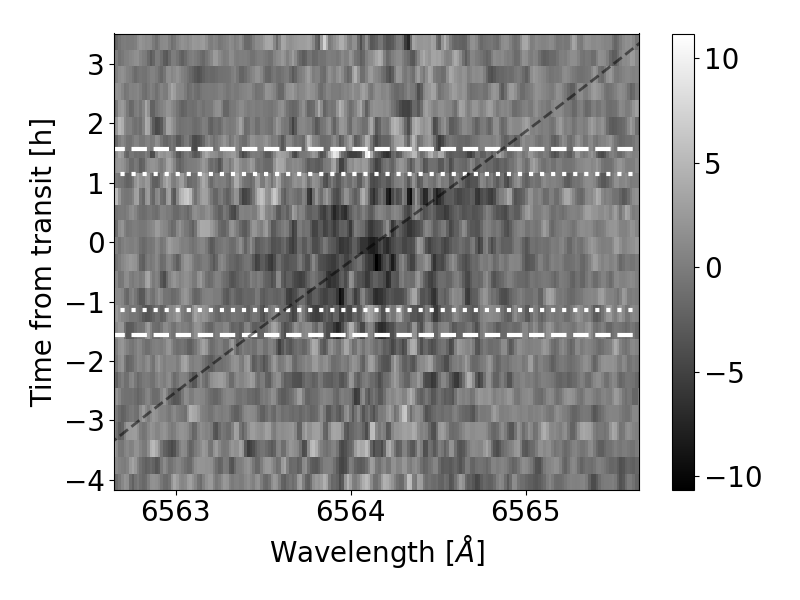}\\
    \caption{Heat maps of the \ha\ line region for night~1 (\textit{top}), night~2 (\textit{middle}), and
    a combination of both (\textit{bottom}). The color bar encodes the amplitude of the residuals in percent. 
    \label{fig:mapsha}}
\end{figure}

\begin{figure}[ht]
        \includegraphics[width=0.49\textwidth]{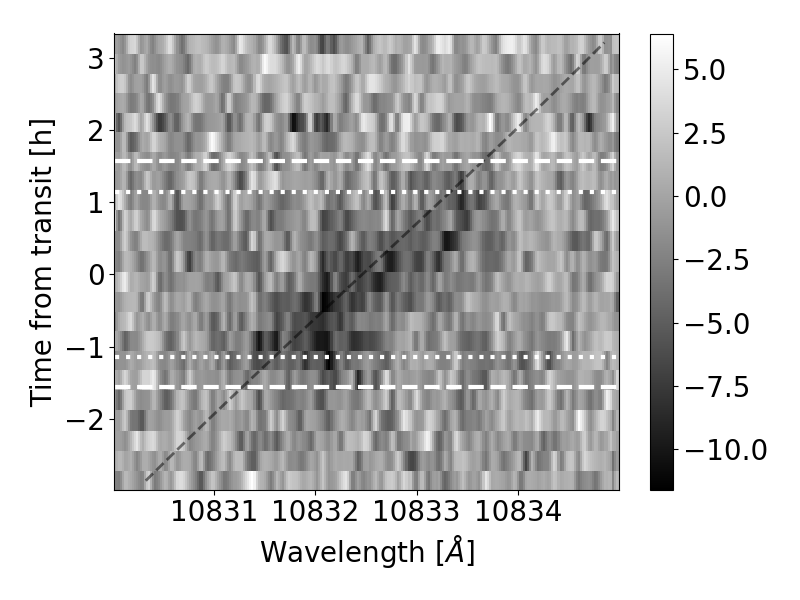}\\
    \includegraphics[width=0.49\textwidth]{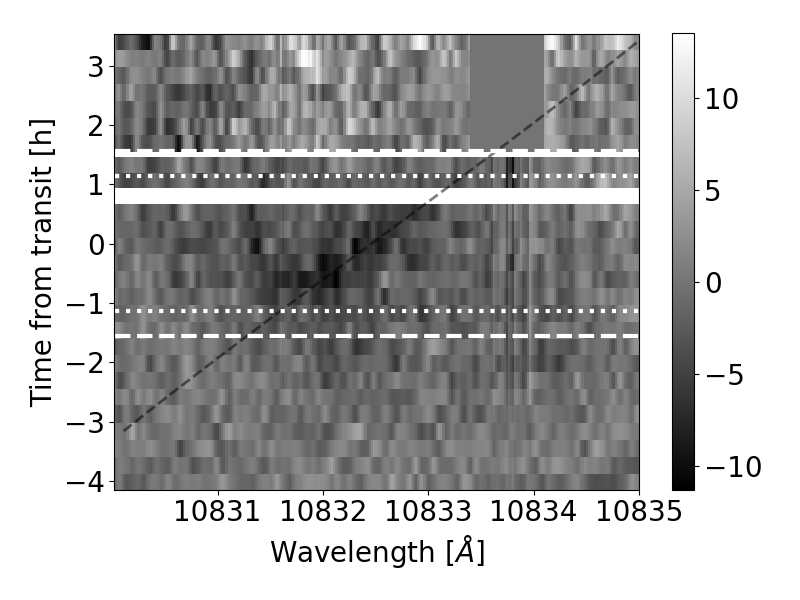}\\
    \includegraphics[width=0.49\textwidth]{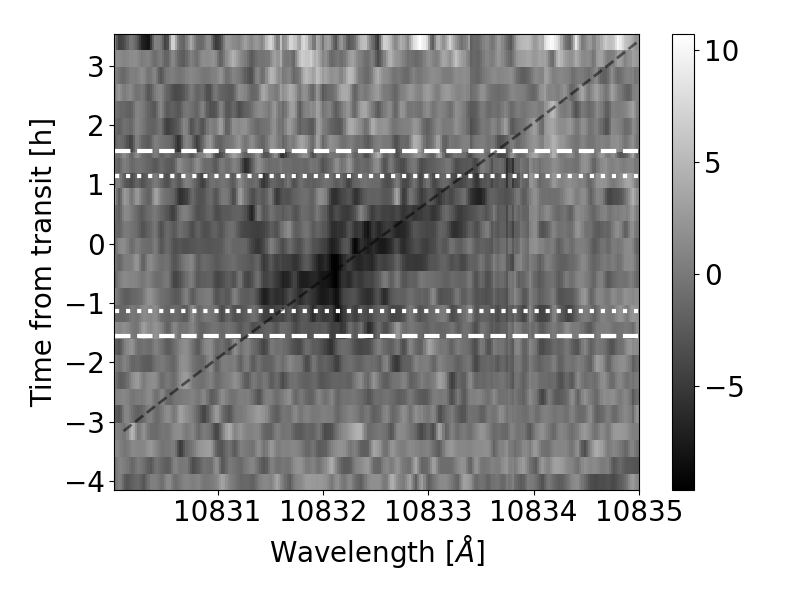}\\
    \caption{Same as Fig.~\ref{fig:mapsha} but for \hei\ lines.
    \label{fig:mapshe}}
\end{figure}

In Figs.~\ref{fig:mapsha}
and \ref{fig:mapshe}, we show the time evolution of the residual spectra, $r_{n,i}$,
for the \ha\ and \hei\ lines on nights~1 and 2 in the form of heat maps along with a map combining the data of the two nights.
In the \hei\ residual map for night~2, we masked the post-transit
residuals associated with the strong $Q_1$ OH emission line doublet (see Table~\ref{tab:OHlines}). 

The \hei\ maps show a pronounced absorption signal, which is associated with the optical
transit in terms of timing and 
moves along with the planetary RV track. A signal with similar properties but less pronounced is
also seen in the \ha\ maps.
The properties of the absorption signals, therefore, point to an origin in the planetary atmosphere.

\subsection{Time-resolved transmission spectra}
\label{sec:timeResolvedTS}

\begin{figure}[ht]
        \includegraphics[width=0.49\textwidth]{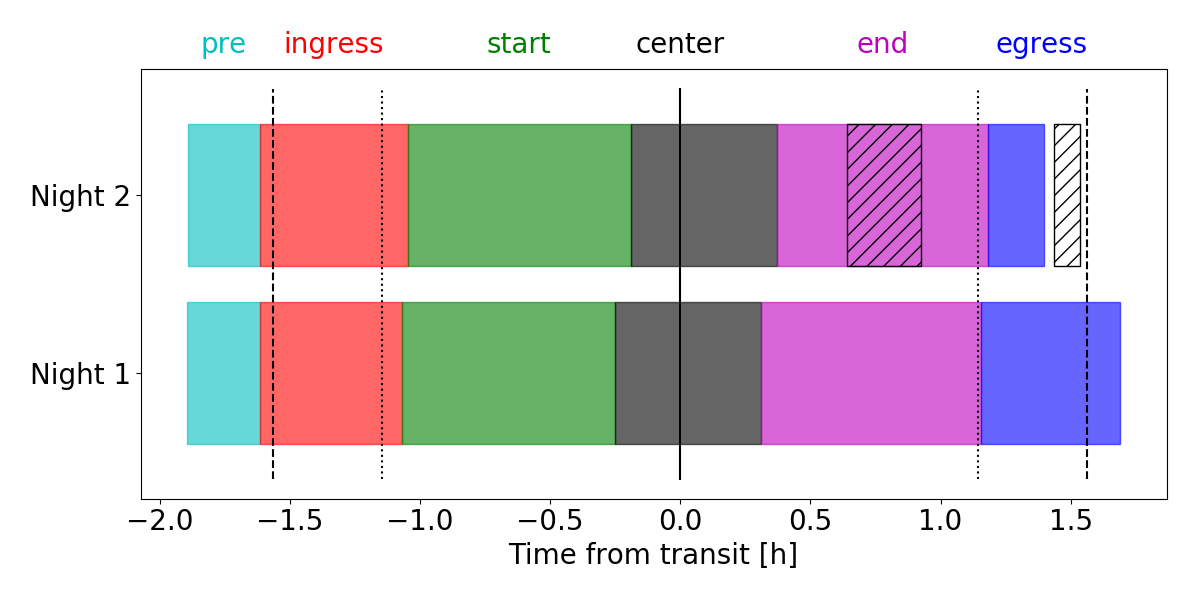}
    \caption{From left to right, the pre, ingress, start, center, end, and egress sections,
    as indicated by the colored areas.
    Hatched sections on night~2 refer to technical dropouts.
    Dashed and dotted lines show the four contact times.
    \label{fig:secs}}
\end{figure}

To study the temporal evolution of the transmission spectrum,
we defined six time intervals, which we call pre-transit, ingress, start, center, end,
and egress and identify the spectra of nights~1 and 2 belonging to the individual
sections. In Fig.~\ref{fig:secs}, we show the temporal coverage resulting from our
attribution scheme; a detailed account for the individual spectra is given in
Tables~\ref{tab:logn1} and \ref{tab:logn2}. 
Our temporal sampling of about $15$~min and cadence shifts due to technical
problems limit the accuracy with which individual sections can be
separated and aligned on nights~1 and 2. 

\begin{figure}[ht]
        \includegraphics[width=0.49\textwidth]{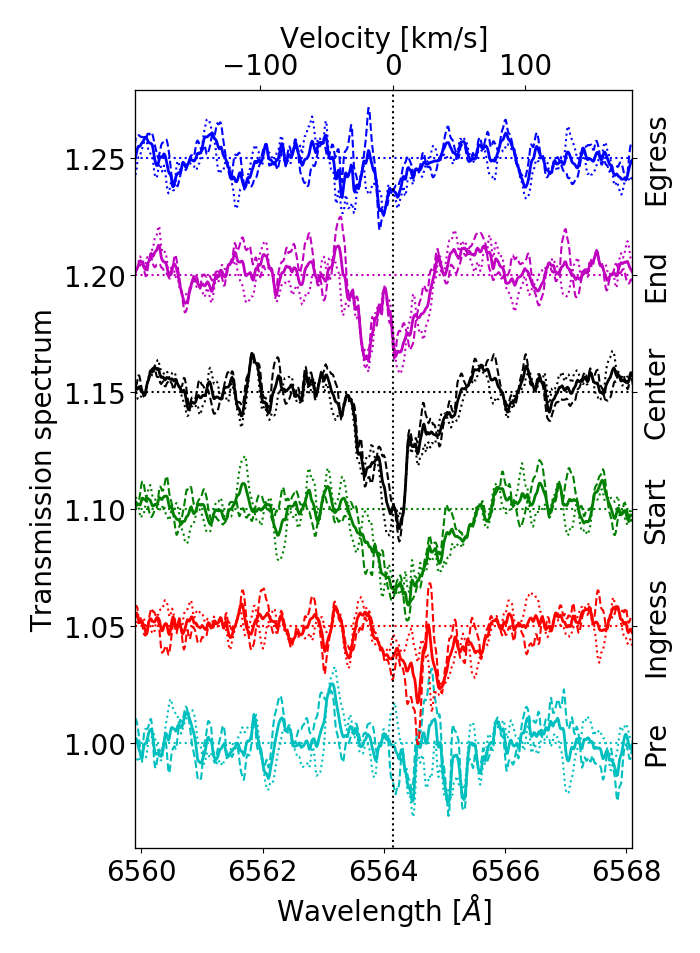}
    \caption{\ha\ transmission spectra for ingress, start, center, end, and egress phase from bottom to top (consecutively offset by $0.05$; comoving planetary frame).
    Dashed lines show night~1 results, dotted lines night~2 results, and solid lines the average of these two.  
    \label{fig:hatresol}}
\end{figure}

\begin{figure}[ht]
            \includegraphics[width=0.49\textwidth]{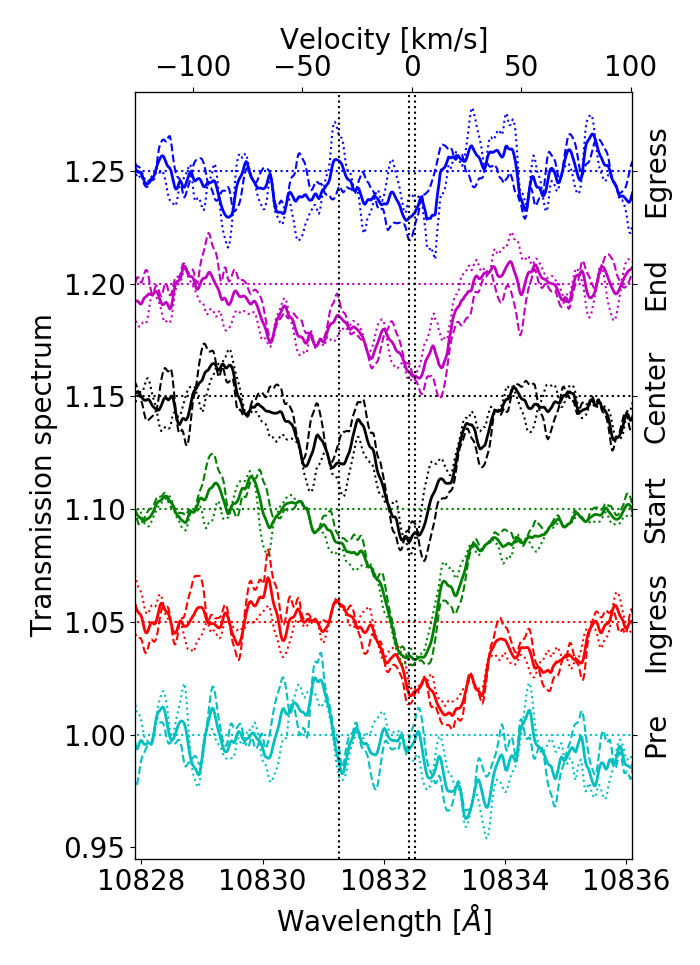}
    \caption{Transmission spectra in the \hei\ line region for ingress, start, center, end, and egress phase from bottom to top (consecutively offset by $0.05$; comoving planetary frame).
    Dashed lines show night~1 results, dotted lines night~2 results, and solid lines the average of these two.  
    \label{fig:hetresol}}
\end{figure}

\begin{figure}[ht]
        \includegraphics[width=0.49\textwidth]{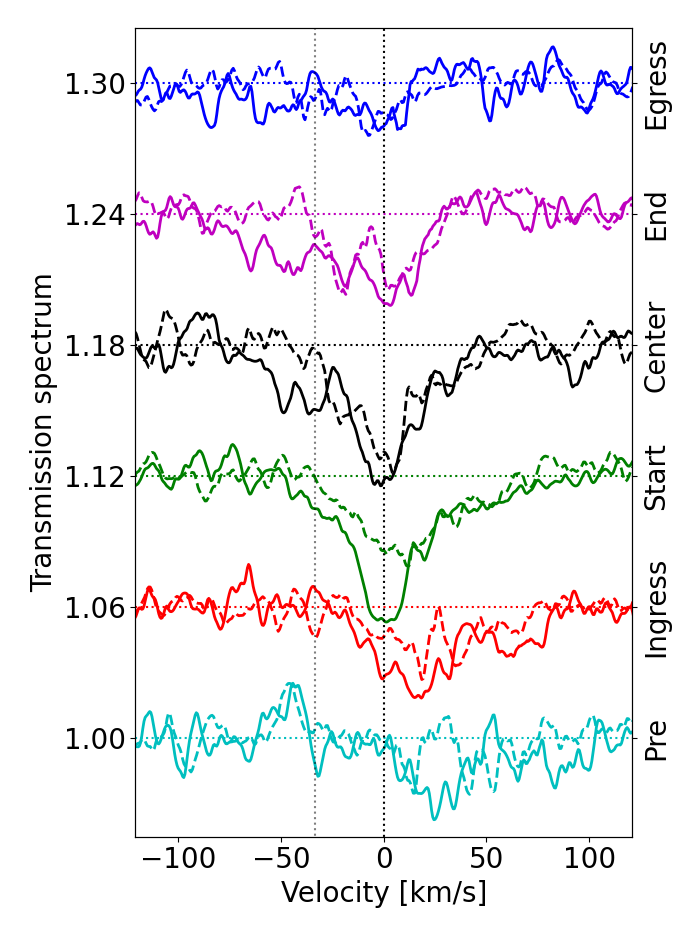}
    \caption{Superimposed transmission spectra in \hei\ (solid) and \ha\ (dashed) for ingress, start, center, end,
    and egress phase from bottom to top (consecutively offset by $0.06$).
    \label{fig:hahetresol}}
\end{figure}

We obtained transmission spectra by coadding the shifted residual spectra, $r^p_{n,i}$, pertaining to the
sections under consideration \citep[see, e.g.,][]{Wyttenbach2015}.
Our transmission spectra were corrected for planetary orbital motion, but not for the systemic RV shift of the whole \hp\ system.
The resulting time-resolved transmission spectra are shown in Figs.~\ref{fig:hatresol} and \ref{fig:hetresol}
for the \ha\ and \hei\ lines. For each line and section, we show
the transmission spectrum obtained on night~1 and night~2 as well as an averaged spectrum. In Fig.~\ref{fig:hahetresol},
we juxtapose the averaged transmission spectra for the \ha\ and \hei\ line regions. 
The shown
transmission spectra were smoothed by a running mean with a window width of $0.165$~\AA\ for the \ha\
and $0.2$~\AA\ for the \hei\ lines.

Both the \hei\ and \ha\ line transmission spectra show
pronounced absorption signals primarily during the start, center, and end
phases. The signals are broad with respect to the instrumental resolution and approximately in the
planetary rest frame.
The central dip of the \hei\ transmission spectrum reaches a depth of around $6$\,\% during the start and center phase.
During the end phase, which is nearly symmetric to the start phase in terms of the geometry of the star-planet
system, the central \hei\ transmission dip is weaker, reaching a depth of about $4$\,\%.
For \ha\ the maximum depth of around $5$\,\% is also observed
during the center phase. In Table~\ref{tab:merit}, we give the EWs of the night-averaged transmission signals in the
$-100$~\kms\ to $+100$~\kms\ range along with an estimate of the uncertainty obtained from repeated reshuffling
of the transmission spectra.

\begin{table}[h]
\center
\caption{Equivalent widths of transmission signals in $-100$~\kms\ to $+100$~\kms\ range
as a function of transit phase.
\label{tab:merit}}
\begin{tabular}{l l l}
\hline \hline
Phase & EW [m\AA] (\ha) & EW [m\AA] (\hei) \\ \hline
Pre & $ 2.8 \pm  1.7$   &   $57.6 \pm  5.3$  \\
Ingress & $23.6 \pm  1.3$   &   $77.3 \pm  4.6$  \\
Start & $42.7 \pm  1.7$   &   $115.5 \pm  6.8$ \\
Center & $44.8 \pm  2.3$   &   $118.4 \pm  7.1$ \\
End & $21.1 \pm  1.4$   &   $74.3 \pm  5.3$  \\
Egress & $ 7.9 \pm  1.1$   &   $36.9 \pm  3.8$  \\
\hline
\end{tabular}
\end{table}

According to the Roche geometry (Sect.~\ref{sec:Roche}), a Roche-lobe filling atmosphere
covers about $2.4$\,\% of the stellar disk, which is less than the depth of absorption observed either in H$\alpha$
or the \hei\ lines. It thus follows that Roche-lobe overflow must play a role in the \hp\ system.

The time series of transmission spectra shows a complex pattern of variability, which may be associated with the
three-dimensional overflow geometry.
Notably, the \hei\ transmission spectrum
displays pronounced absorption, redshifted by about $25$~\kms\ with respect to the planetary body, before
the optical transit commences. In this pre-transit phase, no absorption at the nominal wavelength of rest of the \hei\
lines is detectable. A similar \hei\ absorption component is present during the ingress phase, yet, slightly shifted toward
the line center. Another absorption component redshifted by about $70$~\kms\ can be discerned during ingress,
which is consistently observed in both nights. During the ingress phase, the \hei\ absorption component extends to redshifts
of about $100$~\kms. This component appears to persist into
the start phase, where it is discernible at a weaker level.
The \hei\ egress absorption is weaker than its ingress counterpart. 
Particularly during the end phase, a blueshifted absorption component is visible, which appears in both lines, but
is narrower in the case of the H$\alpha$ line (Figs.~\ref{fig:hatresol} and \ref{fig:hahetresol}). However, absorption
is weak (\hei) or absent (\ha) during the egress phase and no post-transit absorption is detectable. 

Overall, the transmission in the \ha\ line is comparable to or weaker than that in the
\hei\ lines. 
The behavior of the red flank of the \ha\ line
transmission spectrum is similar to that of \hei\ during the ingress and start phases. The \ha\ line transmission spectrum
during the pre-transit phase, however, shows a weaker absorption component, if any (cf., Sect.~\ref{sec:tlc}).

\subsection{Center-to-limb variation}
\label{sec:clv}

During the transit, the opaque planetary disk consecutively covers
different sections of the stellar photosphere, seen at changing viewing angles and rotational shifts, which
produces a signal in the transmission spectrum
\citep[e.g.,][]{Czesla2015, Yan2015}. Following \citet{Salz2018}, we refer to this signal as a pseudo-signal.
It can have far-reaching ramifications for the interpretation of
the transmission spectrum as has been shown, for example, by \citet{Salz2018} and \citet{CasasayasBarris2020}.

\begin{figure}[h]
        \includegraphics[width=0.49\textwidth]{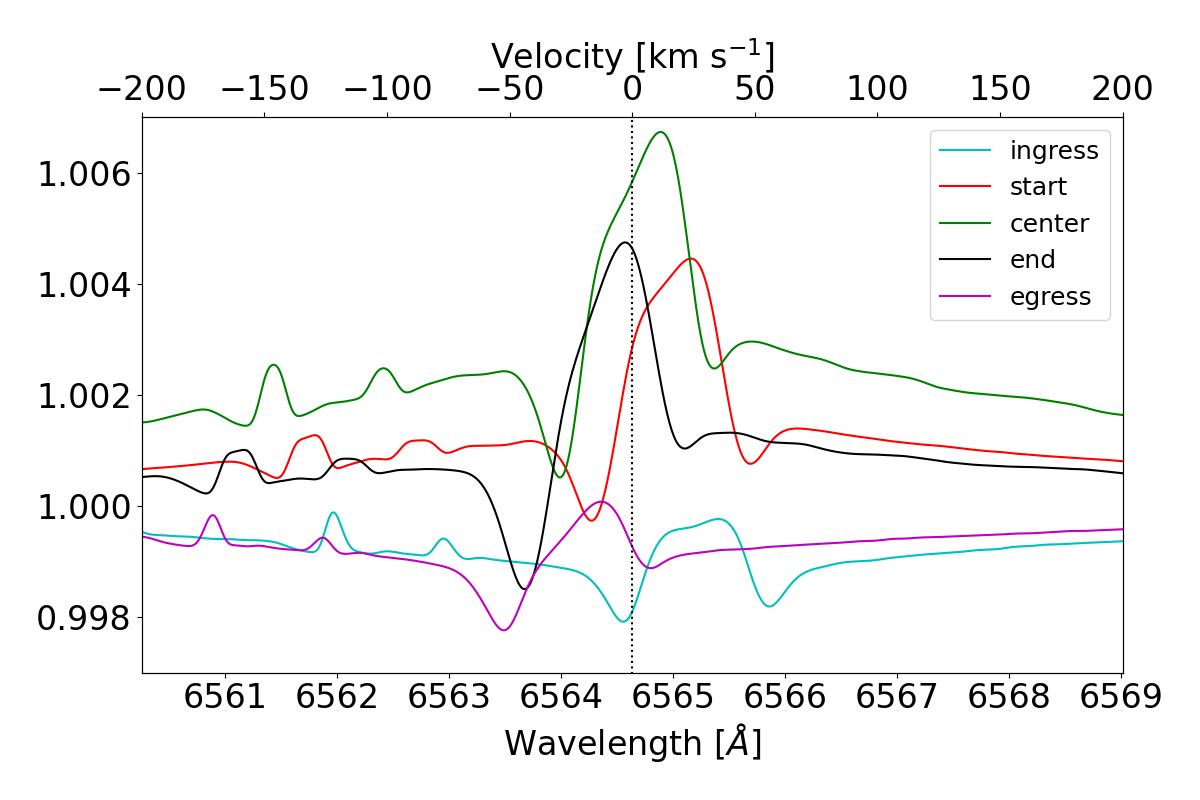}
    \caption{Synthetic (pseudo-)transmission signals caused by the CLV during individual temporal
    sections of the transit. 
    \label{fig:simha}}
\end{figure}

To determine the strength of the pseudo-signal expected to be caused by the center-to-limb variation (CLV) in \hp, we simulated the transmission
spectrum of the opaque planetary disk using the methodology presented by \citet{Czesla2015}. The simulations
are based on a discretized stellar surface and
synthetic specific intensities derived using Kurucz stellar model atmospheres \citep{Castelli2003}
and the \texttt{spectrum} program by R.O. Gray \citep{Gray1994}. Based on these inputs, a spectrum is then
synthesized for every phase of the transit.
In Fig.~\ref{fig:simha} we show the
resulting synthetic transmission spectra produced by the transit of the opaque planetary disk during the
sections indicated in Fig.~\ref{fig:secs}.

As is shown in Fig.~\ref{fig:simha}, the effect of the CLV is mostly one of pseudo-emission, which
is strongest during the center phase of the transit but also then remains $\le 0.7$\,\%, with a
baseline of about $0.2$\,\% belonging to a broad component, attributable to the broad \ha\ line.
During the transit, the
most distinct predicted signal evolves mostly within the $-50$~\kms\ to $50$~\kms\ range in the stellar rest frame.
Analogous predictions for
the \hei\ are naturally limited by the fact that the \hei\ is not included in photospheric synthetic models and has
basically unknown limb-angle dependence. 
However, as the stellar \hei\ line is considerably weaker than the \ha\ line in \hp, we expect also a weaker effect.
Compared to the observed \ha\ and \hei\ transmission signals, the CLV-induced pseudo-signals are a secondary
effect in \hpb\ and do not significantly affect the interpretation of the observed signals.

\subsection{Transmission light curves}
\label{sec:tlc}

\begin{figure*}
        \includegraphics[width=0.99\textwidth]{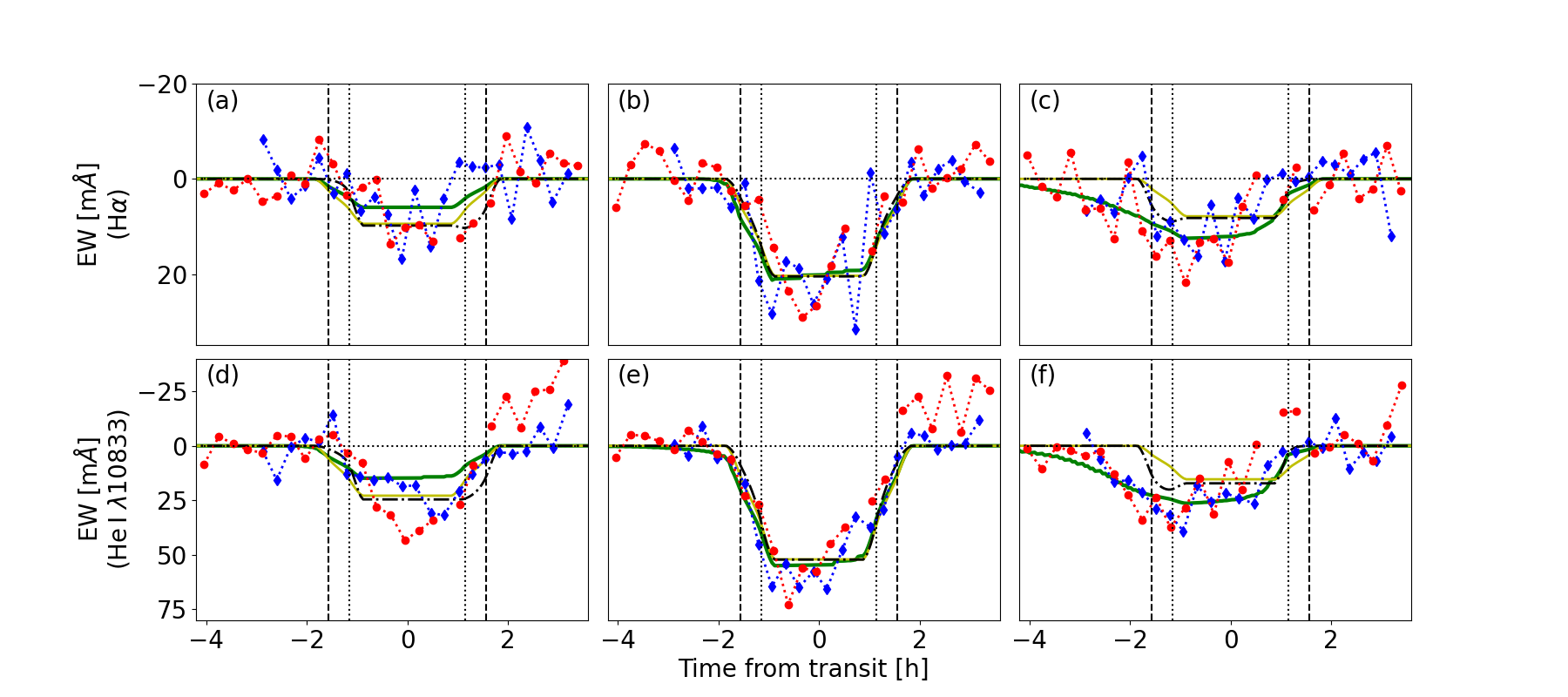}
    \caption{Transmission light curves for H$\alpha$ (top row, panels a$-$c) and \hei\ (bottom row, panels d$-$f) for velocity
    bands $(-50,-15)$~\kms, (-15, +15)~\kms, and $(+15, +50)$~\kms\ from left to right. Blue diamonds indicate night~1
    and red circles indicate night~2. The solid yellow lines represent best-fit light curves from the
    circumplanetary annulus model, dash-dotted black lines are model light curves pertaining to the super-rotating wind model,
    and the green solid lines are those for the combined circumplanetary and up-orbit stream model.
    \label{fig:lcvb}}
\end{figure*}

To investigate the time evolution of the transmission signal from a different perspective,
we constructed transmission light curves by integrating the residual spectra in the co-moving
planetary frame  
\begin{equation}
    l_{n,i} = \int_{\lambda_1}^{\lambda_2} r^p_{n,i}(\lambda)\, d\lambda \; .
\end{equation}
We here chose three wavelength ranges, representing the blue flank, the core, and the red flank
of the transmission spectra shown in Figs.~\ref{fig:hatresol} and \ref{fig:hetresol}.
The adopted ranges correspond to Doppler shifts between
$-50$ and $-15$~\kms, $-15$ and $+15$~\kms, and $+15$ and $+50$~\kms\ 
with respect to the nominal wavelengths of the \ha\ and \hei\ lines, and the resulting light curves
are shown in Fig.~\ref{fig:lcvb}.

The light curve of the \hei\ core (panel~e; dashed lines) shows a pronounced depression during the optical transit with an
EW reaching about $60$~m\AA. The \ha\ line core shows a comparable behavior with a signal of about
$25$~m\AA; as the same velocity band is about $65$\,\% broader around \hei\ than \ha\ in wavelength units,
the difference in signal depth is smaller.
The \hei\ core light curve on night~2 does not return to the pre-transit level after the optical
transit. This behavior is even more pronounced in the blue-flank \hei\ light curve (panel~d), but it does not
occur on night~1. We therefore attribute this
to residual effects caused by the telluric OH emission line doublet, which becomes considerably stronger on night~2
after the optical fourth contact (see Fig.~\ref{fig:q1ew}). Of course, this effect has also to be
present in the transmission spectra shown in Fig.~\ref{fig:hetresol}, and, indeed,
the red flank of the night~2 transmission spectrum shows higher levels than those on night~1.
Compared to the overall strength of the
transmission signal, the difference remains modest, however.

The red-flank \hei\ light curve (panel~f) shows a prominent early ingress, starting about one hour before the first
optical contact. The early ingress is seen in the data of both nights, although the night~1 light curve barely captures
the start of the depression, it is well sampled by the longer pre-transit coverage offered by the night~2
light curve. The red-flank \ha\ and \hei\ light curves are consistent with an early egress, starting before the third optical
contact. The night~2 \hei\ light curve shows a stronger effect here, even turning into apparent residual emission,
which may again be caused by the strong
OH contamination.
Yet, both light curves return to the pre-early-ingress level after the transit. Inspecting the
red-flank \ha\ light curve, a similar pattern can be discerned. These light curves display the early egress
rather well, which commences close to the center of the optical transit. An early ingress may be discernible
as well, although the light curves of both nights show an upward excursion shortly before the first optical contact,
making the situation less clear in this case.

\section{Phenomenological transmission spectrum modeling}
\label{sec:phenomodel}
In both observing nights, we detect strong, consistent absorption signals in both the H$\alpha$ and the \hei\ triplet lines
in \hpb. The lines show substructure and distinct temporal variation, 
which we attribute to the distribution of absorbing material in the system.

In the corotating star-planet frame,
mass is subject to gravitational and centrifugal forces
that can be combined into the scalar Roche potential, and any moving material is
also subject to the Coriolis force. As shown, for instance, by \citet{Bisikalo2013} and
\citet{CarrollNellenback2017}, the combined effect of these forces can deflect a hydrodynamic planetary wind into
a two-stream geometry. According to \citet{CarrollNellenback2017}, an up-orbit stream, launched mainly
from the planetary dayside,
precedes the planet and is directed toward the interior of
the planetary system, where it can lead to the formation of a circumstellar disk by stream-stream interaction
or may eventually be accreted onto the star \citep{Lai2010}. A down-orbit stream, launched mainly
from the planetary nightside, trails the planet, resembling a cometary tail. Such a two-stream geometry
has, indeed, been inferred from optical photometric observations in K2-22\,b, where, however,
the material is in a different physical regime \citep{SanchisOjeda2015}. Further agents such as
stellar wind interaction, radiation pressure, and magnetic fields can significantly modify the geometry
\citep[e.g.,][]{Ehrenreich2015, McCann2019, DaleyYates2019}.

Mostly focusing on the HD~209458 system, \citet{CarrollNellenback2017}
also
show simulations as a function of
the planetary radius expressed in terms of the sonic radius, $r_s$, of the Parker wind, $\Xi_p = R_p\, r_s^{-1}$, and the 
ratio, $\tau$, of Parker time and orbital time $r_s\, (v_s\, \Omega)^{-1}$, where $v_S$ is the speed of
sound and $\Omega$ is $2\pi\, P_{\rm orb}^{-1}$. Adopting $10$\,\kms\ for $v_s$ (Sect.~\ref{sec:atmHe}), we estimate
$(\Xi_p, \tau) \approx (0.3, 1.5)$ for \hpb, which indeed leads to the formation of
up- and down-orbit streams in their model.
The geometry described by \citet{CarrollNellenback2017} shows some of the same features as
the geometry described by \citet{Lai2010} and \citet{Li2010},
who adapted the theory of Roche lobe overflow in semidetached binary stars \citep{Lubow1975} to describe
a stream launched through the first Lagrangian point. 

Apart from the structure of the outermost atmosphere,
other components of the planetary atmosphere that may contribute to the observable
transmission signal include super-rotating winds, which may cause Doppler-broadened, shifted,
or even split transmission signals. A super-rotating atmosphere exhibiting such properties has, for example, been
reported for HD~189733\,b \citep[e.g.,][]{Knutson2007, Louden2015, Salz2018}.

In the following, we model the observed transmission spectra using two alternative assumptions on the
geometry of the absorbing material, inspired both by the morphology of the signals and the models
cited above. First, we assume that
only a circumplanetary atmosphere with potential (super-)rotation is responsible for the observed signals.
Second, we assume that
an up-orbit stream is also present as described by \citet{Lai2010} and \citet{CarrollNellenback2017}.
Dedicated one-dimensional hydrodynamic modeling of the atmosphere of \hpb\ is presented in Sect.~\ref{sec:atmModeling}.

\begin{figure}[ht]
    \includegraphics[width=0.5\textwidth, trim=0 0 0 275, clip=true]{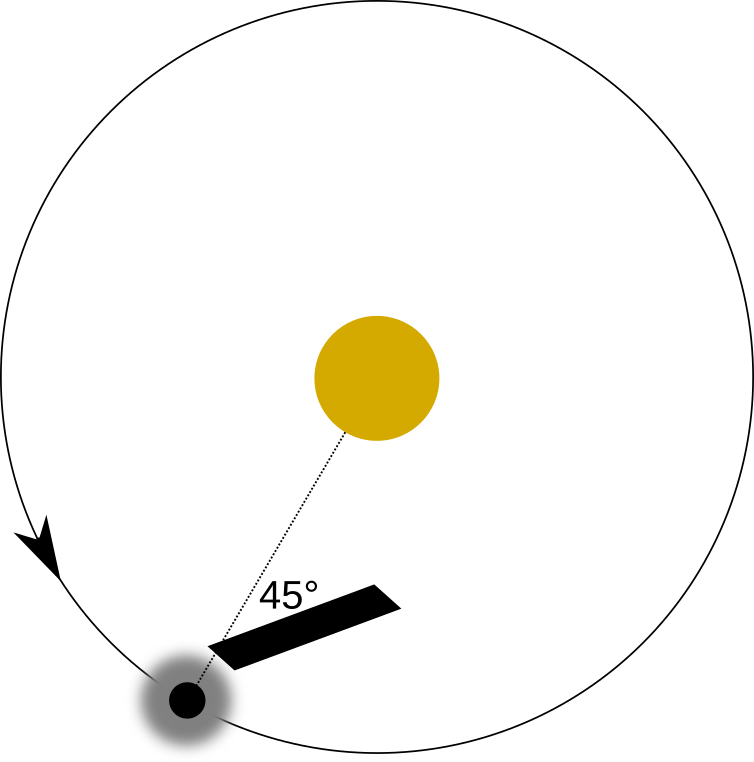}
    \caption{Sketch of the \hp\ system (approximate scales). The planetary body (black circle)
    moves along its orbit (solid line). The direction of orbital motion is indicated
    by an arrow head. The planetary body is surrounded by
    an atmosphere (gray shade). The model up-orbit stream (black rectangle) originates from the first Lagrange point.
    \label{fig:sketch}}
\end{figure}

\subsection{Calculation of synthetic transmission spectra}

We calculated synthetic transmission spectra
with an approach similar to that used by \citet{Salz2018}.
The absorption cross section per atom, $\sigma(\lambda)$, was modeled using Gaussian profiles, $G$,
parameterized by 
the rest wavelength, $\lambda_0$, a velocity shift, $v_s$,
the oscillator strength, $f$, and the velocity dispersion, $v_d$.
The oscillator strength and rest wavelength were adopted from \citet{Drake2006} and
the National Institute of Standards and Technology (NIST, see Table~\ref{tab:linepars}). 
For the \hei\ lines, we used a superposition of three components for the cross section so that
\begin{equation}
    \sigma_{\heion}(\lambda) = \sum_{i=1}^3 G_i(v_d, f_i, \lambda_{\heion,0,i}, v_s) \; .
\end{equation}

\begin{table}
\centering
\caption{Line parameters adopted in the modeling.
\label{tab:linepars}}
\begin{tabular}{l r l}
\hline\hline
Line & Wavelength & Oscillator strength \\
     & [\AA] & \\ \hline
H$\alpha$ & 6564.60 & 0.64108 \\
\ion{He}{I}$_1$ & 10832.0575 & 0.05990 \\
\ion{He}{I}$_2$ & 10833.2168 & 0.17974 \\
\ion{He}{I}$_3$ & 10833.3064 & 0.29958 \\
\hline
\end{tabular}
\end{table}

If (non-overlapping) fractions, $f_j$, of the stellar disk (measured in units of the disk area)
are covered by absorbers with local column densities $n_j$
producing optical depths $n_j\,\sigma_j(\lambda)$,
the empirical transmission spectrum, $T$, is approximated by
\begin{equation}
    T(\lambda) = R_{\rm inst} * \left( 1 - \sum_j f_j \left( 1 - \exp\left(-n_j \, \sigma_j(\lambda)\right)\right)\right) \, , \\
    \label{eq:tmodel}
\end{equation}
where $R_{\rm inst}\,*$ represents convolution with the instrumental profile, which we assume to be Gaussian.
In this approach, the stellar disk is treated as a homogeneous source of light, neglecting CLV
and other sources of inhomogeneity \citep[][]{Czesla2015, Yan2015}.

It is computationally convenient to apply the instrumental broadening directly to the optical depth profile
\begin{eqnarray}
    T(\lambda)  &\approx&  1 - \sum_j \left( f_j - f_j\, \exp\left(-n_j \, R_{\rm inst} * \sigma_j(\lambda)\right)\right) \; .
     \label{eq:tmodelapprox}
\end{eqnarray} 
The first two orders in the Taylor expansion of the resulting expression are identical to that of the
original one (App.~\ref{sec:gaussconv}). The approximation is, thus,
accurate in the optically thin limit. In our phenomenological modeling, optical depths on the order of one
have to be dealt with and the instrumental resolution is high compared to the line width. Therefore,
higher-order corrections terms remain small and can be neglected.

The observationally obtained transmission spectra corresponding to the six orbital phase ranges defined in
Sect.~\ref{sec:timeResolvedTS} all rely on the combination of two or more individual in-transit spectra, which correspond to different
orbit configurations. To obtain synthetic transmission spectra, we first obtain model spectra appropriate
for the mid-exposure time for all individual observations during nights~1 and 2. Subsequently, we average
the synthetic spectra using the same scheme as that for the observations.

\subsection{Treatment of phase smearing}
\label{sec:smearing}

The problem of phase smearing
is caused by the long exposure times \citep[e.g.,][]{RiddenHarper2016}.
A single exposure of $900$~s comprises about $0.5$\,\% of the orbit of \hpb\ or 8\,\% of the optical transit duration.
At mid-transit time, this means that the RV of \hpb\ changes by $5.3$~\kms\ during any single exposure
due to the planetary orbital motion. Even assuming that the profile of the transmission spectrum remains
constant during the exposure, which is not necessarily the case \citep[][]{Deming2017},
this effect distorts the observed line profile in the transmission spectrum
compared to the instantaneous profile.

The effect of phase smearing
can be taken into account in the modeling by applying a convolution with a box-shaped broadening kernel
\citep[][]{Cauley2020, Wyttenbach2020}.
Here we used an effective value for the instrumental resolution, determined by
evaluating the effect of the box-shaped convolution on
the instrumental broadening function. 
An appropriate value can be derived by demanding that the variance of the profile
resulting from time integration is reproduced. The result remains
an approximation in the sense that we reproduce the variance but not necessarily the profile;
an in-depth discussion of the role of instrumental resolution in transmission spectroscopy can be found
in \citet{Pino2018}.  
In the case of \hpb, effective values of $63\,000$ and $58\,000$ for the instrumental resolution
fulfill this condition for the VIS and NIR channels of CARMENES (see Sect.~\ref{sec:tops}).

\subsection{Circumplanetary atmosphere and super-rotation}
\label{sec:circToyModel}
First, we approximated the planetary atmosphere by a face-on annulus, which surrounds the
opaque disk of the planet, reaching from the planetary surface at \rpp\ to
an outer radius $R_{a,\rm out}$. We assumed constant
surface column densities, $N_a$, of absorbers in the H$\alpha$ and \hei\ lines. 
As shown in Sect.~\ref{sec:atmModeling}, the true atmospheres show a distribution of column densities. Therefore,
the value derived here should be understood as an effective number, representing the value reproducing the line profile
best.
The material in the atmospheric annulus is dragged along with the planetary frame. In a first step, we allowed
for broadening with velocity $v_{t,a}$, which captures all broadening mechanisms and, in a second step, we
considered (super-)rotation of the planetary atmosphere with equatorial wind velocity $v_{\rm w}$ and the
planetary equator lying in the orbital plane.  
To get an idea of the amount of absorbing material and the broadening of the lines, 
we here assumed an atmosphere with a fixed maximum stellar disk coverage fraction of $10$\,\%, corresponding
to an outer radius of $2.3$~\rpp.
As long as no strong saturation effects
or timing effects due to the extent of the atmosphere occur, the values of the atmospheric fill factor and
surface column density remain largely degenerate.  
Treating the surface column density
and the turbulent velocity as free parameters,
we carried out a fit by minimizing $\chi^2$
using the transmission spectra for each of the phases simultaneously.
The fit results for the individual sections of the transit, as well as a comparison by means
of the Bayesian information criterion (BIC) are given in Tables~\ref{tab:fitResults} and \ref{tab:bic}.
The best-fit spectral models
are shown in Fig.~\ref{fig:TSplusModel} and the
corresponding transmission light curves are indicated in Fig.~\ref{fig:lcvb}.

Allowing also the planetary equatorial wind velocity, $v_{\rm w}$, to vary, yields an overall better
fit, which is mainly driven by the ingress phase (Table~\ref{tab:fitResults}). This may be expected, because
this phase is where the asymmetry in the atmospheric motion is most pronounced. The weak signal in the egress phase
does not yield a strong lever to distinguish the models. 
The best-fit wind velocity is around $23$~\kms\ for the \ha\ and \hei\ lines independently, and the turbulent
velocity is diminished because some of the broadening is absorbed by the atmospheric wind motion (Table~\ref{tab:atman}).

\begin{table}
\centering
\caption{Best-fit parameters for annulus, super-rotating wind, and up-orbit stream models. Reduced $\chi^2$ values,
$\chi^2_r$,
are calculated over the $-100 \mbox{ to } +100$~\kms\ range.
\label{tab:atman}
}
\begin{tabular}{l l l}
\hline\hline
                            &  H$\alpha$        & \hei \\ \hline
\multicolumn{3}{c}{Annulus atmosphere model} \\
$N_a$ [$10^{12}$ cm$^{-2}$]   & $1.8 \pm 0.14$     & $2.1 \pm 0.15$ \\
$v_{t,a}$ [\kms]                & $19.3\pm 1.6$     & $15.2 \pm 3.1$ \\
$\chi^2_r$ & 3.48 & 3.06 \\ \hline
\multicolumn{3}{c}{\textbf{Annulus with rotating atmosphere}} \\
$N_a$ [$10^{12}$ cm$^{-2}$]   & $1.8 \pm 0.14$     & $2.2 \pm 0.15$ \\
$v_{t,a}$ [\kms]                & $16.0\pm 1.7$     & $11.2 \pm 3.1$ \\
$v_{w}$ [\kms] & $23.2\pm 4$ & $22.8\pm 5.1$ \\
$\chi^2_r$ & 3.17 & 2.63 \\ \hline
\multicolumn{3}{c}{Up-orbit stream model} \\
$N_a$ [$10^{12}$ cm$^{-2}$]   & $1.5 \pm 0.18$     & $1.8 \pm 0.18$ \\
$v_{t,a}$ [\kms]                & $15.5\pm 2.1$     & $11.7 \pm 1.0$ \\
$v_{\rm max}$ [\kms]        & $90\pm 11$          & $131 \pm 15$ \\
$N_s$  [$10^{33}$]          & $1.7\pm 0.4$       & $3.1\pm 0.5$ \\
$\chi^2_r$ & 2.85 & 2.14 \\ \hline
\end{tabular}
\end{table}

Clearly, this model reproduces the main depression in the transmission spectrum in both lines, but it can neither
explain the difference between the observed central depression during the start and end phases, as the star--planet geometry is symmetric, nor does it provide
an explanation for the pre-transit absorption signal or the line wings. This is also reflected by the associated
transmission light curves in Fig.~\ref{fig:lcvb}, which are reasonable for the line center, but significantly
off for the line wings (see panels~c and f in particular).
Taking the coverage fraction of the atmosphere into account, we converted
the column densities from Table~\ref{tab:atman} into a total number
of $(4\pm 0.3)\times 10^{33}$ H$\alpha$ absorbers
and $(4.7\pm 0.4)\times 10^{33}$ \hei\ absorbers, which corresponds to
$(6.8\pm 0.6)\times 10^{9}$~g and $(3.2\pm 0.3)\times 10^{10}$~g, respectively.

\subsection{Circumplanetary atmosphere plus up-orbit stream}
In a next step,
we extended the circumplanetary (annulus) atmosphere model by a semi ad hoc version of the up-orbit stream model
presented by \citet{Lai2010}. The geometry is depicted in Fig.~\ref{fig:sketch}. From the first
Lagrange point, the up-orbit stream is launched toward the interior of the planetary system. 
In the theory of \citet{Lubow1975} adopted by \citet{Lai2010}, the launching angle is about $30^{\circ}$ with respect
to the radius vector of the planetary body, but Coriolis forces tend to increase that angle. 
The final geometry depends on the physical stream conditions as well as
the interaction with the environment.

\begin{figure*}
            \includegraphics[width=0.49\textwidth]{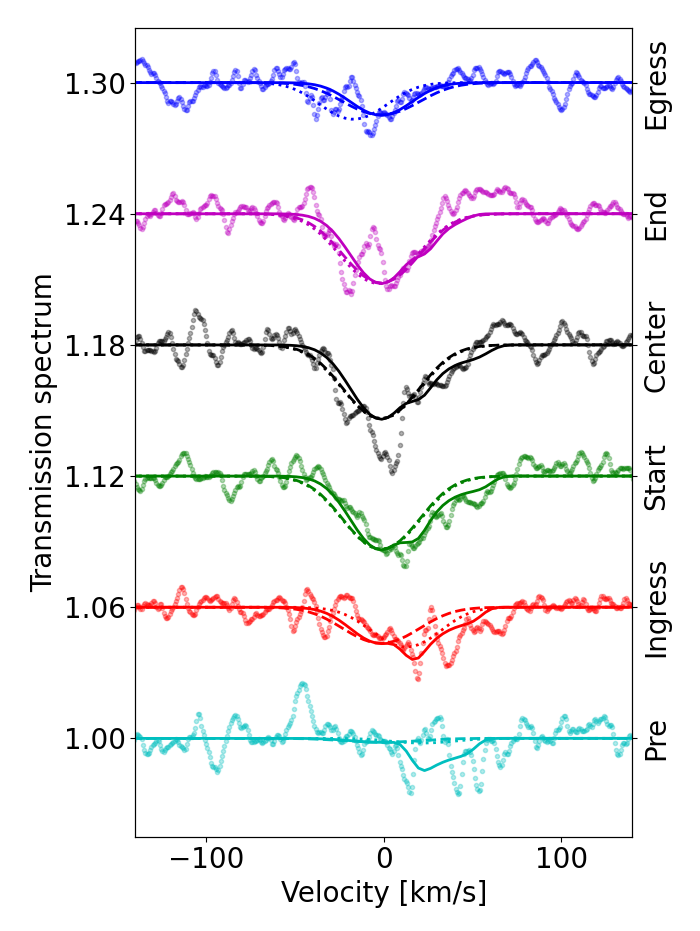}
    \includegraphics[width=0.49\textwidth]{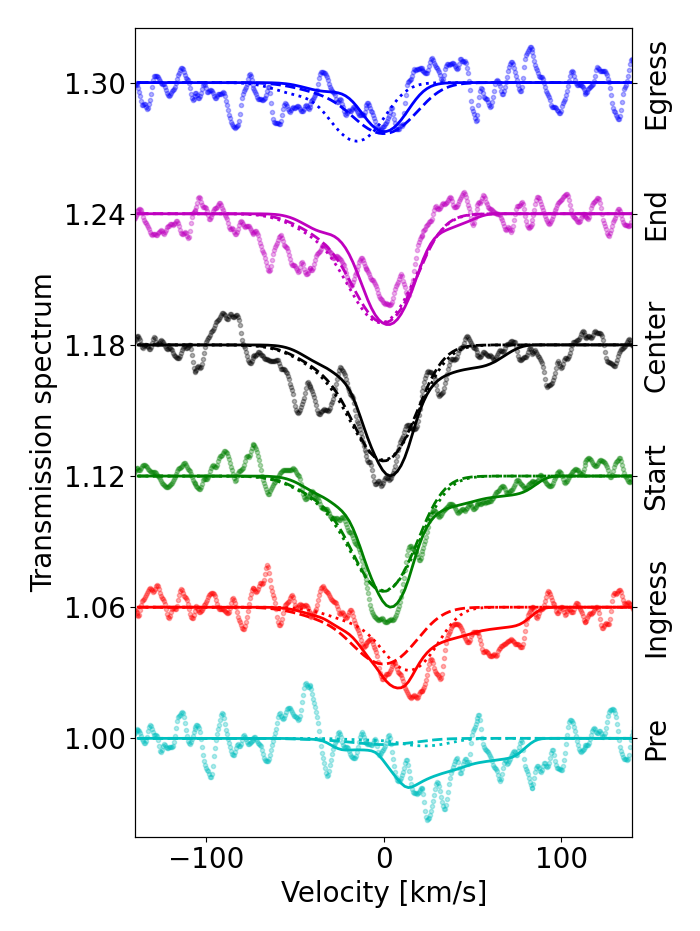}
    \caption{Average transmission spectra along with best-fit models
    for annulus atmosphere (dashed), super-rotating wind (dotted), and up-orbit stream (solid)
    for the H$\alpha$ line (left) and the \hei\ lines (right).
    \label{fig:TSplusModel}}
\end{figure*}

As detailed physical stream modeling is beyond our scope, we assumed a fixed, intermediate value of
$45^{\circ}$ for the stream angle in our phenomenological modeling.
For the stream, we assumed a quadratic cross section, constant along the stream, with an edge length
equal to the diameter of the planet (\mbox{$2$\,\rpp}). We fixed the total length, $L_s$, of
the stream at $1.5$~stellar radii in our modeling. 
While we did not treat the stream length as a free parameter, we note that it is constrained by the ingress timing.
Conceivable physical factors limiting the stream length are
the presence of a circumstellar disk formed by self-interaction of the stream or stellar wind interaction \citep[e.g.,][]{Lai2010}.
The streaming velocity was assumed to increase linearly with the distance, $l$, from the L1 point along the stream,
starting from a minimal velocity, $v_{\rm min}$, of $5$~\kms\ \citep[set by the speed of sound,][]{Lai2010}
to a maximum velocity, $v_{\rm max}$. If $\dot{n}_k$ is the number of particles of type $k$ streaming
through any perpendicular cross section of the
stream with area $A$ per unit time, then continuity requires that
\begin{equation}
    \dot{n_k} = \rho_k(l)\, A\, v_s(l) \; ,
\end{equation}
where $\rho_k(l)$ is the volume particle density.
Here, we neglect that the number density of H$\alpha$ and \hei\ absorbers may change
due to (de-)excitation along the stream. Integration yields the total number of particles in the stream
\begin{equation}
    n_{s,k} = \frac{ L_s\, \dot{n}_k}{(v_{\rm max} - v_{\rm min})} \ln\left(\frac{v_{\rm max}}{v_{\rm min}} \right) \; .
    \label{eq:stream_part_num}
\end{equation}
In our modeling, we only treated $v_{\rm max}$ and $n_{s,k}$ as free parameters. The best-fit parameters are
given in Table~\ref{tab:atman}, and the best-fit spectral models and associated transmission light
curves are shown in Figs.~\ref{fig:lcvb} and \ref{fig:TSplusModel}.

The up-orbit stream model considerably improves the fit to the transmission spectra. In particular, the
pre-transit depression can now be modeled and the red-wing line profile is better approximated
compared to the circumplanetary atmosphere model alone. This is also clearly seen in the
associated transmission light curves shown in Fig.~\ref{fig:lcvb}.
Formally, the improvement in $\chi^2_r$ is highly significant with an F-test yielding $p$-values smaller
than $10^{-6}$. While the fit is significantly better, we caution that this does not prove the
correctness of the model.

Compared to the previous model, both the best-fit values of the surface column densities and the turbulent
velocities are decreased, because some absorption and broadening is now accounted for by the stream component.
The maximum stream velocities of $90$~\kms\ and
$131$~\kms\ for the \ha\ and \hei\ components are roughly compatible with the
model prediction of less than half of the planetary orbital speed of
about $90$~\kms\ for \hpb\ given by 
\citet{Lai2010}.
Based on our
modeling, we obtain streaming rates of $(6.6\pm 1.7)\times 10^4$~\gs\ and $(6.3\pm 1.1) \times 10^5$~\gs\
of \ha\ and \hei\ absorbers.

\subsection{Limits of the model and further components}
Although our phenomenological modeling accounts for the main characteristics of the observed transmissions spectra,
not all aspects are captured. In the following, we outline those aspects along with some speculative
explanations.

The depth of the central part of the \hei\ transmission spectrum
decreases between the start and end phase of the transit (Fig.~\ref{fig:hetresol}), which may be accounted for
by an asymmetric model component such as the streaming funnel. In our implementation,
however, the observed decrease in depth of the central \hei\ component is not reproduced.
Nonetheless, effects caused by the changing line of sight through a potentially asymmetric and at least partially
optically thick atmosphere as well as occultation of parts of the atmosphere by the opaque planetary disk
provide a conceivable explanation for the observation.
Also heterogeneities in an hypothetical circumstellar disk formed by self-interaction of the up-orbit stream and its
impact on the disk may contribute to the observed transmission spectrum \citep{Lai2010},
including the changing depth of the central absorption over the transit.

Particularly during the end phase of the transit,
a blueshifted absorption component at a velocity of around $-50$~\kms\ appears to be present, which is not
accounted for by our modeling. This signal may be attributable to the weaker, bluest component of
the \hei\ triplet, the contribution of which becomes stronger through absorption by larger column densities
\citep[][]{Salz2018}. Conceivably, material might also streams across the second Lagrange point
after which it is accelerated away from the star, potentially forming a comet-like tail
\citep[e.g.,][]{CarrollNellenback2017, Nortmann2018}. 
A high-velocity \ion{He}{i} population produced by charge
exchange with the stellar wind is also conceivable; however, we would expect that to be observable during the
entire transit, which does not seem to be the case.
 
In the central transit phase, the \ha\ transmission line shows a central absorption component, which is
deeper than that of the model.
During the subsequent end phase of the transit, the H$\alpha$ line transmission spectrum shows
a marked double-peak structure, which
is not observed during the start phase of the
transit (Fig.~\ref{fig:hatresol}).
Such a structure may possibly be
caused by a circumplanetary disk, seen nearly edge-on. However, the planetary Roche lobe is almost filled
by the planetary body already (Sect.~\ref{sec:Roche}), leaving limited space for a gravitationally bound disk,
and the signature would somehow have to be suppressed in the remaining transit phases.

\section{Hydrodynamic atmospheric modeling}
\label{sec:atmModeling}

We now turn to hydrodynamic modeling of the atmosphere of \hpb\ to further explore the properties of the outflow.
To that end, we applied the two separate methodologies 
of \citet{GarciaMunoz2019} for the \ha\ line and that of \citet{Lampon2020} to investigate the \hei\ lines.
A joint analysis of the lines and their temporal variability based on hydrodynamic modeling is
deferred to future work.
Throughout this section,
we refer to the center-phase transmission spectrum for the comparison with the models. 

\subsection{Modeling of the \ha\ transmission signal}
\label{sec:atmHa}

We used the model developed by \citet{GarciaMunoz2019} to predict the population of 
hydrogen atoms excited into the H(2) state that causes the \ha\ absorption in the upper atmosphere of {\hpb}. 
The model builds upon past approaches to hydrodynamic escape \citep{GarciaMunoz2007} by incorporating a non-local thermodynamic equilibrium (NLTE) treatment of the hydrogen atom, together with the description of the direct (stellar) and diffuse radiation components within the gas. 
The simultaneous solution to the continuity, momentum, and energy conservation equations allows us to trace the transition from a hydrostatic gas at the $\sim$1 $\mu$bar pressure level to a rapidly escaping and much hotter gas at pressures of $0.02-0.03$~$\mu$bars, where the {\ha} lines originate. The model is one-dimensional and takes the radial distance to the planet center as the only spatial variable in the equations.
It takes as input the stellar spectrum described in Sect.~\ref{sec:irradiation}. Particularly important are the energy fluxes incident upon the planet in the XUV ($<\,912$~{\AA}) and the NUV ($912-3646$~{\AA}).

We set the base of the upper atmosphere in our models at the pressure level 
of 1~$\mu$bar. 
It is not straightforward to translate this into a radius R$_{p,1\mu\rm{bar}}$ because the transition from the $\sim$mbar pressure level probed at optical and NIR wavelengths to the $\mu$bar level depends on the temperature and the dissociation/ionization state of the intervening gas, which is not well constrained. 
We estimated that R$_{\rm{p},1\mu\rm{bar}}$/R$_p$=1.1 and used this choice for our hydrodynamic calculations. 
We confirmed that the solution is not strongly sensitive to the specific choice.
Our current version of the NLTE-hydrodynamic model includes atomic hydrogen 
(with excitation states up to the principal quantum number $n$=5), protons and electrons but not molecular hydrogen, helium or other heavier atoms. 
Although the main features of the flow are dictated by the dominating hydrogen gas, as shown by \citet{GarciaMunoz2007}, the omission of helium prevents for the time being
the direct comparison with the \hei\ line measurements.

\begin{figure}
        \includegraphics[width=0.49\textwidth]{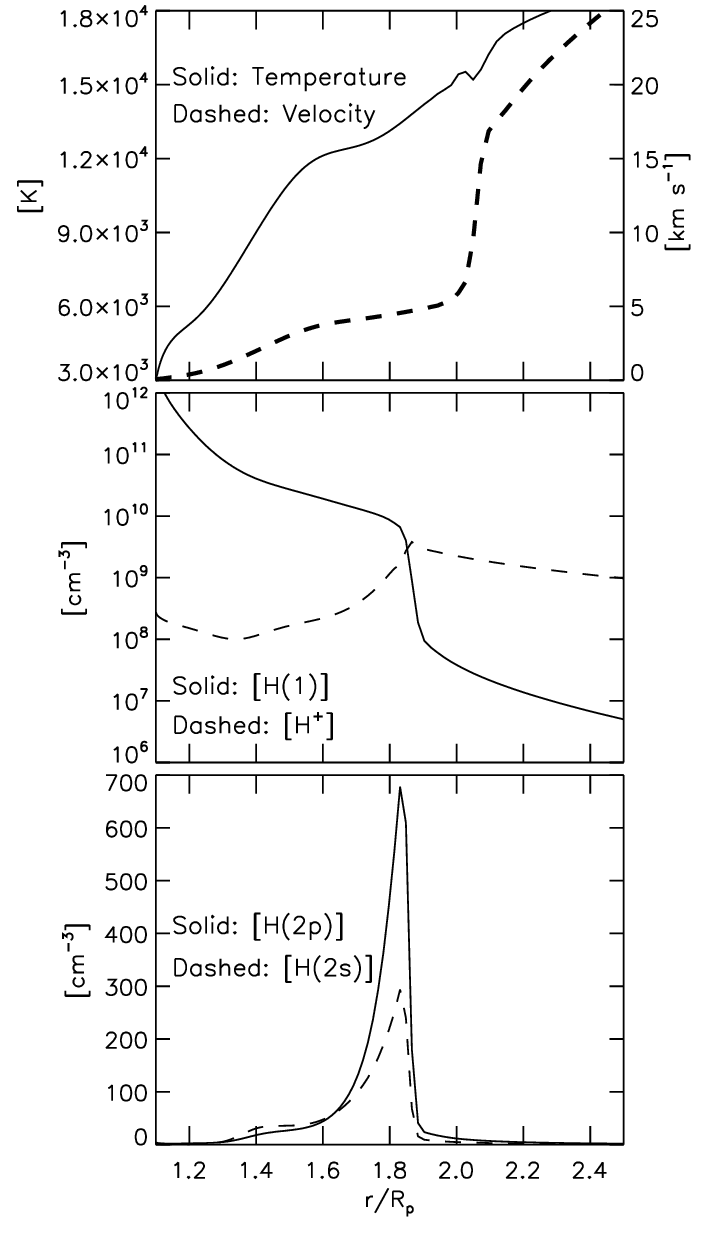}
    \caption{Solution to the hydrodynamic problem for the investigation of the {\ha} line.
    \label{fig:panel_ha}
    }
\end{figure}

\begin{figure}
            \includegraphics[width=0.49\textwidth]{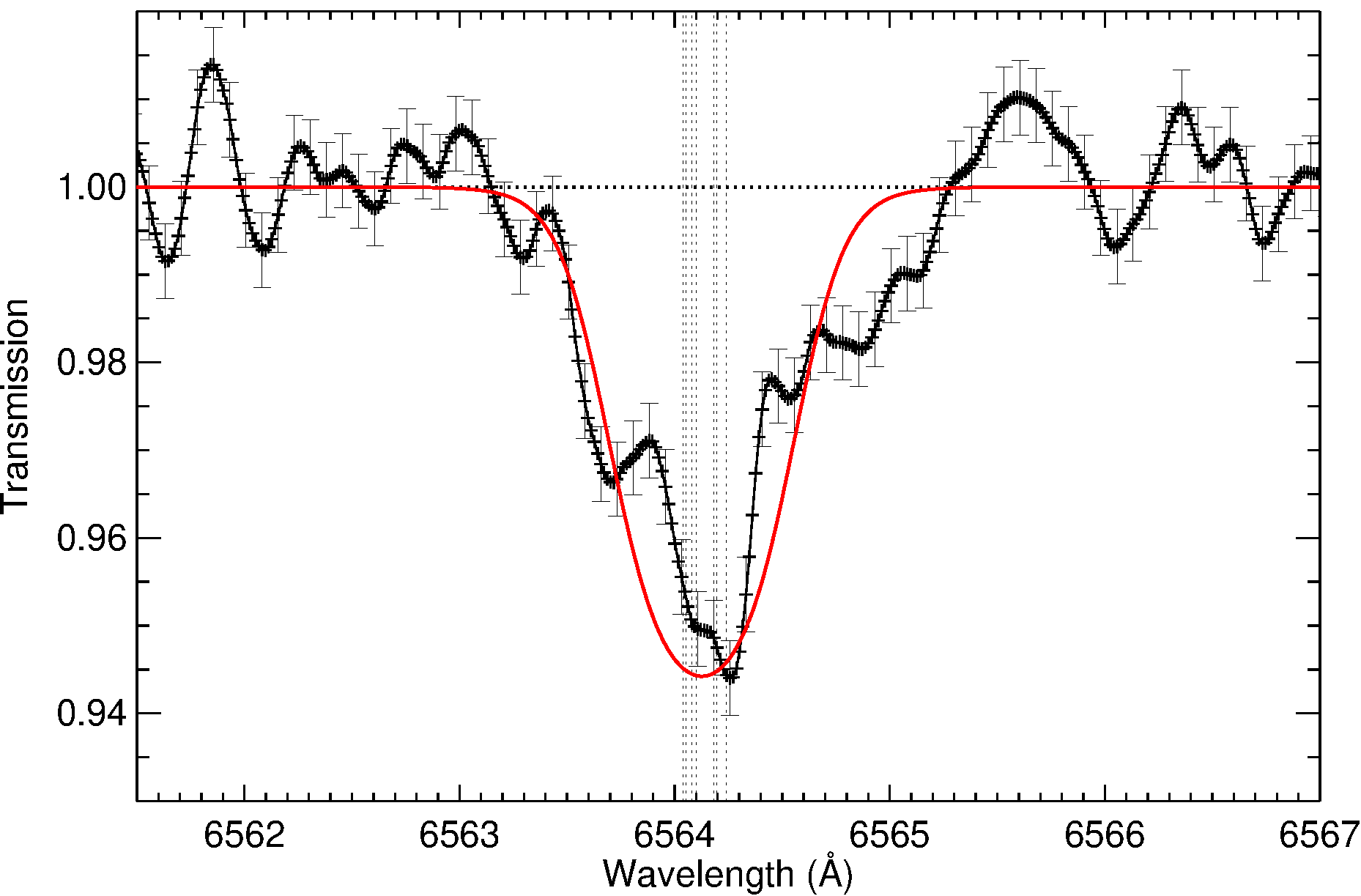}
    \caption{Synthetic transmission spectrum of {\hpb} (red) in the {\ha} line for the 
    conditions of Fig. \ref{fig:panel_ha}. Wavelengths are shifted by systemic motion of \hp.
    \label{fig:fig_ha}}
\end{figure}

Figure \ref{fig:panel_ha} (top panel) shows that 
temperatures of up to 18\,000~K and supersonic velocities as high as 25~\kms\ are reached in the vicinity of the planet. 
From these simulations, 
we estimate that the planet is losing mass at a rate $\pi\rho u r^2$=1.4$\times$10$^{13}$~\gs, where $\rho$ is the
density, $u$ the velocity, and $r$ the respective radius. 
Correspondingly, the middle and bottom panels show that most of the H(2) state is formed within a relatively narrow layer 
at $\sim 1.8$ planetary radii. 
In this region, the gas transitions rapidly from mostly neutral to mostly ionized, a fact that tends to maximize the excitation of H(2) from collisions of H(1) and electrons.
This layer has an optical thickness at the line core well in excess of one and causes the
high-altitude layer of {\ha} detected around the planet.
Although the altitude and strength of this layer can be sensitive to a number of factors that will be explored in future work, it is apparent that both observations and model predictions are consistent (Fig. \ref{fig:fig_ha}).

\subsection{Modeling of the \hei\ transmission signal}
\label{sec:atmHe}

To analyze the \hei\ absorption signal we used the model described by \citet{Lampon2020}.
The model has two major components, first, the solution of the hydrodynamical equations and, second, the computation of the NLTE population of the \het\ excited state responsible for the planetary atmospheric absorption.

The hydrodynamical model is a variation of the isothermal spherically symmetric Parker wind approach \citep{Parker1958}, with the main difference being that not the temperature but the speed of sound, $v_s$, is assumed to be constant. This is given by
\begin{equation}
v_s \,=\, \sqrt{k\,T(r)/\mu(r)} \; ,
\label{eq:vs}
\end{equation}
where $k$ is the Boltzmann constant, $r$ is altitude, $T(r)$ is temperature, and $\mu(r)$ denotes the mean molecular weight 
of the thermosphere. 
By construction, the thermosphere shows an altitude-independent $T(r)/\mu(r)$ ratio.
The value of the speed of sound is given by $v_s= \sqrt{k\,T_0/\bar{\mu}}$, where $\bar{\mu}$ is the averaged mean molecular weight, calculated in the model, and $T_0$ is a free model parameter that is similar to the maximum of the thermospheric temperature profile calculated by comprehensive hydrodynamic models that solve the energy balance equation \citep[e.g.,][]{Salz2016, GarciaMunoz2019}. A novelty of this hydrodynamical model is the way in which the averaged mean molecular weight is computed \citep[see Eq.\,A.3 in][]{Lampon2020}, which makes the convergence of the solution fast. In addition to the temperature $T_0$, the model includes two other free parameters, the hydrogen-to-helium number ratio, and the atmospheric mass-loss rate, \mlr, which here refers
to the substellar value scaled by planetary surface area.

Along with the density and RV profile, the model yields the distribution of the species H, H$^+$, \hes\, He$^+$, and \het\ by solving the hydrodynamical equations and their respective continuity equations. The production and loss terms of those species are detailed in Table~2 of \citet{Lampon2020}. The computation of those quantities requires as input the stellar flux received at the top of the planetary atmosphere (see Sect.~\ref{sec:irradiation}).

Based on the \het\ radial distribution, the model computes the \het\ absorption by using a radiative transfer code for the primary transit geometry \citep[see Sect.\,3.3 in][]{Lampon2020}. The absorption coefficients and wavelengths for the three metastable helium lines were taken from the NIST Atomic Spectra Database\footnote{\url{https://www.nist.gov/pml/atomic-spectra-database}} (see Table\,\ref{tab:linepars}). 
The lines are assumed to have Gaussian Doppler line shapes with two broadening contributions, viz., thermal broadening and a second optional component, produced by turbulence with a velocity scale given by the speed of sound ${\rm v}_{turb}$\,=\,$\sqrt{5kT/(3m)}$, where $m$ is the mass of a He atom. In addition to microscopic broadening, the model also accounts for the broadening of the lines caused by the bulk wind motion of the material in the atmosphere along the observer's line of sight \citep[see Eq.\,15 in][]{Lampon2020}.

\begin{figure}
\includegraphics[angle=0,width=0.49\textwidth]{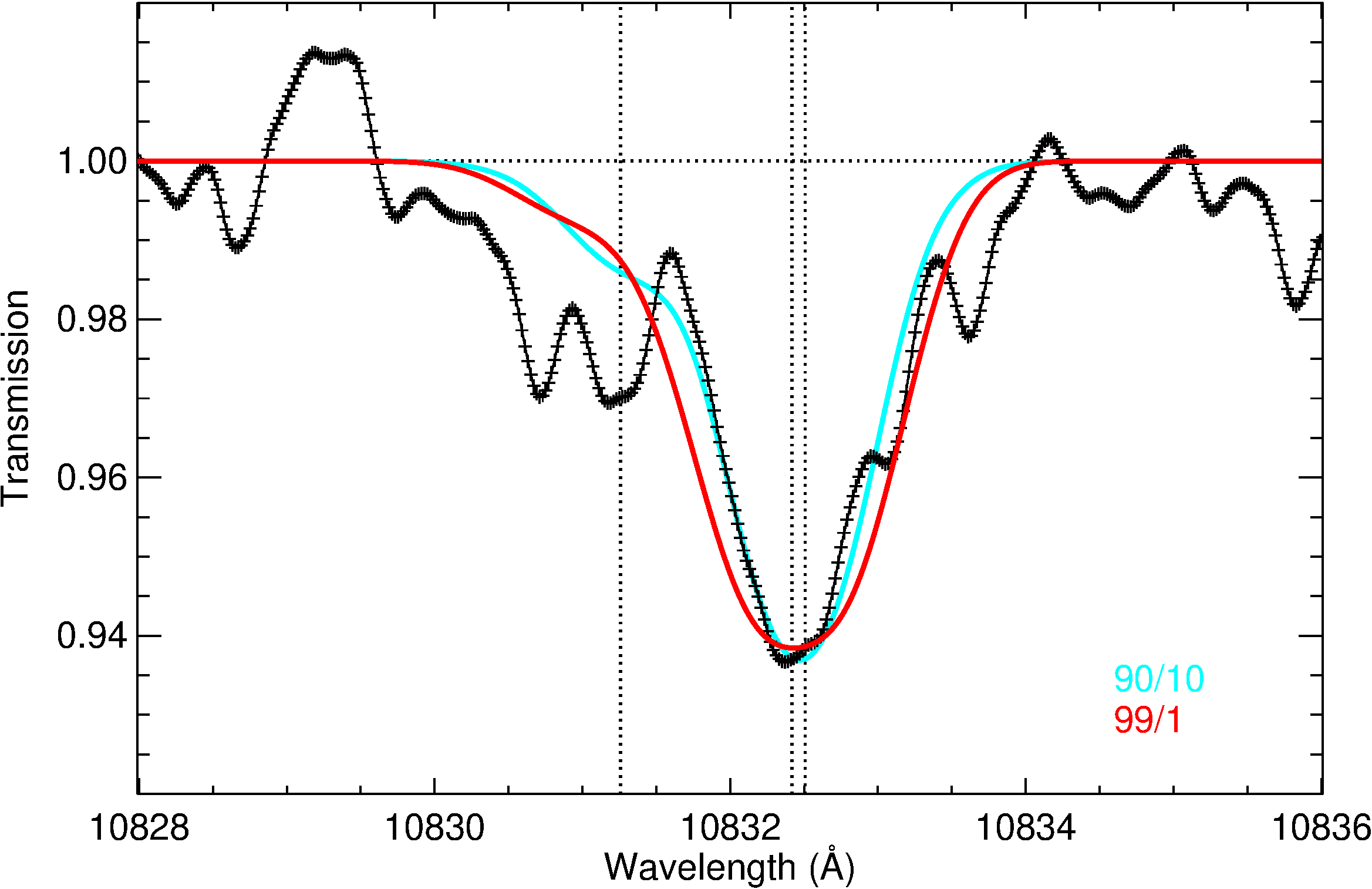}
\caption{Planetary \hei\ absorption profile for the ``center" phase (mid-transit,  see Fig.\,\ref{fig:hahetresol}) compared to two model absorption profiles for H/He ratios of 90/10 (cyan) and 99/1 (red) (see Fig.\,\ref{fig:manuel_den}) The 
vertical dotted lines indicate the positions of the lines.  
    \label{fig:manuelprof}}
\end{figure}

\begin{figure}
\includegraphics[angle=90,width=0.49\textwidth]{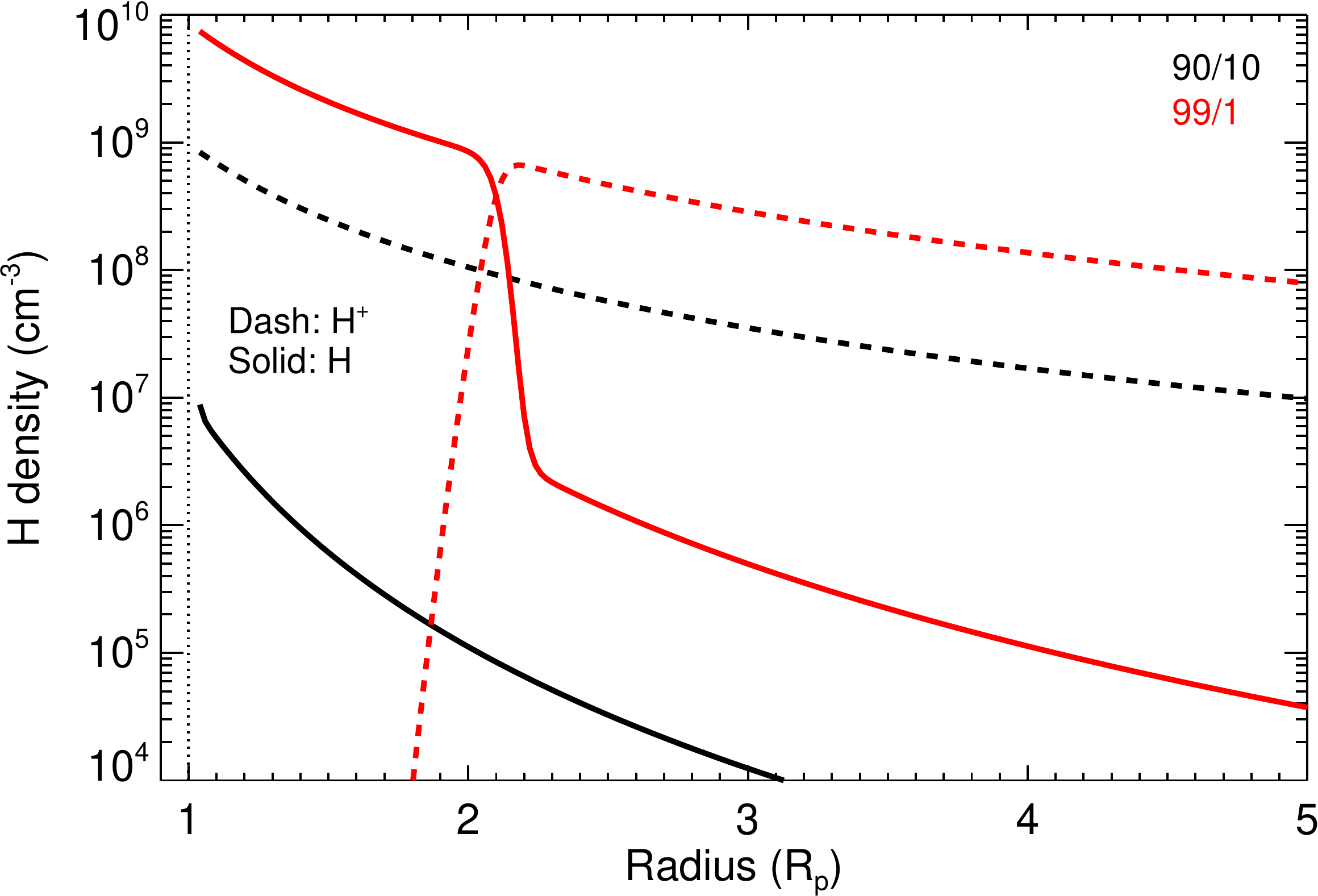}
\includegraphics[angle=90,width=0.49\textwidth]{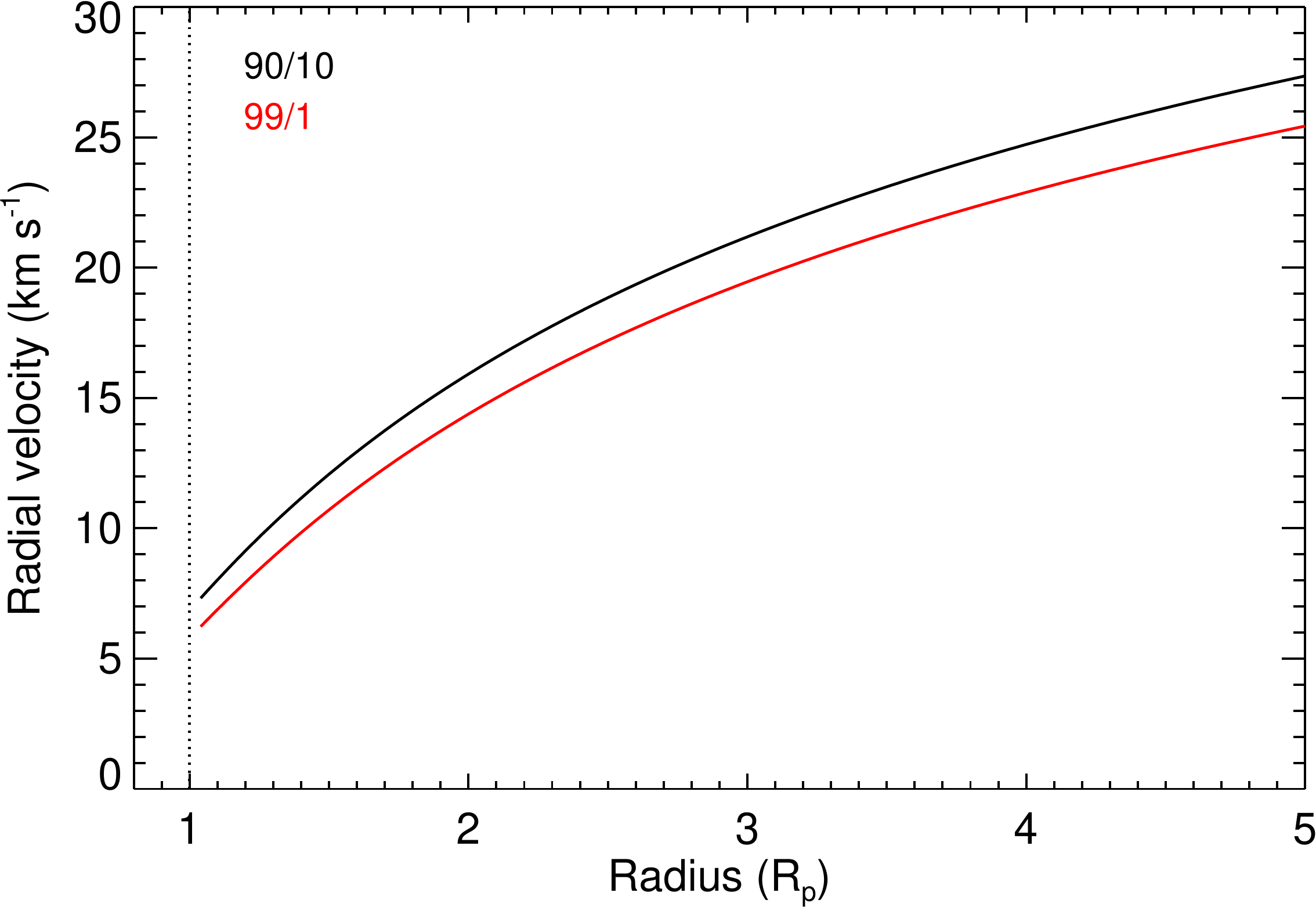}
\includegraphics[angle=90,width=0.49\textwidth]{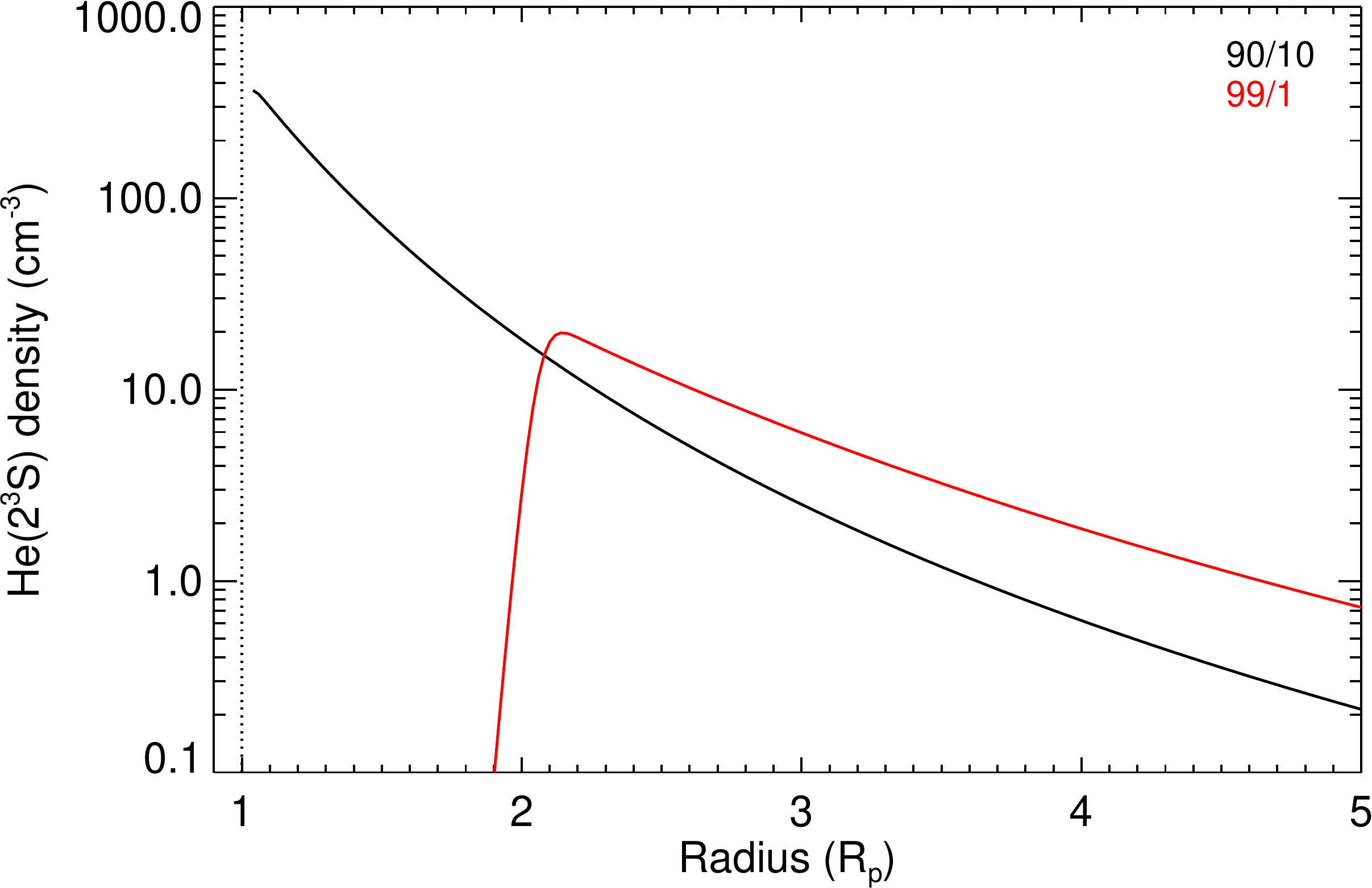}
\caption{Outputs of the He hydrodynamics model showing the H and H$^+$ densities (top), the radial wind velocity (middle), and the \het\ density (bottom) obtained from the fit of the observed \het\ absorption (see Fig.\,\ref{fig:manuelprof}). Profiles are shown for the canonical H/He ratio of 90/10 (black) and for a high H/He ratio of 99/1 that resembles the model of H$\alpha$ (red). The temperature was 14,000\,K in both cases and the mass-loss rate of 3.6$\times 10
  ^{12}$\,\gs\ and 1.6$\times 10^{13}$\,\gs, respectively. }
    \label{fig:manuel_den}
\end{figure}

In Figure~\ref{fig:manuelprof}, we show the resulting absorption profiles adopting two H/He ratios. First, the canonical cosmic value of 90/10 and, second, a value of 99/1, which resembles the assumptions of the  H$\alpha$ model, which does
not currently account for helium (see Sect.\,\ref{sec:atmHa}).
We adopted a value of $14\,000$~K for $T_0$ in accordance with the temperature obtained by the H$\alpha$ model at the altitude where the ionization front is predicted (see Fig.~\ref{fig:panel_ha} and Sect.~\ref{sec:atmHa}).
The model was run for different mass-loss rates to fit the measured \het\ absorption profile (see Fig.~\ref{fig:manuelprof}). In this way, we estimated 
mass-loss rates of $3.6\times 10^{12}$~\gs\ and $1.6\times 10^{13}$~\gs\ for H/He ratios of 90/10 and 99/1, respectively.

The \het\ density profiles depend significantly on the H/He ratio (see Fig.\,\ref{fig:manuel_den}, lower panel). For a H/He ratio of 90/10 it exhibits a moderately compressed shape with a peak density of about 300\,atoms\,cm$^{-3}$ close to the lower boundary of the model at 1\,$\mu$bar (1.02\,\rp). The density then decreases exponentially. At 3\,\rp\ it has fallen by a factor of about 100. For a H/He ratio of 99/1,
the ionization front occurs at about 2\,\rp\ (see Fig.\,\ref{fig:manuel_den}, upper panel), where
the density profile peaks. Only above that altitude are electrons, mainly produced by hydrogen ionization,
are available to form \het\ through recombination.

For a H/He ratio of 90/10, the two stronger \het\ lines, which account for the major absorption peak, are saturated at radii smaller than $\sim$1.5\rp, leading to a relative strengthening of the absorption of the weaker triplet line compared to optically thin conditions \citep[][]{Salz2018}. This effect partially accounts for the underestimation of the absorption in the weak \het\ line component (see Fig.~\ref{fig:manuelprof}). An even more compressed atmospheric component could possibly explain the measured relative absorption between the \hei\ triplet components \citep{Lampon2021a}.
Absorption in the stronger triplet lines remains significant up to radii of 4\,\rp.

The RVs computed by our hydrodynamical model reach from about $6-7$\,\kms\ at low altitudes, from where they steadily increase up to approximately 25\,\kms\ at 5\,\rp. This velocity field broadens the absorption lines and appropriately explains the observed broadening of the stronger line cores for an H/He value of 90/10 (cyan curve in Fig.~\ref{fig:manuelprof}). Thermal broadening alone remains insufficient to explain the observed line width, even if the turbulence term is included.
For the larger H/He ratio of 99/1, we observe that the broadening is slightly overestimated, because absorption takes place at higher altitudes, where the velocities are larger (bottom panel of Fig.\,\ref{fig:manuel_den}, red curve).

Overall, our model reproduces the observations reasonably well. Because of the symmetry of the model, it is clear that the model transmission profile is also symmetric. Nonetheless, the model provides a plausible physical interpretation of the observed overall shape of the absorption profile.

\section{Discussion}

We analyzed two transit time series of the hot Jupiter \hpb\ obtained with CARMENES and found
prominent transmission signals in both the \ha\ and \hei\ lines, which we attribute to the planetary
atmosphere.

\subsection{Putting the transmission signals in context}

None of our results contradicts the presence of a cloud deck, which has been suggested
by several studies (Sect.~\ref{sec:intro}). We speculate that the non-detection of \ha\ absorption
by \citet{Mallonn2016b} is caused by a lack of sensitivity in that study, which was not based on
high-resolution spectra. However, time variability in the \ha\ absorption cannot be excluded as a
confounding factor. 

In Table~\ref{tab:contrastcomp}, we list the planetary semimajor axis, mass, radius, and density along with the 
transmission contrasts of the \ha\ and \hei\
lines for several planets discussed in the literature in order of ascending host star effective temperature.
For \hpb\ the contrast is time dependent, so we report an average
value derived from our phenomenological modeling presented in Sect.~\ref{sec:circToyModel}.
To the best of our knowledge, no simultaneous detection of \ha\ and \hei\ transmission in a single planetary
atmosphere has so far been reported in the literature. In HD~189733\,b, however, separate detections of
both lines have been published. 

Among the planets in Table~\ref{tab:contrastcomp}, \hpb\ shows the lowest density estimate, which is,
however,
not exceptionally low as comparison with WASP-107\,b or WASP-121\,b demonstrates.
In terms of reported contrast, the atmosphere of \hpb\ is only second to the atmospheres of WASP-12\,b
and WASP-107\,b in the \ha\ and \hei\ lines. While detections of \hei\ transmission have
predominantly been found in planets orbiting stars with lower effective temperature than
\hp, reports of planetary \ha\ transmission signals
tend to be associated with planets orbiting stars with higher effective temperatures.
While the photospheric NUV continua intensify toward higher effective
temperatures, magnetically driven activity phenomena, such as coronal X-ray emission, are thought to
vanish at about spectral type A (in main sequence stars), where the outer convection zone disappears
\citep[$T_{\rm eff} \approx 8000$~K, e.g.,][]{Guedel2004}.
It seems that
the mass and effective temperature of the host star \hp\ is in an intermediate region, where
the mass remains sufficiently low for the star to drive
a magnetic dynamo in an outer convection zone, which is thought to power
the EUV emission required to populate the metastable helium level in the planetary atmosphere, and its effective temperature
is high enough to foster conditions favoring the population of
the H(2) ground level of \ha\ there.

Our results show no significant difference between the observations carried out on night~1 and night~2. This
is consistent with the findings of \citet{Kirk2020} and \citet{Allart2019} in WASP-107\,b
and \citet{Salz2018} in HD~189733\,b,
who also find no detectable inter-transit variation in the \hei\ transmission signal.
\citet{Borsa2021} report \ha\ line absorption in two transits of WASP-121\,b with potentially variable contrast 
between the transits, which may be attributed to changes in the atmospheric blueshift or stellar activity levels.
\citet{YanDongdong2021} reproduce the \ha\ absorption lines using one-dimensional hydrodynamic simulations with detailed radiative transfer modeling and find that different XUV irradiation levels caused by a change in the stellar activity state during the two observations are a
viable explanation for the observation.

\citet{Allart2019} and \citet{Kirk2020} find
evidence for blueshifted \hei\ absorption after the optical transit, which may be indicative of escaping material
trailing WASP-107\,b. Similarly, a prolonged \hei\ transit was found by \citet{Nortmann2018} in WASP-69\,b.
In \hpb, we find no convincing evidence for post-transit absorption in the \ha\ or \hei\ lines. However,
we see marked pre-transit absorption. \citet{Salz2018}
report redshifted \hei\ line absorption during ingress in HD~189733\,b, which may be caused by a
super-rotating atmosphere \citep{Lampon2021a}, but no pre-transit absorption was reported. \citet{Cauley2015} and \citet{Cauley2016}
report on time-variable pre- and post-transit \ha\ line absorption in HD~189733\,b. The cause is, however,
confounded by activity phenomena \citep{Barnes2016, Cauley2017, Kohl2018}.

\begin{table*}
\caption{Compilation (possibly incomplete) of \ha\ and \hei\ transmission line contrasts, listed
in order of ascending effective host star temperature.
\label{tab:contrastcomp}}
\centering
\begin{tabular}{l r l l l l l l l}
\hline\hline
System & $T_{\rm eff}$$^a$ [K] & a$^a$ [$10^{-2}$~au] & $M_p$$^a$ [$M_{\rm Jup}$] & $R_p$$^a$ [$R_{\rm Jup}$] & $\rho_p$ [$\rho_{\rm Jup}$]  & C(\ha) & C(\hei) & Refs.$^e$ \\
\hline
 GJ 3470 & 3600 & 3.6 & 0.04  & 0.37 & 0.84      &  &  $1.5\pm 0.3$\,\% & 1 \\
 WASP-107 & 4430 & 5.5 & 0.12  & 0.97 & 0.13      &  & $7.26\pm 0.24$\,\% & 2 \\
 WASP-69 & 4700 & 4.5 & 0.26  & 1.06 & 0.22      &  &  $3.6\pm 0.2$\,\% & 3 \\
 HAT-P-11 & 4780 & 5.3 & 0.08  & 0.42 & 1.10      &  &  $1.2$\,\% & 4 \\
 HAT-P-18 & 4803 & 5.6 & 0.20  & 0.99 & 0.20    &  & $0.46 \pm 0.12$\,\%$^f$ & 5 \\
 WASP-52 & 5000 & 2.7 & 0.46  & 1.27 & 0.22     & $0.86\pm 0.13$\,\% &  & 6 \\
 HD 189733 & 5040 & 3.1 & 1.14  & 1.14 & 0.78      & $1.9$\,\%$^b$  & $1.04\pm 0.09$\,\% & 7 \\
 HD 209458 & 6065 & 4.7 & 0.69  & 1.36 & 0.27      & & $0.9\pm 0.1$\,\% & 8 \\
 HAT-P-32 & 6269 & 3.4 & 0.59 & 1.79 & 0.11       & $3.3$\,\% & $5.3$\,\% & TW$^d$ \\
 WASP-12 & 6300 & 2.3 & 1.36  & 1.79 & 0.24      & $6$\,\%$^c$ & & 9 \\
 WASP-121 & 6460 & 2.5 & 1.18  & 1.81 & 0.20      & $1.87\pm 0.11$\,\% & & 10 \\
 WASP-33 &  7430 & 2.6 & 2.8 & 1.60 & 0.84        & $0.99\pm 0.05$\,\% & & 11 \\
 KELT-20 & 8720 & 5.4 & 1.00  & 1.78 & 0.18     & $0.6\pm0.1$\,\% & & 12 \\
 KELT-9 & 10170 & 3.4 & 2.88  & 1.94 & 0.40     & $1.15$\,\% &  & 13 \\
\hline
\end{tabular}
\tablefoot{References: 1) \citet{Palle2020}, \citet{Ninan2020}; 2) \citet{Kirk2020}, \citet{Allart2019}, \citet{Kasper2020};
3) \citet{Nortmann2018}, \citet{Vissapragada2020};
4) \citet{Allart2018};
5) \citet{Paragas2021};
6) \citet{Chen2020};
7) \citet{Salz2018}, \citet{Guilluy2020}, \citet{Jensen2012}, \citet{Barnes2016},
\citet{Cauley2015}, \citet{Cauley2016}, \citet{Cauley2017}, \citet{Kohl2018};
8) \citet{AlonsoFloriano2019};
9) \citet{Jensen2018};
10) \citet{Cabot2020}, \citet{Borsa2021};
11) \citet{Yan2021}, \citet{Cauley2021}, \citet{Borsa2021};
12) \citet{CasasayasBarris2019}, \citet{CasasayasBarris2018};
13) \citet{Yan2018}, \citet{Cauley2019}, \citet{Wyttenbach2020};
$^{a}$ Values in columns two through five adopted from \url{exoplanets.org} \citep{Han2014};
$^{b}$ Estimated from \citet{Jensen2012} (Fig.~3), detection controversial. $^c$ Estimated from \citet{Jensen2018}
(Fig.~9). $^d$ For this work (TW), the contrast is time variable. We here quote the numbers estimated from the
phenomenological model in Sect.~\ref{sec:circToyModel}.; $^{e}$ If more than one reference is given, the listed results
refer to the first one.; $^{f}$ In 6.35~\AA\ band
}
\end{table*}

\subsection{Hydrodynamical modeling}

The one-dimensional hydrodynamical models presented in Sect.~\ref{sec:atmModeling} can reproduce the
observed \ha\ and \hei\ transmission line cores during the center phase reasonably well. This refers to both the strength
and width of the core transmission. We thus conclude that absorption profiles in the core are likely dominated by the
radial planetary wind component of \hpb. This result is consistent with the findings for the atmosphere
of the warm Neptune GJ~3470\,b \citep{Palle2020, Lampon2021a}, which also shows broadened \hei\ absorption. 

In our \ha\ line modeling, the distribution of H(2) and, therefore, the \ha\ line opacity,
is strongly concentrated within a comparatively narrow shell around the planet. This behavior is reproduced
by our \hei\ line modeling if a high H/He ratio is assumed. A fully integrated hydrodynamical
treatment of \ha\ and \hei\ line transmission was not attempted here and
our hydrodynamical modeling does not, in its current form,
allow the temporal variability of the transmission signal to be addressed.

Under the energy-limited escape hypothesis, a mass-loss rate of about $10^{13}$~\gs\ was predicted 
for \hpb\ (Sect.~\ref{sec:irradiation}).
This rate is consistent to within a factor of a few with the values obtained from our hydrodynamical atmospheric
modeling. As far as our \hei\ modeling is concerned, however,
the solution is not unique, because the mass-loss rate is a free parameter and
different model temperatures and H/He number ratios yield different best-fit estimates.
While an exhaustive exploration of the parameter space and the hydrodynamic escape regime
is beyond this paper \citep[]{Lampon2020, Lampon2021a, Lampon2021b},
consistent values are obtained
for plausible physical assumptions, as demonstrated in Sect.~\ref{sec:atmModeling}.

In terms of \ha\ absorption and modeling,
it is worth comparing the cases of {\hpb} and the ultra-hot Jupiter KELT-9\,b, which also exhibits strong {\ha} absorption \citep{Yan2018,GarciaMunoz2019}. 
For {\hpb}, the modeling presented here shows that the H(2) layer is geometrically narrow and the outflow is principally driven by stellar irradiation in the XUV.
In contrast, the H(2) layer of KELT-9\,b extends significantly toward the lower atmosphere, and stellar NUV irradiation plays a key role at heating the upper atmosphere, thereby driving the outflow and producing a sizable population of H(2). Because the corresponding NUV/XUV ratios of their host star emissions differ considerably ($\sim$140 for {\hpb} and $\sim$7.5$\times$10$^6$ for KELT-9\,b), the specifics of H(2) excitation are also different in these two \mbox{(ultra-)hot} Jupiters. 

\subsection{The shape of the atmosphere}

The transmission signal of \hpb\ shows a complex pattern of temporal variability with
detectable pre-transit absorption in the red flank. We attempted to model the variation in the signal
using a phenomenological model incorporating, first, a circumplanetary annulus, second, a super-rotating
wind, and third,
an up-orbit stream with velocity components directed in the orbital forward direction
and toward the star (Sect.~\ref{sec:phenomodel}). 
The overall fit quality is best for the up-orbit stream model, followed by the super-rotating wind model
and the annulus model (Table~\ref{tab:atman}).
As is clear from the consideration of the fit quality in the individual phases (Tables~\ref{tab:fitResults}
and \ref{tab:bic}), the main differences are observed in the pre, ingress, and start phases of the transit.
Both the super-rotating wind and up-orbit stream models improve the ingress-phase fit in the \ha\ and \hei\ lines with respect
to the annulus atmosphere model, and the up-orbit stream model provides additional
improvement over the super-rotating wind model in the \hei\ lines. In the start phase, the up-orbit stream model outperforms
the annulus and super-rotating wind models both in the \ha\ and \hei\ lines, because neither of the latter models can produce an
asymmetry in the line profile to compete with the stream model. The same holds true for the pre phase, where only the
up-orbit stream model can provide an improvement. During and after the central phase, it is the annulus model, which tends to
provide the best fit, although the differences are not as pronounced.

Taking the parameters of the up-orbit stream model at face value, the streaming rates of H(2) and \het\ are
about one and four times $10^{5}$~\gs. If all
mass lost from the planet were transported through this stream,
the average fraction of H(2) atoms in the stream would be about $10^{-8}$ and that of He atoms in
the \het\ state would be around $10^{-7}$. The
numbers decrease if less mass is funneled through the stream and 
they depend on the composition. By comparison with our hydrodynamic models, we find that a fraction of H(2)
of $10^{-8}$ corresponds to that found at about two planetary radii. Likewise, the radius corresponding to
a \mbox{He($2^3$S)} fraction of $10^{-7}$ is found at six planetary radii (H/He 90/10). While a detailed hydrodynamic
modeling is clearly required to understand the physical regime, this demonstrates that the order of magnitude of
these numbers are not at odds with those of the outer planetary atmosphere.  
As the circumplanetary environment around \hpb\ is complex, further atmospheric components, such as a circumplanetary
disk, may be present, which have not been accounted for in our phenomenological modeling so far. 

\section{Conclusion}

\hp\ is a highly active F-type star with an X-ray luminosity of
$2.3\times  10^{29}$~\ergs ($5$--$100$~\AA\ band) as measured by \xmm. The star is orbited by the low-density
planet \hpb\ on an oblique, circular orbit.
Energy-limited escape considerations lead to an estimate on the order of $10^{13}$~\gs\ for
the mass-loss rate.

Our \carm\ high-resolution optical and NIR transmission spectra show pronounced absorption signals in the
\ha\ and \hei\ triplet lines, which we attribute to the planetary atmosphere. To the best of our knowledge,
\hpb\ is the first exoplanet in which both \ha\ and \hei\ signals could simultaneously be detected and
both absorption
signals are among the strongest yet observed in planetary atmospheres. The transmission signals are
time dependent. In particular, an early ingress of redshifted
absorption is observed in both lines, though more clearly in the \hei\ triplet lines. We used phenomenological
transmission models to demonstrate that this signal is plausibly caused by an up-orbit stream.
Such a component is in agreement with predictions of
hydrodynamical models by of comparable systems \citet{Lai2010, Bisikalo2013, CarrollNellenback2017}, and
the configuration resembles the type~III
star-planet interaction scenario described by \citet{Matsakos2015}.

Among the sample of planets with known atmospheric \ha\ or \hei\ absorption signals, \hpb\ is situated on
the low end of the density range (Table~\ref{tab:contrastcomp}). While detections
of \ha\ absorption have predominantly been reported in systems with host stars of higher
effective temperature, those of \hei\ absorption tend to be associated with systems with
cooler host stars. Although the number of systems remains low,
the presumed dichotomy in the distribution of \ha\ and \hei\ signals can plausibly be attributed
the evolution of magnetic activity and NUV photospheric continua as a function of spectral type.

Hydrodynamic one-dimensional models of the mid-transit absorption signals of the \ha\ and \hei\ lines presented in
this paper reproduce the observed line profiles well, indicating that the absorption in the line cores
is explained by a radial planetary wind. The mass-loss rates predicted by the hydrodynamic models are
consistent with $10^{13}$~\gs\ to within factors of a few, which is in line with predictions from
energy-limited escape. The assumption of a constant mass-loss rate yields
a lifetime of about $3.5$~Gyr for \hpb,
comparable to the estimated system age of $2.7\pm 0.8$~Gyr \citep[][]{Hartman2011}.
Clearly, this makes 
hydrodynamic escape a key factor for the planetary evolution of \hpb, which may have
started out with about twice its current mass. 
In terms of host star activity and system geometry, \hpb\ is one of the most peculiar systems observed to date,
making \hp\ a testbed for our understanding of planetary mass-loss and evolution.

\begin{acknowledgements} 
We thank Norbert Schartel for providing \xmm\ DDT time.
We thank M. Salz, F. Bauer, and F.~J. Alonso-Floriano for highly valuable contributions to this paper.
CARMENES is an instrument for the Centro Astron\'omico Hispano-Alem\'an (CAHA) at Calar Alto (Almer\'{\i}a, Spain), operated jointly by the Junta de Andaluc\'ia and the Instituto de Astrof\'isica de Andaluc\'ia (CSIC).
  CARMENES was funded by the Max-Planck-Gesellschaft (MPG), 
  the Consejo Superior de Investigaciones Cient\'{\i}ficas (CSIC),
  the Ministerio de Econom\'ia y Competitividad (MINECO) and the European Regional Development Fund (ERDF) through projects FICTS-2011-02, ICTS-2017-07-CAHA-4, and CAHA16-CE-3978, 
  and the members of the CARMENES Consortium 
  (Max-Planck-Institut f\"ur Astronomie,
  Instituto de Astrof\'{\i}sica de Andaluc\'{\i}a,
  Landessternwarte K\"onigstuhl,
  Institut de Ci\`encies de l'Espai,
  Institut f\"ur Astrophysik G\"ottingen,
  Universidad Complutense de Madrid,
  Th\"uringer Landessternwarte Tautenburg,
  Instituto de Astrof\'{\i}sica de Canarias,
  Hamburger Sternwarte,
  Centro de Astrobiolog\'{\i}a and
  Centro Astron\'omico Hispano-Alem\'an), 
  with additional contributions by the MINECO, 
  the Deutsche Forschungsgemeinschaft through the Major Research Instrumentation Programme and Research Unit FOR2544 ``Blue Planets around Red Stars'', 
  the Klaus Tschira Stiftung, 
  the states of Baden-W\"urttemberg and Niedersachsen, 
  and by the Junta de Andaluc\'{\i}a.
   Based on data from the CARMENES data archive at CAB (CSIC-INTA).
    We acknowledge financial support from the Agencia Estatal de Investigaci\'on of the Ministerio de Ciencia, Innovaci\'on y Universidades and the ERDF through projects 
  PID2019-109522GB-C5[1:4]/AEI/10.13039/501100011033      PGC2018-098153-B-C33                       and the Centre of Excellence ``Severo Ochoa'' and ``Mar\'ia de Maeztu'' awards to the Instituto de Astrof\'isica de Canarias (SEV-2015-0548), Instituto de Astrof\'isica de Andaluc\'ia (SEV-2017-0709), and Centro de Astrobiolog\'ia (MDM-2017-0737), the Generalitat de Catalunya/CERCA programme.
D. Yan acknowledges support by the Strategic Priority Research Program of Chinese Academy of Sciences, Grant No. XDB 41000000 and National Natural Science Foundation of China (Nos. 11973082).
SC and EN acknowledge DFG support under grants CZ~222/3-1 and CZ~222/5-1.
This research has made use of the Exoplanet Orbit Database
and the Exoplanet Data Explorer at exoplanets.org.
This work has made use of data from the European Space Agency (ESA) mission
{\it Gaia} (\url{https://www.cosmos.esa.int/gaia}), processed by the {\it Gaia}
Data Processing and Analysis Consortium (DPAC,
\url{https://www.cosmos.esa.int/web/gaia/dpac/consortium}). Funding for the DPAC
has been provided by national institutions, in particular the institutions
participating in the {\it Gaia} Multilateral Agreement.
\end{acknowledgements}

\bibliographystyle{aa}
\bibliography{helib}

\clearpage
\newpage

\appendix

\section{Observing log}

Tables~\ref{tab:logn1} and \ref{tab:logn2} give the observational logs for nights~1 and 2.

\begin{table}[ht]
\caption{Observational log for night~1.
\label{tab:logn1}}
\begin{tabular}{l l r l l l} \hline\hline
LN & BJD$^a$ & \multicolumn{1}{l}{$\Delta t^b$} & EXPT$^c$ & Overlap$^d$ & Section$^e$ \\
   & [d] & \multicolumn{1}{l}{[h]}        & [s]       &  & \\ \hline
 1 & 0.44186 & $-2.8712$ & 898 &  0.000& \\
 2 & 0.45333 & $-2.5959$ & 898 &  0.000& \\
 3 & 0.46492 & $-2.3179$ & 898 &  0.000& \\
 4 & 0.47683 & $-2.0320$ & 898 &  0.000& \\
 5 & 0.48839 & $-1.7545$ & 898 &  0.000& pre\\
 6 & 0.50000 & $-1.4758$ & 898 &  0.177& ing \\
 7 & 0.51154 & $-1.1989$ & 898 &  0.872& ing\\
 8 & 0.52246 & $-0.9369$ & 898 &  1.000& sta\\
 9 & 0.53400 & $-0.6599$ & 898 &  1.000& sta\\
10 & 0.54493 & $-0.3977$ & 898 &  1.000& sta\\
11 & 0.55709 & $-0.1057$ & 898 &  1.000& cen\\
12 & 0.56803 & $ 0.1568$ & 898 &  1.000& cen\\
13 & 0.58089 & $ 0.4654$ & 898 &  1.000& end\\
14 & 0.59182 & $ 0.7277$ & 898 &  1.000& end\\
15 & 0.60418 & $ 1.0243$ & 898 &  1.000& end\\
16 & 0.61509 & $ 1.2863$ & 898 &  0.679& egr\\
17 & 0.62602 & $ 1.5486$ & 898 &  0.065& egr\\
18 & 0.63773 & $ 1.8297$ & 898 &  0.000& \\
19 & 0.64866 & $ 2.0919$ & 898 &  0.000& \\
20 & 0.65959 & $ 2.3542$ & 898 &  0.000& \\
21 & 0.67136 & $ 2.6367$ & 898 &  0.000& \\
22 & 0.68229 & $ 2.8989$ & 898 &  0.000& \\
23 & 0.69443 & $ 3.1903$ & 898 &  0.000& \\
\hline
\end{tabular}
\tablefoot{\tablefoottext{a}{Barycentric Julian Date minus $2458363$~d corresponding to center of observation},
\tablefoottext{b}{Time from center of optical transit}, \tablefoottext{c}{Exposure time}, \tablefoottext{d}{Time-averaged overlap fraction of planetary with stellar disk}, \tablefoottext{e}{Section: pre-transit (pre), ingress (ing), start (sta), center (cen), end (end), egress (egr)}}
\end{table}

\begin{table}[ht]
\caption{Observational log for night~2.
\label{tab:logn2}}
\begin{tabular}{l l r l l l} \hline\hline
LN & BJD$^a$ & \multicolumn{1}{l}{$\Delta t^b$} & EXPT$^c$ & Overlap$^d$ & Section$^e$ \\
   & [d] & \multicolumn{1}{l}{[h]}        & [s]       &  & \\ \hline
 1 & 0.29350 & $-4.0409$ & 898 &  0.000& \\
 2 & 0.30571 & $-3.7479$ & 898 &  0.000& \\
 3 & 0.31776 & $-3.4587$ & 898 &  0.000& \\
 4 & 0.32953 & $-3.1762$ & 898 &  0.000& \\
 5 & 0.34186 & $-2.8804$ & 898 &  0.000& \\
 6 & 0.35378 & $-2.5943$ & 898 &  0.000& \\
 7 & 0.36539 & $-2.3157$ & 898 &  0.000& \\
 8 & 0.37721 & $-2.0319$ & 898 &  0.000& \\
 9 & 0.38886 & $-1.7524$ & 898 &  0.000& pre\\
10 & 0.40043 & $-1.4747$ & 898 &  0.180& ing \\
11 & 0.41204 & $-1.1961$ & 898 &  0.877& ing\\
12 & 0.42442 & $-0.8989$ & 898 &  1.000& sta\\
13 & 0.43639 & $-0.6116$ & 898 &  1.000& sta\\
14 & 0.44803 & $-0.3322$ & 898 &  1.000& sta\\
15 & 0.45983 & $-0.0492$ & 898 &  1.000& cen\\
16 & 0.47146 & $ 0.2300$ & 898 &  1.000& cen\\
17 & 0.48341 & $ 0.5169$ & 898 &  1.000& end\\
\sout{18}$^f$  & 0.49531 & $ 0.8025$ & 637 &  1.000&end \\
19 & 0.50563 & $ 1.0502$ & 898 &  0.998&egr \\
20 & 0.51655 & $ 1.3122$ & 898 &  0.607&egr \\
\sout{21}$^f$  & 0.52355 & $ 1.4802$ & 220 &  0.136&egr \\
22 & 0.53106 & $ 1.6605$ & 898 &  0.001& \\
23 & 0.54395 & $ 1.9699$ & 898 &  0.000& \\
24 & 0.55577 & $ 2.2535$ & 898 &  0.000& \\
25 & 0.56778 & $ 2.5418$ & 898 &  0.000& \\
26 & 0.57961 & $ 2.8257$ & 898 &  0.000& \\
27 & 0.59136 & $ 3.1076$ & 898 &  0.000& \\
28 & 0.60348 & $ 3.3985$ & 898 &  0.000& \\
\hline
\end{tabular}
\tablefoot{\tablefoottext{a}{Barycentric Julian Date minus $2458462$~d corresponding to center of observation},
\tablefoottext{b}{Time from center of optical transit}, \tablefoottext{c}{Exposure time}, \tablefoottext{d}{Time-averaged overlap fraction of planetary with stellar disk}, \tablefoottext{e}{Section: pre-transit (pre), ingress (ing), start (sta), center (cen), end (end), egress (egr)}, \tablefoottext{f}{Disregarded because of technical problems.}}
\end{table}

\section{Telluric correction}
\label{sec:TC}
The spectral regions of the optical H$\alpha$ and infrared \hei\ triplet lines are affected by water
absorption lines. The infrared is additionally contaminated by OH emission lines. All of these must
be corrected to obtain reliable results from transmission spectroscopy.   

\subsection{Absorption lines}
\begin{figure}[ht]
        \includegraphics[width=0.49\textwidth]{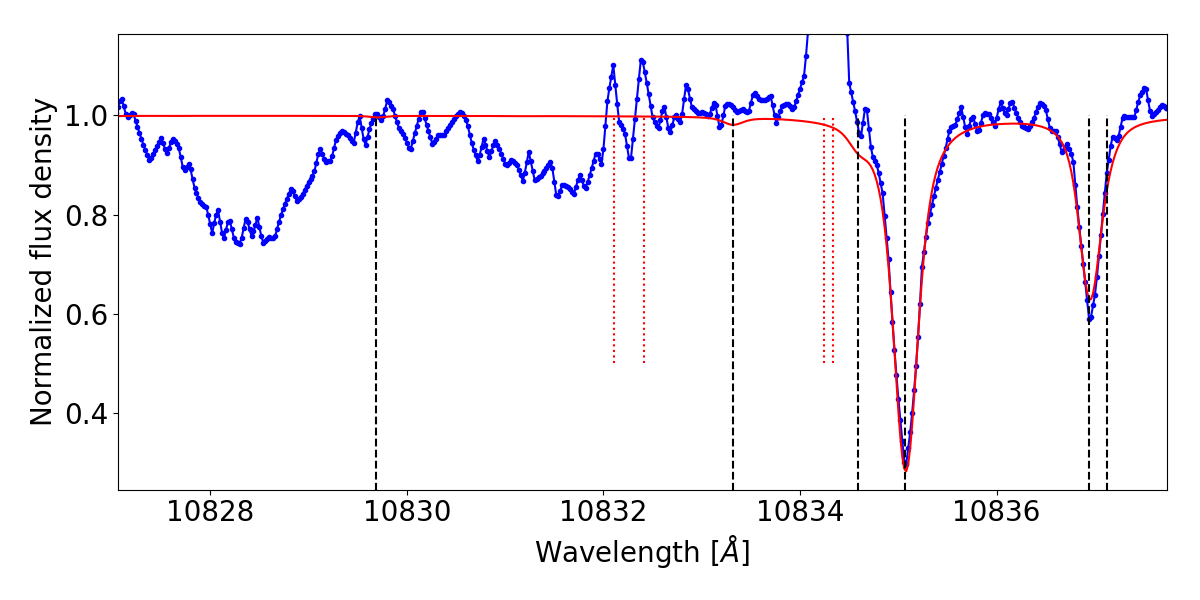}
    \caption{Excerpt of the first spectrum on night~1 (blue dots) along with telluric absorption model (solid red)
    and locations of relevant water absorption lines (vertical dashed black). Hydroxide emission lines are 
    indicated by shorter, dotted red lines.
    \label{fig:h2osp}}
\end{figure}

The telluric transmission spectrum in the region of the \hei\ lines is dominated by water absorption lines.
In Fig.~\ref{fig:h2osp}, we plot as an example the first spectrum obtained on night~1 along with a telluric
transmission model obtained with \mf. The locations of the most relevant water absorption lines
at 10\,829.69, 10\,833.32, 10\,834.59, 10\,835.07, 10\,836.94, 10\,837.12~\AA\ are indicated
\citep[see also][for the line data bases]{aer, HITRAN2016}.

\begin{figure}[ht]
        \includegraphics[width=0.49\textwidth]{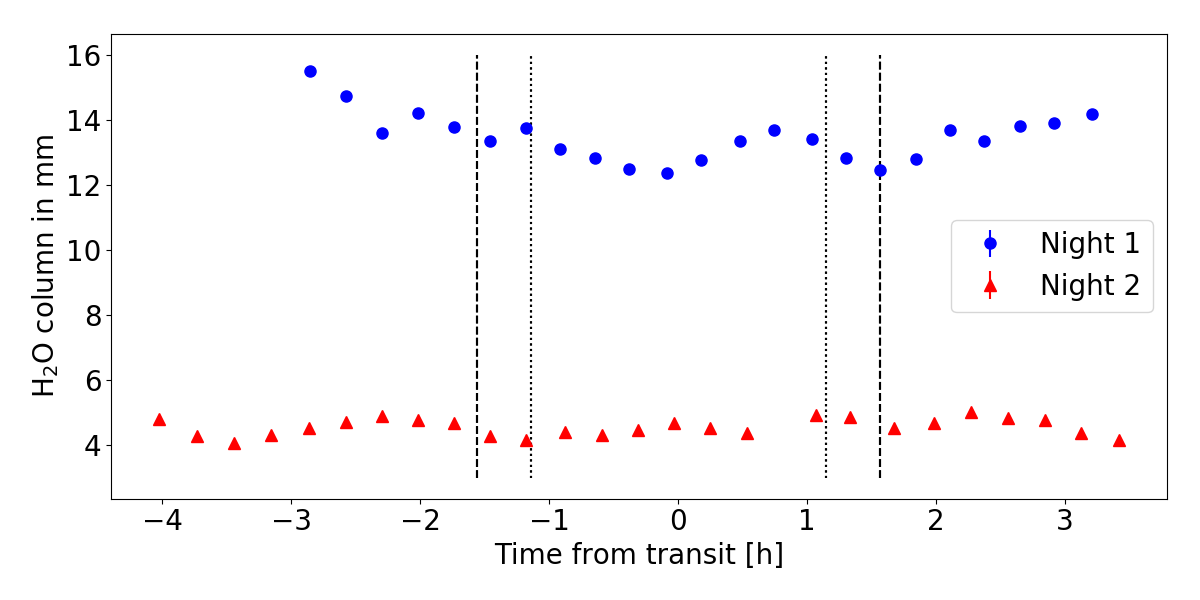}
    \caption{The depth of the water column obtained with \texttt{molecfit} by fitting the NIR channel data as a function of time.
    \label{fig:h2ocol}}
\end{figure}

We fitted the depth of the atmospheric water vapor column using several parts of the observed spectrum using \mf.
Stellar lines are not a serious interference in the NIR in \hp, because they are sparse and broad. We then obtained
the model telluric transmission spectrum using \mf\ and divided by it to correct for the water vapor contamination.
In Fig.~\ref{fig:h2ocol}, we show our results for the depth of the atmospheric water vapor column as a function of
time. Clearly, the first night is more heavily affected by telluric water absorption than the second. In both nights,
the depth of the water column remains rather stable during the respective observing run.

\subsection{Emission lines}
\begin{figure}
            \includegraphics[width=0.49\textwidth]{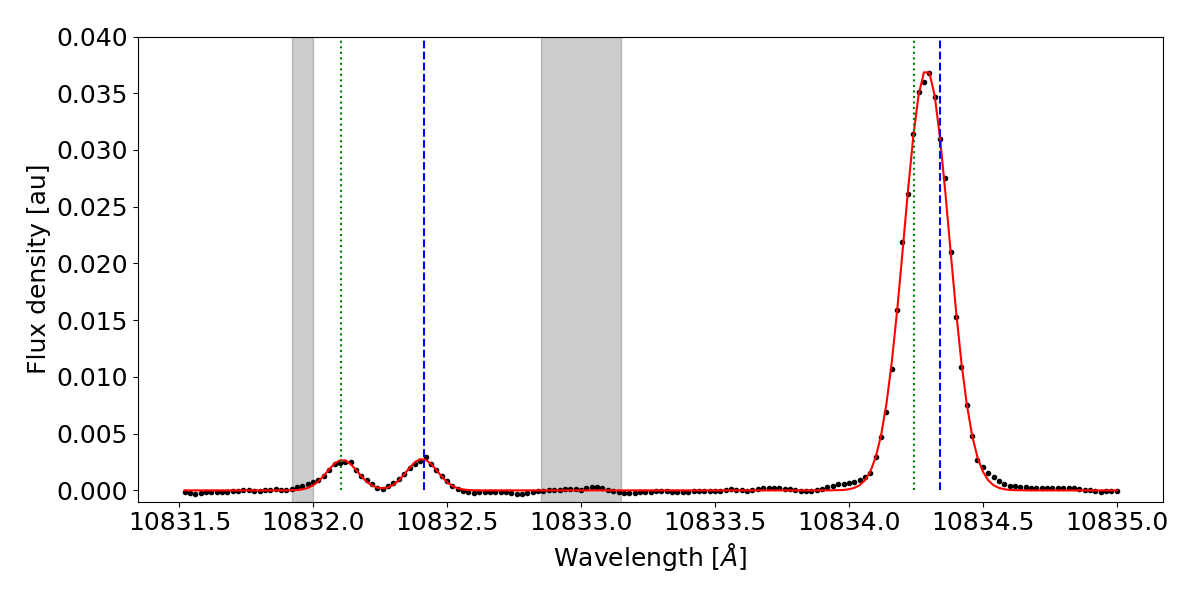}
    \caption{Mean sky spectrum during night~1 (black points) along with best-fit model (solid red line)
    and wavelengths of the Q$_1$ and Q$_2$ doublet lines according to \citet{Oliva2015}
    (f-components green dotted, e-components dashed blue vertical lines). Gray-shaded regions are exempted from the fit.
    \label{fig:meansky}}
\end{figure}

In the vicinity of the \hei\ lines, four emission lines of hydroxide (OH) are prominent \citep[][]{Oliva2015}, the vacuum wavelengths
and identifications of which are reproduced in Table~\ref{tab:OHlines}. We here use the same nomenclature for line
identification as \citet{Phillips2004}.
In the table, the lines are arranged in two rows, each representing the lines of
a so-called $\Lambda$-doublet, resulting from spin--orbit coupling. Their components are designated by letters e and f, indicating
lower state parity.
All lines considered here derive from a transition between vibrational states with quantum numbers five and two.
Both doublets belong to the Q-branch, where transitions
do not involve a change in the rotational quantum number.
The Q-subscript indicates the spin quantum number and
the second number in parenthesis the orbital angular momentum, pertaining to both the upper and lower state.
In Fig.~\ref{fig:meansky}, we show the mean sky spectrum observed by the sky-fiber during the night~1 run
along with the line locations listed in Table~\ref{tab:OHlines}. 
While the Q$_2$ doublet is easily resolved by \carm, the Q$_1$ doublet is not.

\begin{table}
\caption{Telluric emission lines of OH with line identifications as specified by \citet{Oliva2015}.
\label{tab:OHlines}}
\begin{tabular}{l l l l} \hline\hline
Wavelength & Line & Wavelength & Line \\
$[$\AA$]$ & identification & [\AA] & identification \\ \hline
10832.412 & (5-2)Q$_2$(0.5)e & 10832.103 & (5-2)Q$_2$(0.5)f \\    
10834.338 & (5-2)Q$_1$(1.5)e & 10834.241 & (5-2)Q$_1$(1.5)f \\ \hline
\end{tabular}
\end{table}

\subsubsection{Emission line behavior}
\label{sec:elb}

To produce a correction scheme for the sky emission lines, we studied the strength and evolution of the
three resolved OH components using a total of 1836 \carm\ sky-fiber spectra obtained in the context of
our two observing runs and several other observing runs obtained by the \carm\ consortium. 
To model the sky emission spectrum around the \hei\ triplet lines,
we used a model consisting of three Gaussian components and an offset parameter, which
was fitted to all individual sky-emission spectra. 
In the fit, we vary the
areas of the three Gaussians independently and adapt the offset parameter. While we assumed the instrumental width for the
resolved lines of the Q$_2$ doublet, the width of the component representing the blended Q$_1$ doublet was left as a free
parameter. We also allowed for a small shift in the line wavelengths.
Although the CARMENES pipeline \citep[][]{Zechmeister2018} is excellent in dealing with detector defects,
we chose to mask the two gray-shaded spectral regions indicated in Fig.~\ref{fig:meansky}
in the fit because they can be affected by known bad pixels.  
As an example, we also show the best-fit model to the mean night~1 sky spectrum in Fig.~\ref{fig:meansky}.

\begin{figure}
        \includegraphics[width=0.49\textwidth]{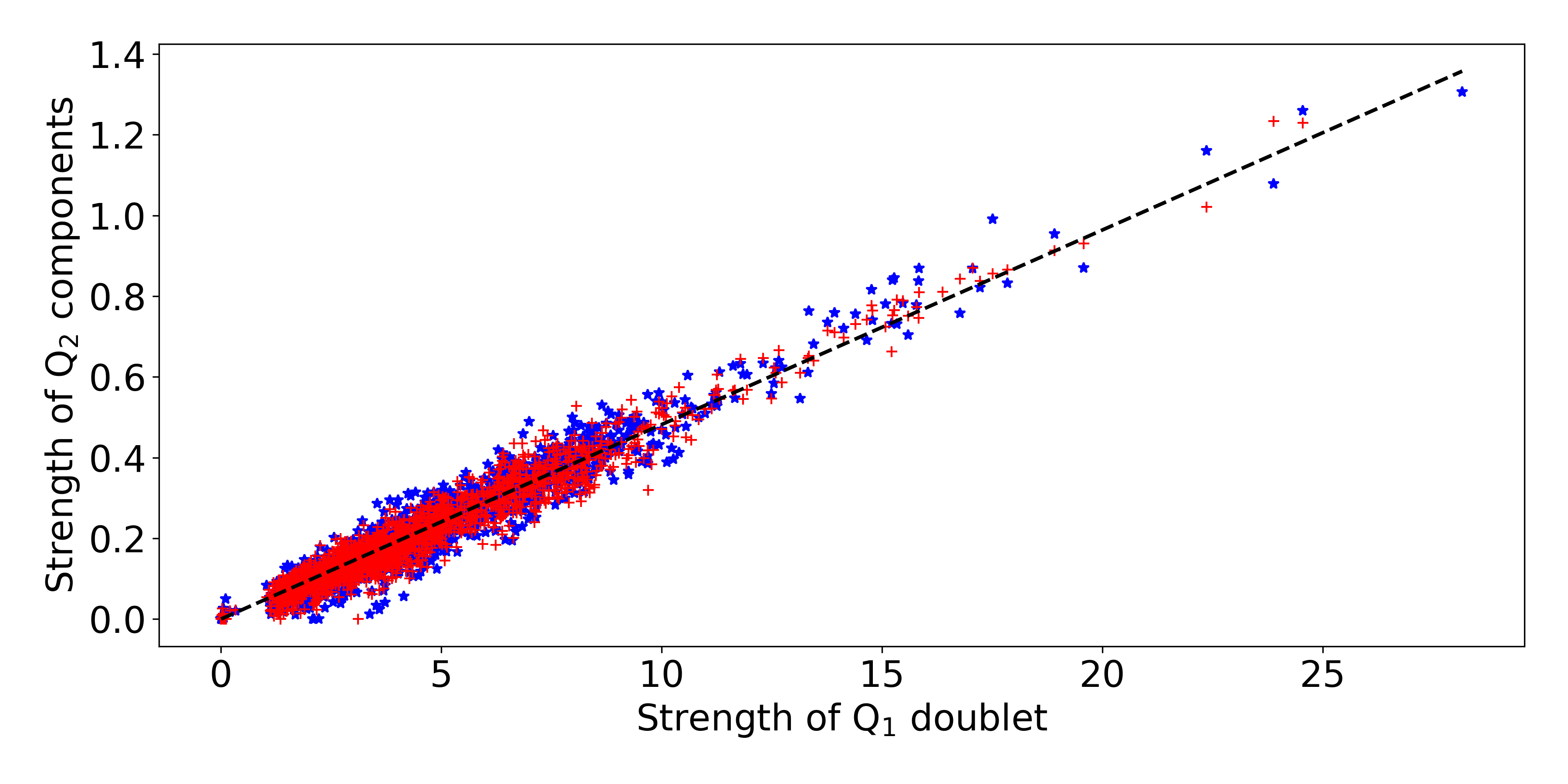}
    \caption{Ratio between strengths of the Gaussian model components representing the Q$_{2}$e (blue stars) and Q$_2$f (red pluses) lines and
    the strength of the combined Q$_1$ components along with best-fit relation (dashed black). 
    \label{fig:slr}}
\end{figure}

In Fig.~\ref{fig:slr}, we show the fitted strengths of the individual Q$_2$ emission lines as a function of that of the Q$_1$
line components. Prior to the analysis, a small number of 13 outliers were rejected based on the Generalized Extreme Studentized
Deviate (ESD) test \citep{Rosner1983}.
The e and f components of the doublets arise from spin--orbit coupling. They should therefore behave essentially
identically, which is consistent with our findings. The two considered $\lambda$-doublets also arise from closely related transitions
and are therefore expected to behave similarly. Our results are consistent with a linear relation between the strength of the
$Q_1$ doublet and the Q$_2$ doublet components. Thus, we obtained the best-fit ratio by regression through the origin.
The resulting ratio,
\begin{equation}
        S_{\rm Q_2\; e,f} =  (0.0482 \pm 0.0002) \times S_{\rm Q_1} \; ,  
\end{equation}
is also shown in Fig.~\ref{fig:slr}. 
The uncertainty was derived using the Jackknife method \citep{Efron1981}.
The strength of the combined Q$_1$ doublet lines is larger than its Q$_2$ equivalent by about an order of magnitude,
which is consistent with the results obtained by \citet{Oliva2015}.

Similar ratios between OH emission lines can be used to study the conditions of the upper atmosphere
\citep{Phillips2004}. However,
the ratio derived above is biased with respect to the physical value because the spectral sampling is not
taken into account in the modeling. The ratio is appropriate in the correction of our science spectra, because they
were treated identically.

\subsubsection{Emission line correction}

To correct the emission line contribution in the science spectra, we made use of the modeling in Sect.~\ref{sec:elb}.
The unresolved Q$_1$ doublet lines are prominent also in the science spectra (Fig.~\ref{fig:h2osp}). As \hp\ is
a fast rotator, the strength of the comparatively narrow Q$_1$ doublet can be determined by a Gaussian fit with relative ease
also in the science spectra. Based on this value and the now known relation between the strengths of the Q$_1$ and weaker
Q$_2$ components, we constructed a model emission spectrum based on the Gaussian modeling and subtract it from the science spectrum.

\begin{figure}[ht]
        \includegraphics[width=0.49\textwidth]{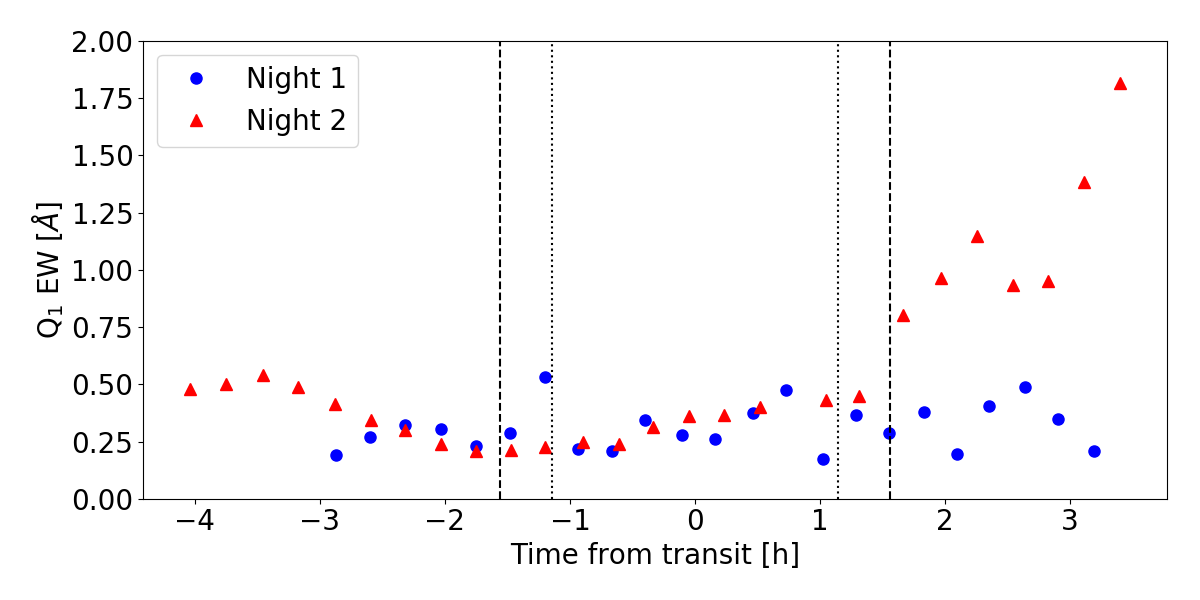}
    \caption{Equivalent width of the Q$_1$ doublet as a function of time.
    \label{fig:q1ew}}
\end{figure}

\begin{figure}[ht]
        \includegraphics[width=0.49\textwidth]{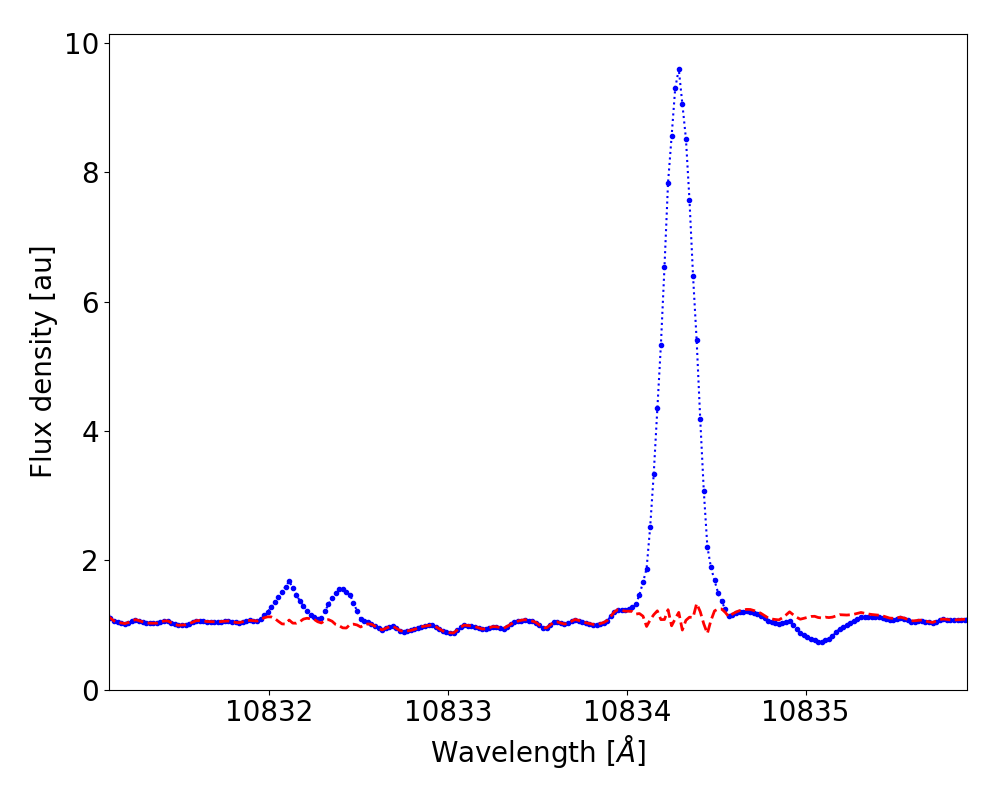}
    \caption{Excerpt of normalized last spectrum from night~2 before (blue, dotted) and after (red, dashed)
    correction for OH emission lines. The correction for telluric absorption lines is also included.
    \label{fig:emlc}}
\end{figure}

The resulting EW of the Q$_1$ component during nights~1 and 2 is shown in Fig.~\ref{fig:q1ew}. Until about
fourth contact, the OH contamination of the spectra is relatively similar on nights~1 and 2.
After the transit on night~2, the EW of OH emission rises and has approximately tripled at the end of
the observing run compared to its start. In Fig.~\ref{fig:emlc}, we show the effect of the removal of a telluric
emission lines using the most heavily contaminated spectrum, viz., the last spectrum of night~2, as an example.
This figure also displays an example for the removal a telluric water absorption line at $10\,835.07$~\AA.
Although the emission lines are very prominent in the spectrum, the correction reduces them substantially.
Nonetheless, some residuals remain (e.g., Sect.~\ref{sec:tlc}).

\section{Reference spectra}
\label{sec:refspec}
To study the presence of an additional absorption component superimposed on the stellar spectrum during
transit, a reference spectrum representing the stellar spectrum, $F_{\rm r}$ , needs to be
known as accurately as possible for comparison.
As the stellar spectrum may be variable, for example, because of stellar activity \citep[e.g.,][]{Klocova2017}
and the observed spectrum is
affected by variable telluric contamination, spectra obtained near the transit event under consideration were preferred
for the construction of such a reference. The combination of several spectra
increases the S/N of the reference. Incorporation of spectra obtained prior and past the time range is often desirable
to diagnose spectral changes and potentially counterbalance small changes in the observed spectrum evolving
linearly in time.

In the following, $\mathcal{S}_r$ denotes a suitable subset of the observed spectra and $f_i(\lambda)$ refers to
the normalized flux density after the barycentric velocity correction, the telluric correction,
and possible further pertinent treatment.  
We proceeded to construct a reference spectrum by
averaging the resulting spectral flux density 
\begin{equation}
    F_{\rm r} = \frac{1}{\#\mathcal{S}_r} \sum_{i \in \mathcal{S}_r} f_i \; ,
\end{equation}
where $\#\mathcal{S}_r$ is the number of spectra in the set. If the RV shift of the star is relevant across
the subset $\mathcal{S}_r$ or considerable changes in the S/N take place, such effects may also have to be taken into
account, but we consider this unnecessary here.

\subsection{Reference spectrum for night~1}
\label{sec:rsn1}
\begin{figure}[ht]
        \includegraphics[width=0.49\textwidth]{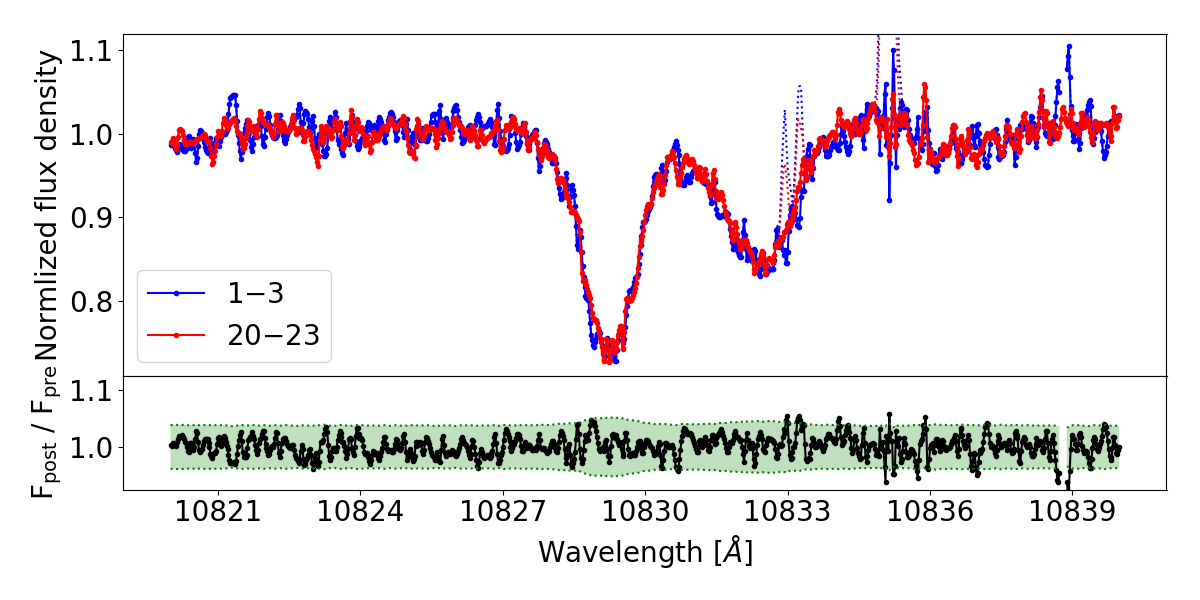}\\
    \includegraphics[width=0.49\textwidth]{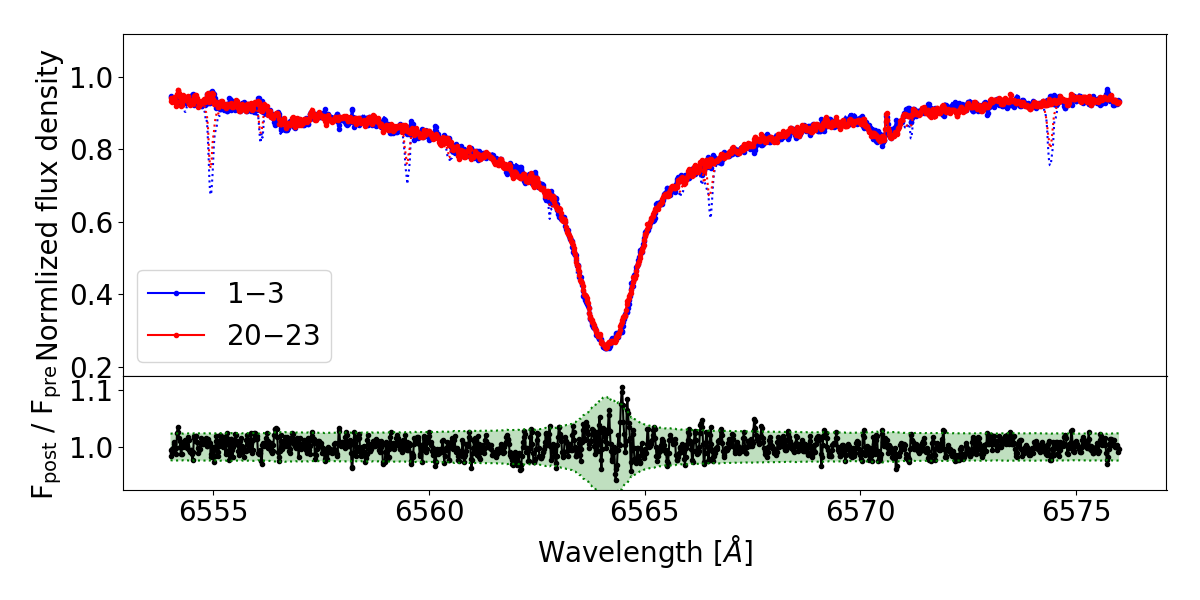}
    \caption{Comparison of pre- and post-transit reference spectra.
    Top plot, top panel: Averaged, normalized pre-transit spectra (nos. 1,2,3) and post-transit spectra (nos. $20-23$)
    around the \hei\ lines during night~1.
    Dotted lines indicate the averaged spectra obtained without our telluric emission line correction. Top plot, bottom panel
    shows the ratio of the averaged post- and pre-transit spectra along with an estimate of the $95$\,\% confidence range
    resulting from noise (green shades). The bottom panels show the same for the \ha\ line region.
    \label{fig:rspec_n1}}
\end{figure}

On night~1, a suitable reference spectrum could be constructed by
averaging the spectra with numbers $1-3$ and $20-23$ as defined in Table~\ref{tab:logn1}, which
cover the exposures obtained 
about $2.2$~h or more before and after the transit. This choice incorporates
pre- and post-transit coverage and uses spectra comfortably set apart in time from the actual
optical transit.
In Fig.~\ref{fig:rspec_n1}, we show the reference spectra in the spectral region around
the \ha\ and \hei\ lines separately for the
indicated ranges of pre- and post-transit spectra. Additionally, we show their ratio
and estimated $95$\,\% confidence range of variation resulting from noise.  
Naturally, the core of the deep \ha\ line suffers from an elevated noise level caused by
the flux depression.
We found the two ratios to be consistent within the expected noise variation. Therefore, we used
the combined spectra as final reference spectra ($\mathcal{S}_{r, \rm night~1} = \{1,2,3,20,21,22,23\}$).

\subsection{Reference spectrum for night~2}
\label{sec:rsn2}
On night~2 we have more pre- and post-transit coverage in our data set. Unfortunately,
the post-transit NIR data suffer from two deteriorating and likely related effects, viz.,
a decrease in S/N in the stellar spectrum, accompanied by a strengthening of telluric emission
compared to the stellar signal. This is not the case for the VIS data.

\begin{figure}[ht]
        \includegraphics[width=0.49\textwidth]{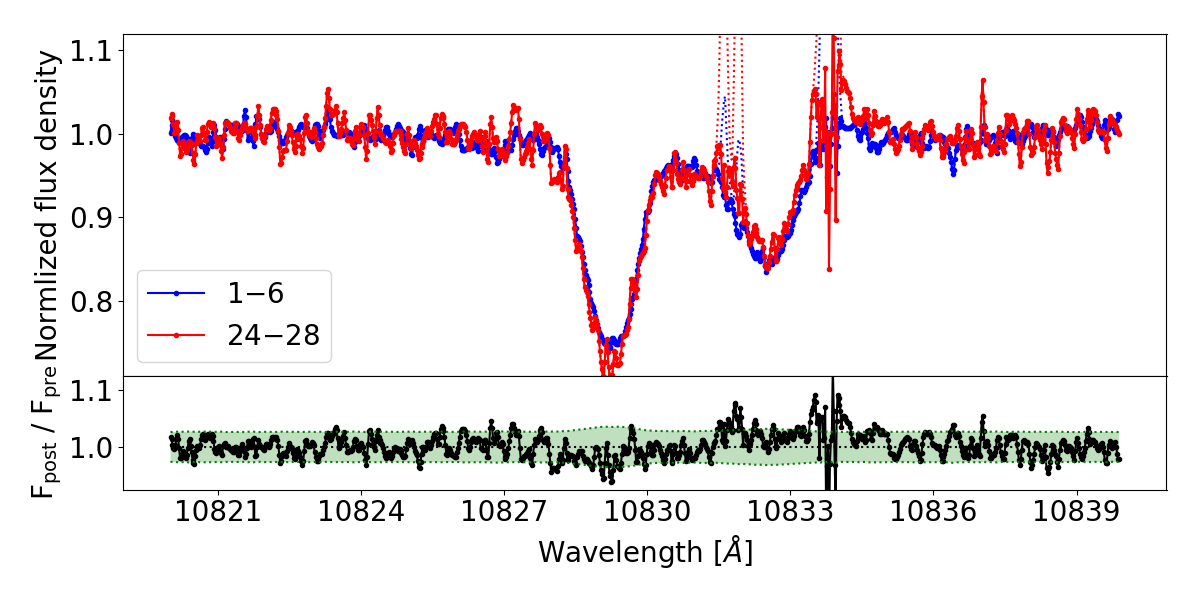}\\
    \includegraphics[width=0.49\textwidth]{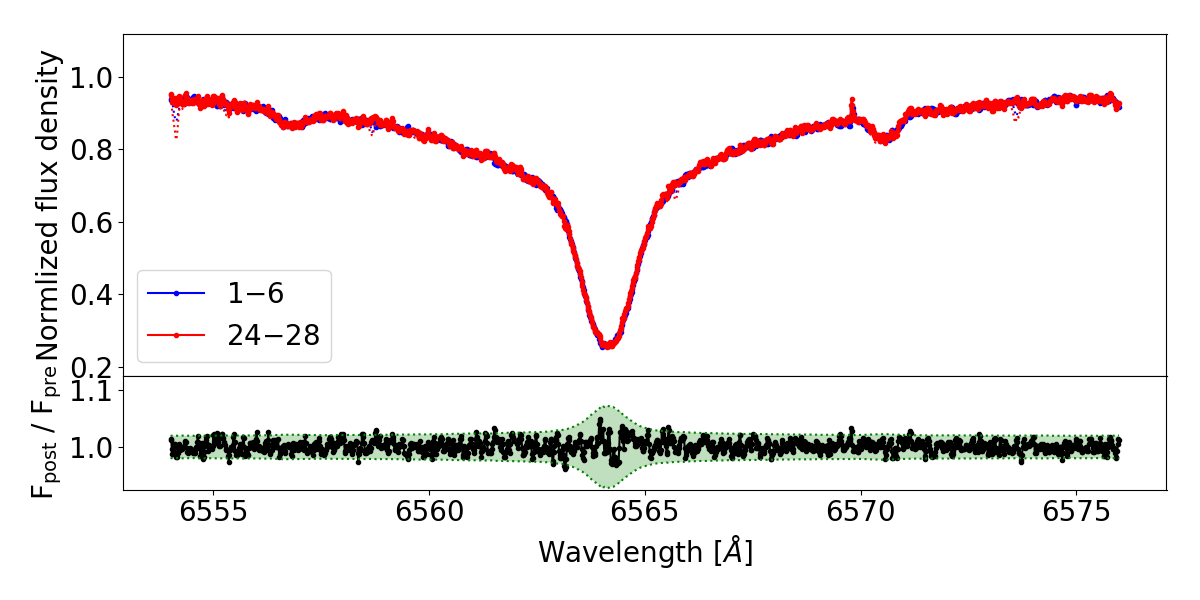}
    \caption{As Fig.~\ref{fig:rspec_n1} but for night~2.
    \label{fig:rspec_n2}}
\end{figure}

Combining the spectra with the log numbers $1-6$ for the pre-transit and $24-28$ for the post-transit
reference fulfills about the same timing requirements as used on night~1, each range being
offset by more than $2.2$~h from the actual optical transit center. In Fig.~\ref{fig:rspec_n2},
we again show the resulting reference spectra along with their ratio around the \ha\ and the \hei\
lines. The situation around the \ha\ line is satisfactory. Around the \hei\ lines, we found a less
satisfying situation. The effect of the lower S/N obtained post-transit was clearly reflected by
larger scatter in the associated reference spectrum.
The ratio of the pre- and post-transit spectra shows systematic
deviations certainly related to the telluric emission lines and perhaps other effects. We note that
the emission line corresponding to the Q$_1$ doublet reaches a peak height of around $6.5$ in the
post-transit reference compared to about $2.5$ in the pre-transit reference. It appears that the
line develops non-Gaussian wings, which were not corrected at the required level. Also, the \ion{Si}{i}
line shows deviations of unknown origin. As neither the \ha\ line shown in Fig.~\ref{fig:rspec_n2}
nor the \cairt\ lines provide evidence for significant variation related to activity, we decided to
only use the pre-transit reference for the spectral region around the \hei\ lines, while the pre-
and post-transit spectra can be used for the \ha\ line, in particular, $\mathcal{S}_{r, \rm n2, NIR} = \{1,2,3,4,5,6\}$
and  $\mathcal{S}_{r, \rm n2, VIS} = \{1,2,3,4,5,6,24,25,26,27,28\}$.

\section{Treatment of phase smearing}
\label{sec:tops}

When $g(x)$ is a continuous distribution function with variance $\sigma_g^2$ and mean $\mu_g$
then
\begin{equation}
    \sigma_g^2 = \int x^2 g(x) \, dx - \mu_g^2 \; .
\end{equation}
Here, we take $g(x)$ to describe the profile of an instrumentally broadened spectral line. 
A new distribution, $f(x)$, is generated by shifting the original distribution between
$-\frac{T}{2}$ and $+\frac{T}{2}$ around its mean (i.e., by convolution with a box-like profile) so that
\begin{equation}
    f(x) = \frac{1}{T} \int_{-\frac{T}{2}}^{+\frac{T}{2}} g(x-t) \, dt \; .
\end{equation}
This is equivalent to a convolution with a box-shaped profile.
Symmetry implies identical mean values $\mu_g = \mu_f$.
The variance of $f(x)$ is given by
\begin{eqnarray}
    \sigma_f^2 &=& \int (x-\mu_f)^2 f(x) \, dx \\
    &=& \int (x-\mu_f)^2 \frac{1}{T} \int_{-\frac{T}{2}}^{+\frac{T}{2}} g(x-t) \, dt \, dx \\
    &=& \frac{1}{T} \int_{-\frac{T}{2}}^{+\frac{T}{2}} \int (x^2 - 2x\mu_f + \mu_f^2) g(x-t) \, dx \, dt \\
    &=& \frac{1}{T} \int_{-\frac{T}{2}}^{+\frac{T}{2}} \sigma_g^2 (\mu_g+t)^2 - 2\mu_g (\mu_g+t) + \mu_g^2 \, dt \\
    &=& \frac{1}{T} \int_{-\frac{T}{2}}^{+\frac{T}{2}} \sigma_g^2 + t^2 \, dt \\
    &=& \sigma_g^2 + \frac{T^2}{12} \; .
\end{eqnarray} 
The additional, second variance term can be absorbed in the instrumental resolution.
 
CARMENES provides a spectral resolution of $80\,400$, which corresponds to a line profile with
full-width at half-maximum (FWHM) of $3.7$~\kms\ 
in the NIR arm. For Gaussian profiles, the FWHM and the standard deviation, which is the square root of
the variance, are related by a factor of $2\sqrt{2\ln(2)} \approx 2.35$.
In the case of \hpb, 
phase smearing adds another $2.3$~km$^2$\,s$^{-2}$ to the variance.
The effective resolution, including phase smearing, becomes $58\,000$.
In the optical arm, an effective resolution of $63\,000$ applies.

\section{Uncertainty of transmission spectra}
\label{sec:TSunc}
To estimate the uncertainty of the data points in the combined transmission spectra
from night~1 and night~2, we used the $\beta\sigma$ estimator as given by \citet{Czesla2018}.

The transmission spectra are oversampled so that consecutive data points are not independent.
We find a mean autocorrelation scale, $\tau_a$, of $11$ data points in the transmission spectra. Using 
a ``jump parameter'' of 22 data points samples independent points to estimate the noise. Combined with
a second-order approximation of the underlying variation, we obtain the standard deviations, $\sigma$,
for the transmission spectra obtained for the \ha\ and \hei\ lines in the individual phases.
In estimating $\chi^2$ and BIC values, we use an effective number of independent data points given by
\begin{equation}
    N_{\rm eff} = \frac{N}{\tau_a} \; ,
\end{equation}  
where $N$ is the number of points in the oversampled transmission spectrum \citep{Bayley1946, Priestley1981}.

\begin{table}
\centering
\caption{Standard deviation, $\sigma$, of uncertainty for individual data points (in units of $10^{-2}$).
\label{tab:TSunc}
}
\begin{tabular}{l l l}
\hline\hline
     Phase & $\sigma_{\rm H\alpha}$ & $\sigma_{\rm \ion{He}{i}}$ \\
\hline
       pre & 0.569 & 0.753\\
   ingress & 0.364 & 0.546\\
     start & 0.359 & 0.449\\
   central & 0.447 & 0.722\\
       end & 0.391 & 0.554\\
    egress & 0.408 & 0.705\\
\hline
\end{tabular}
\end{table}

\section{Fit quality in phenomenological modeling}
\label{sec:phenoFitQuality}

\begin{table}
\centering
\caption{Fit results of phenomenological modeling ($(-100,+100)$~\kms\ region). $\chi^2$ and
reduced $\chi^2$ values for annulus (a), super-rotating wind model (w),
and up-orbit stream model (s).
\label{tab:fitResults}}
\begin{tabular}{l r r r}
\hline \hline
Phase & $\chi^2_a$/$\chi^2_{r,a}$ & $\chi^2_w$/$\chi^2_{r,w}$ & $\chi^2_s$/$\chi^2_{r,s}$ \\ \hline
\multicolumn{4}{c}{\ha} \\
pre & 781.1/2.7 &  757.8/2.6 &  788.1/2.7 \\
ingress & 1534.9/5.3 &  939.2/3.2 &  952.8/3.3 \\
start & 1175.9/4.0 &  1153.3/3.9 &  532.5/1.8 \\
center & 880.7/3.0 &  863.5/3.0 &  938.5/3.2 \\
end & 1239.6/4.2 &  1262.9/4.3 &  1316.7/4.5 \\
egress & 487.6/1.7 &  575.6/2.0 &  461.3/1.6 \\
\hline
\multicolumn{4}{c}{\hei} \\
pre & 1137.8/3.1 &  1054.0/2.9 &  678.4/1.9 \\
ingress & 1882.0/5.2 &  1099.8/3.0 &  752.4/2.1 \\
start & 1123.7/3.1 &  1103.4/3.0 &  473.7/1.3 \\
center & 597.9/1.7 &  587.8/1.6 &  638.4/1.8 \\
end & 1219.3/3.4 &  1204.0/3.3 &  1494.0/4.1 \\
egress & 682.0/1.9 &  661.0/1.8 &  606.2/1.7 \\
\hline
\end{tabular}
\end{table}

\begin{table}
\centering
\caption{Values of Bayesian information criterion for comparison of annulus model (a),
super-rotating wind model (w), and up-orbit stream model (s). Negative values indicate
preference for the first-mentioned model.
\label{tab:bic}}
\begin{tabular}{l r r r}
\hline \hline
Phase & BIC (w/a) & BIC (s/w) & BIC (s/a) \\ \hline
\multicolumn{4}{c}{\ha} \\
    pre & $  4.4$ & $  9.3$ & $  7.2$ \\
ingress & $-47.6$ & $  7.8$ & $-46.4$ \\
  start & $  4.5$ & $-49.9$ & $-51.9$ \\
 center & $  5.0$ & $ 13.4$ & $ 11.8$ \\
    end & $  8.7$ & $ 11.5$ & $ 13.6$ \\
 egress & $ 14.6$ & $ -3.8$ & $  4.2$ \\
\hline
\multicolumn{4}{c}{\hei} \\
    pre & $ -1.1$ & $-27.6$ & $-35.2$ \\
ingress & $-64.5$ & $-25.0$ & $-96.1$ \\
  start & $  4.7$ & $-50.7$ & $-52.5$ \\
 center & $  5.6$ & $ 11.2$ & $ 10.2$ \\
    end & $  5.2$ & $ 32.9$ & $ 31.5$ \\
 egress & $  4.6$ & $  1.6$ & $ -0.3$ \\
\hline
\end{tabular}
\end{table}

\section{Exchanging convolution and exponentiation}
\label{sec:gaussconv}
If $g(\lambda; \lambda_0, \sigma^2)$ denotes the normal density with mean $\lambda_0$ and variance $\sigma^2$,
we observe that for positive integer $n$
\begin{equation}
    g(\lambda; \lambda_0,\sigma^2)^n = c_n(\sigma^2) \times g\left(\lambda; \lambda_0, \frac{\sigma^2}{n}\right) 
    \label{eq:gn}
\end{equation}
where
\begin{equation}
    c_n(\sigma^2) = n^{-\frac{1}{2}} \left(2\pi\sigma^2 \right)^{-\frac{n-1}{2}} \; .
    \label{eq:cn}
\end{equation}
For positive, real $x$, the exponential function can be expanded into
\begin{equation}
    e^{-x} = 1 + \sum_{n=1}^{\infty} (-1)^n \frac{x^n}{n!} \; .
    \label{eq:ex}
\end{equation}
If the instrumental profile, $R$, is normal such that $R=g(\Delta \lambda; 0,\sigma_R^2)$ and the optical
depth profile, $\tau(\lambda)$, is
proportional to a normal density such that $\tau(\lambda) = A\,g(\lambda; \lambda_0,\sigma_{\tau}^2)$. This
implies that the optical depth at the line center, $d_{\tau}$, is $A\,(2\pi\stau^2)^{-1/2}$.
The optical depth profile, $\tau(\lambda)$, leads to an observed spectral line of the form $s(\lambda) = R * \exp(-\tau(\lambda))$,
where $*$ denotes the convolution.
As the result of the convolution of two normal functions is conveniently also normal,
the question as to when it is appropriate to absorb the instrumental profile in the optical depth arises. To that end,
we investigate the difference
\begin{equation}
    \delta(\lambda) = R*\exp\left(-\tau(\lambda) \right) - \exp(-R*\tau(\lambda)) \; .
\end{equation}
Combining Eqs.~\ref{eq:gn} and Eq.~\ref{eq:ex}, one obtains
\begin{eqnarray}
    \delta(\lambda) &=& \left(1+\sum_{n=1}^{\infty} (-1)^n\,A^n\frac{c_n(\sigma_{\tau})^2 g\left(\lambda_0,\frac{\stau^2}{n}+\sigma_R^2\right)}{n!} \right) - \nonumber \\
    & &  \left(1+\sum_{n=1}^{\infty} (-1)^n\,A^n \frac{c_n(\sigma_R^2+\stau^2) g\left(\lambda_0,\frac{\stau^2+\sigma_R^2}{n}\right)}{n!} \right) \; .
\end{eqnarray}
The first two terms (constant and linear) in the expansion cancel and, therefore,
\begin{eqnarray}
    \delta(\lambda) &=& \sum_{n=2}^{\infty} \frac{(-1)^n\,A^n}{n!} \left[ c_n(\sigma_{\tau}^2) g\left(\lambda_0,\frac{\stau^2}{n}+\sigma_R^2\right) \right. - \nonumber \\
    & & \left. c_n(\sigma_R^2+\stau^2) g\left(\lambda_0,\frac{\stau^2+\sigma_R^2}{n}\right) \right] \; .
\end{eqnarray}
Focusing on the line center, and defining $\sigma_R^2 = \alpha \stau^2$ with $\alpha>0$, we find
\begin{eqnarray}
    \delta(\lambda_0) &=& \sum_{n=2}^{\infty} \frac{(-1)^n\,A^n}{n!} \left[ c_n(\sigma_{\tau}^2) \left(2\pi\left(\frac{\stau^2}{n}+\sigma_R^2\right) \right)^{-\frac{1}{2}}  \right. - \nonumber \\
    & & \left. c_n(\sigma_R^2+\stau^2) \left(2\pi \left(\frac{\stau^2+\sigma_R^2}{n}\right)\right)^{-\frac{1}{2}} \right] \; \nonumber \\
    &=& \sum_{n=2}^{\infty} \left(\frac{A}{\sqrt{2\pi\stau^2}}\right)^n  \frac{(-1)^n}{n!} \left( \frac{1}{\sqrt{(1+n\alpha)}} - \frac{1}{(1+\alpha)^{\frac{n}{2}}} \right) \nonumber \\
    &=& \sum_{n=2}^{\infty} d_{\tau}^n \,  \frac{(-1)^n}{n!} \left( \frac{1}{\sqrt{(1+n\alpha)}} - \frac{1}{(1+\alpha)^{\frac{n}{2}}} \right) \label{eq:deltacenter} \; . 
\end{eqnarray}
The sum in Eq.~\ref{eq:deltacenter} consists of terms with alternating sign, as the term in parentheses
is always positive. The sum converges by construction.
Although the summands tend to decrease quickly in absolute
value, owing to the inverse dependence on the factorial of $n$, monotonous decrease, which is
required for convergence, does not necessarily set in immediately. Nonetheless, a good approximation can
usually be obtained, considering only a few leading summands.

In the case of $d_{\tau} = 1$ and $\alpha = 0.1$, which is typical of values encountered in this paper, the
difference $\delta(\lambda_0)$ is about $10^{-3}$.

\end{document}